\documentclass[12pt, journal, onecolumn, draftcls]{IEEEtran}
\usepackage[utf8]{inputenc}
\usepackage{amsmath}
\usepackage{amsthm}
\usepackage{mathtools}
\usepackage{amssymb}
\usepackage{dsfont}
\usepackage{stmaryrd,color}
\usepackage{array}

\newtheorem{theorem}{Theorem}
\newtheorem{lemma}{Lemma}
\newtheorem{remark}[theorem]{Remark}
\newtheorem{definition}[theorem]{Definition}
 \newtheorem{proposition}{Proposition} 
 \newtheorem{corollary}{Corollary}
 
\newcommand{\mkv}{-\!\!\!\!\minuso\!\!\!\!-}
\newcommand{\argmax}[1]{\underset{#1}{\operatorname{argmax}}} 
\newcommand{\cB}{{\mathcal B}}
\newcommand{\cC}{{\mathcal C}}

\newcommand{\cJ}{{\mathcal J}}

\newcommand{\cL}{{\mathcal L}}
\newcommand{\cM}{{\mathcal M}}
\newcommand{\cN}{{\mathcal N}}
\newcommand{\cO}{{\mathcal O}}
\newcommand{\cP}{{\mathcal P}}
\newcommand{\cQ}{{\mathcal Q}}
\newcommand{\cR}{{\mathcal R}}
\newcommand{\cS}{{\mathcal S}}
\newcommand{\cT}{{\mathcal T}}
\newcommand{\cU}{{\mathcal U}}
\newcommand{\cV}{{\mathcal V}}
\newcommand{\cW}{{\mathcal W}}
\newcommand{\cX}{{\mathcal X}}
\newcommand{\cY}{{\mathcal Y}}
\newcommand{\cZ}{{\mathcal Z}}

\newcommand{\dsE}{{\mathds E}} 
\newcommand{\bN}{{\mathds N}} 
\newcommand{\dsP}{{\mathds P}} 
\newcommand{\dsT}{{\mathds T}}

\newcommand{\bB}{{\mathbf B}}
\newcommand{\bh}{{\mathbf h}} 
\newcommand{\bg}{{\mathbf g}}
\newcommand{\bK}{{\mathbf K}}
\newcommand{\bx}{{\mathbf x}}
\newcommand{\bX}{{\mathbf X}}

\newcommand{\typ}[1]{T_\delta^n(#1)} 

\newcommand{\toas}[1]{\xrightarrow[#1]{}}

\providecommand{\abs}[1]{\left|#1\right|}

\providecommand{\norm}[1]{\lVert#1\rVert}

\begin{document}
\sloppy

\title{\huge On the Compound Broadcast Channel:  Multiple Description Coding and Interference Decoding} 

\author{
  \IEEEauthorblockN{Meryem~Benammar and Pablo~Piantanida and Shlomo~Shamai~(Shitz)}\\
\thanks{The material in this paper was partially published in the IEEE Information Theory Workshop, Seville, September 9-13, 2013 and  in the IEEE Information Theory Workshop, Tasmania, Australia,  2-5 November 2014. This research was partially supported by the FP7 Network of Excellence in Wireless COMmunications NEWCOM\#. }
\thanks{Meryem~Benammar and Pablo~Piantanida are with the
Department of Telecommunications, SUPELEC, 91192 Gif-sur-Yvette, 
France (e-mail: \{meryem.benammar, pablo.piantanida\}@supelec.fr).}
\thanks{Shlomo Shamai (Shitz) is with the Department of Electrical Engineering, Technion, Haifa, Israel, 
 (e-mail: sshlomo@ee.technion.ac.il).}
 \thanks{}
 }

\maketitle 

\begin{abstract}
This work investigates the general two-user Compound Broadcast Channel (BC) where an encoder wishes to transmit common and private messages  to two receivers while being oblivious to two possible channel realizations controlling the communication. The focus is on the characterization of the largest achievable rate region by resorting to more evolved \emph{encoding} and \emph{decoding} techniques than the conventional coding for the standard BC.  The role of the decoder is first explored, and an achievable rate region is derived based on the principle of ``Interference Decoding" (ID) where each receiver decodes its intended message and chooses to (non-uniquely) decode or not the interfering message. This inner bound is shown to be capacity achieving for a class of non-trivial compound BEC/BSC broadcast channels while the worst-case of Marton's inner bound --based on ``Non Interference Decoding" (NID)-- fails to achieve the capacity region. The role of the encoder is then studied, and  an achievable rate region is derived  based on ``Multiple Description" (MD) coding  where the encoder  transmits a common as well as multiple dedicated private descriptions to the many instances of the users channels.  It turns out that MD coding outperforms the single description scheme --Common Description (CD) coding-- for a class of compound Multiple Input Single Output Broadcast Channels (MISO BC). 
\end{abstract}

\section{Introduction} 
The two-user Broadcast Channel (BC) --as first introduced by Cover in~\cite{CoverBroadcastChannels72}-- consists of an encoder  transmitting both a common message and two private messages to two users. Following this seminal work, intensive research was undertaken to characterize the capacity region of this setting for which the key feature in designing optimal codes is to allow for an efficient interference mitigation. In this work, we study the general two-user Compound BC where an encoder wishes to communicate one common message and two private messages to two users who can each observe one of many output channel statistics. The actual channel controlling the communication is unknown at the transmit side but assumed to remain constant during the communication and belongs to a known set of possible channels. Coding successfully for such a setting requires that the encoder must guarantee --whatever the channel realizations-- reliable communication. Thus, it is well understood that the compound BC is equivalent to a BC with multiple users and common information. Our aim is to improve the understanding of how interference should be dealt within the current setting where both channel uncertainty and interference are coupled. To this end, we study alternative encoding and decoding techniques to the usual coding schemes that were proved to be capacity achieving for some broadcast channels.

Let us first briefly discuss the optimal coding schemes for the two-user BC, reported also partly in \cite{CoverComments}. Although the capacity region of the BC still remains an open problem to this day, Marton established in~\cite{MartonInnerBound79} an inner bound on the general two-user BC based on the notion of \emph{random binning} and \emph{superposition coding} with common and private messages, commonly referred to as ``Marton's coding". This inner bound remains the best hitherto known in literature while the best outer bound on the capacity region of the BC is due to Nair \& El Gamal~\cite{NairELGamal2006}. These two bounds were shown to coincide for several classes of ``ordered" channels, citing here: degraded, less noisy, and more capable BCs (see~\cite{Gamal1979LNMCDEG} and references therein) and more recently~\cite{NairBECBSC2010}, for essentially less noisy and essentially more capable BCs, the key feature being the use of \emph{superposition coding} as an encoding strategy. Marton's inner bound also proved to be capacity-achieving for some non-ordered channels: the deterministic and semi-deterministic BC in~\cite{MartonInnerBound79} and \cite{GelPins1980OneDetrministicComponent}, the MIMO BC in \cite{WeingartenMIMO} while the capacity region of a BC consisting of the product and sum of two unmatched channels is also reported in~\cite{ElGamalProductSum}. In these cases, it is \emph{random binning} that proves to be crucial for interference management. 

In the above mentioned works, the channel statistics are perfectly known to the transmitter and thus the encoder can exploit this knowledge to allow for an efficient interference mitigation scheme. In all the cases where Marton's inner bound is tight, the construction of the optimizing auxiliary code depends on the prior knowledge of either the channel output statistics (e.g. deterministic and semi-deterministic BCs) or a function of these statistics (e.g. users' ordering in ordered single antennas BCs). When the encoder is oblivious to any such information about the channel state --no channel state information (CSIT)--, the effect of interference coupled with channel uncertainty on Marton's coding technique can be more stringent. This arises the necessity to explore encoding and decoding schemes that are powerful enough to deal with the effects of channel uncertainty. 
 
 \subsection{Related Work}
 
It is worth mentioning here that few works dealt lately with alternate decoding techniques. We cite here first~\cite{Bandemer2012}, where the authors characterized the maximum rate region for general interference networks under a given code constraint. This work generalizes the technique of ``Interference Decoding" (ID), which was already used in~\cite{Bacelli}, and consists in an alternate strategy for treating interference at receive terminals. More precisely, ID combines \emph{non-unique decoding} with the possibility at each receiver to decode or not the interfering  messages intended to the others users. As a matter of fact, the gain of ID does not result from non-unique decoding~\cite{Bidokhti2014} as much as it follows from decoding interference. Yet, the straight-forward extension of the results of this work \cite{Bandemer2012} to the BC is not strong enough for it encompasses only \emph{superposition coding} but not \emph{random binning}. Nevertheless, it provides an interesting insight on how to recover a \emph{superposition coding} like inner bound with alternative decoding strategies, while keeping a symmetric encoding which will be useful for ordered channels. 

Later, authors in~\cite{CosetCodes} derived an inner bound based on ``Coset Codes" for the three users BC possibly enlarging the best-known known inner bound. Coset codes are \emph{structured codes} that allow the destinations to decode a ``compressive" function of the interfering messages and thus  a complete cancellation of interference with less impediment to the information rates than fully decoding the interfering messages. A class of 3 users BCs is proposed where two links are interference free and for which the straightforward extension of Marton's coding scheme, stays strictly suboptimal compared to the suggested rate region. Such a coding technique based on Coset Codes, proves to be useful for three user BC, however, it does not enlarge Marton's inner bound in the two user's case. Yet this work presents the first class of 3 users BC for which Marton's inner bound, with many common layers, is strictly sub-optimal. 
 
When the channels are not ordered, e.g, MISO BC, the effect of channel uncertainty on the ``Degrees of Freedom'' (DoF) --insightful to understand how interference should be managed with no CSIT-- is rather well understood. For finite state compound settings, Weingarten~\emph{et al.} had first derived both inner and outer bounds on the DoF region and on the sum-DoF of the compound MISO BC~\cite{WeingartenDoF} with some cases of optimality. The outer bound derived therein was conjectured to be loose, but later Gou~{\it{et.al}}~\cite{Gou2011} and Maddah-Ali~\cite{MaddahAli} proved the optimal DoF region of the generic compound MISO BC, both in the complex and in the real settings, to perfectly match this outer bound. The achievability of the optimal DoF relies on either a Linear or a non-Linear coding scheme combined with ``symbol extensions" in~\cite{WeingartenDoF} while the proof made in~\cite{MaddahAli} resorts to number theory tools and consists in interference alignment over rational dimensions of the real numbers (see also~\cite{JafarBook}). When the states span an infinite set, i.e., in the ergodic setting, DoF can experience severe loss. In~\cite{Huang2012}, it is shown that with Rayleigh fading channels, the sum-DoF collapses to the number of transmit antennas: time-sharing is optimal. A few more works deal with alternate settings where various models of the amount and accuracy of CSI available at the transmitter are considered, e.g.~\cite{Tandon2013}. It turns out that richer encoding strategies, like \emph{Interference Alignment} (IA) along with block expansion (coding over many time slots) are crucial in dealing with interference, and thus, any optimal scheme for the finite power limited MISO BC should encompass such coding strategies. 

 
 \subsection{Our Contribution}
In this work, we explore the role that two main interference mitigation techniques can play in the compound BC setup, and show that, by operating clever optimization either on the encoding or on the decoding side, we can alleviate the effect of uncertainty when coupled with interference in two different ways. We first start by deriving a rate region that takes advantage of the combination of each of ID, Marton's \emph{random binning}, and \emph{superposition coding}. We prove that for the compound BC --unlike the standard two-user BC-- ID can strictly outperform its antagonist ``simpler" strategy, i.e., ``Non Interference Decoding" (NID). The gain is due to the fact that ID allows for a symmetric encoding, and thus deals better with the source's uncertainty while relegating the ``clever decoding" to the receive terminals. To illustrate clearly the role of this decoding, we investigate a class of discrete ordered compound BCs for which this improvement is strict and where ID is crucial to recreate superposition-coding like rate regions without specifying prior coding hierarchy and decoding orders. 

However, if the channels are not ordered, then the ID gain is less explicit, and thus, more evolved encoding schemes need to be investigated. For this reason, we look at the role that ``Multiple Description" (MD) coding can play in the non-ordered compound BC, where we allow each possible instance of the same user to decode a ``private description" unintended for the other channels instances.  We follow a similar approach to that in~\cite{Piantanida2010} where MD coding had been already proven to be useful over compound state-dependent channels. Such a scheme allows the encoder to treat differently the many channels instances of each user, and the resulting decoding constraints are therefore less stringent than the ``Common Description" (CD) coding scheme~\cite{MartonInnerBound79}. Indeed, the introduction of several private descriptions results in a cost tantamount to their overall correlation. Therefore, the primary question that we aim to address here is whether this correlation is more harmful than the channel uncertainty. Our answer is mostly  negative and this is stated by a class of compound MISO BC where we show that, under a specific ``Dirty-Paper Coding" (DPC) scheme~\cite{1056659}, MD coding can strictly outperform CD coding. By using a fraction of the power intended to superimpose private descriptions, each aligned for an instance of each user, can be strictly useful. Finally, we discuss the relative behavior of ID and MD coding techniques and present a brief example to support their exclusive inclusion. 
 
The remainder of this paper is organized as follows. Section~\ref{Definitions} plots the system model and provides basic definitions as well as a simple outer bound on the capacity region of a general compound BC. In Section~\ref{ID}, we study the utility of ID for the Compound BC. We start by deriving the ID inner bound in~Section \ref{IDinnerBound} and show in Section~\ref{Application} that ID is capacity achieving for a class of discrete Compound BCs while NID stays strictly sub-optimal. Next, in Section~\ref{MDcompound}, we introduce MD coding for a Compound BC setup which is studied  for the Compound Gaussian MISO BC in Section~\ref{MDinnerBounds}. The performances of these two inner bounds are then compared to the outer bound presented in Section~\ref{OuterBoundMD}. Last, we compare the relative behavior of both the ID and MD inner bounds in Section~\ref{MDVsID} and end with summary and discussion in Section~\ref{Summary}. 

\subsubsection*{Notations}
The term pmf will refer to probability mass function. Random variables and their realizations are denoted by upper resp. to lower case letters. Vectors are denoted by bold font characters and RV stands for random variable while FME stands for Fourier Motzkin Elimination. 

For any sequence~$(x_i)_{i\in\mathbb{N}_+}$, notation $x_k^n$ stands for the collection $(x_k,x_{k+1},\dots, x_n)$. $x_1^n$ is simply denoted by $x^n$. Entropy is denoted by $H(\cdot)$, and mutual information by $I(\cdot;\cdot)$ while differential entropy is denoted by $h(\cdot)$. $\mathds{E}$ resp. $\mathds{P}$ denote the expectation resp. the generic probability measure while the notation $P$ is specific to the pmf of a RV. $\|\cX\|$ stands for the cardinality of the set $\cX$. We denote typical and conditional typical sets by $\typ{X}$ and $\typ{Y|x^n}$, respectively (see Appendix~\ref{sec:typical} for details). Let $X$, $Y$ and $Z$ be three RVs on some alphabets with probability distribution~$p$. If $p(x|yz)=p(x|y)$ for each $x,y,z$, then they form a Markov chain, which is denoted by $X\mkv Y\mkv Z$. The \emph{binary entropy function} $H_2$ is defined $\forall x \in [0:1]$ by  $H_2(x) \triangleq - x \log_2(x) - (1-x) \log_2 (1-x)$,  and the binary convolution operator $(\star)$ as: $x \star y \triangleq x (1-y) + (1-x) y$ for all $(x,y) \in [0:1]^2$. For two channels with outputs $Y_1$ and $Y_2$, $\preccurlyeq$ stands for $Y_1$ is less noisy than $Y_2$.  On the other hand, $\bh^t$ is to be understood as the transpose of the real valued vector $\bh$. Let $\bB_u$ be a unit norm $2\times1$ column vector. We denote the scalar product between vectors $\bB_u$ and $\bh_j$ by $h_{j,u} = \bh_j^t \, \bB_u$.

\section{Problem Definition}\label{Definitions}

The Compound Broadcast Channel model consists in one source terminal and two distinct receivers each observing one of many possible channel outputs. The source wishes to communicate two private messages, one to each receiver, while a common message is intended to both of them. This setup is equivalent to a setting where each user is represented by multiple users that are interested in the same message. As a matter of fact, this model is also equivalent to pairing up users from distinct groups, leading to a compound setup whose class of channels consists of all possible BCs created with possible pairs of users, and where the source is oblivious to the channel controlling the communication. 

\subsection{Definition of the Compound Broadcast Channel (BC)}
\begin{itemize}
	\item Consider a collection of $n$-th extensions of discrete memoryless BCs:
\begin{equation}
	 \{ \cW^n_j\}_{j\in \cJ} = \big\{P_{Y^{n}_j Z^{n}_j|X^n }: \cX^n \longmapsto \cY^n \times \cZ^n \big\} \ , 
\end{equation} 
defined by the conditional pmfs: 
\begin{equation}
 P_{Y^{n}_j Z^{ n}_j|X^n } = \prod^{n}\limits_{i=1} P_{Y_{j,i} Z_{j,i }|X_i} \ . 
\end{equation}
	\item Users' pair of index $j$ takes values in the finite set of indices $\cJ= [1:N]$. 
	\item An ${(M_{0n},M_{1n}, M_{2n}, n)}$-code for this channel consists of: three sets of messages $\cM_0$, $\cM_1$ and $\cM_2$, an encoding function that assigns an n-sequence $x^n (w_0, w_1, w_2) $ to each triple of messages $(w_0, w_1, w_2) \in \cM_0 \times \cM_1 \times\cM_2 $ and decoding functions, one at each receiver, that assign to the received signal an estimate message pair $(\hat{w}_{0k}, \hat{w}_k)$ in $\cM_0 \times \cM_k$, for $k\in \{ 1,2\} $ or an error.
	
	 The probability of error is given by: 
\begin{IEEEeqnarray}{rCl}
 P_e^{(n)}(j) &\triangleq& \dsP \left( \bigcup_{k\in\{1,2\}} \bigl\{ \big(\hat{W}_{0k}(j), \hat{W}_k(j)\big) \neq (W_0, W_k) \bigr\}\right). 
\end{IEEEeqnarray}
	\item A rate tuple $(R_0, R_1, R_2)$ is said to be achievable if there exists an ${(M_{0n},M_{1n}, M_{2n}, n)}$-code satisfying: 
\begin{IEEEeqnarray}{rCl}
	 \liminf\limits_{n \rightarrow \infty} \frac{1}{n} \log_2 M_{kn} &\geq & R_k \,\,\,\textrm{ $\forall \,k=\{0,1,2\}$}\,,\\
	 \limsup\limits_{n \rightarrow \infty} \,\max_{j\in \cJ} \,P_e^{(n)}(j) &=& 0\ . 
\end{IEEEeqnarray}
The capacity region is the set of all achievable rate tuples.
	\end{itemize}
	
\subsection{Outer Bound of the Capacity of the Compound BC}\label{SimpleConverse}
We derive in this section a simple and intuitive outer bound on the capacity region of the compound BC. This outer bound results from a straightforward extension to the compound setting of the best-known outer bound on the capacity of the BC. It will be useful in the examples we shall study later.  

Let the rate region $\cR^{(j)}_{\textrm{NEG}}$ denote the outer bound derived in~\cite{NairELGamal2006} applied to each pair of users with index ``$j$". For the private message setup, the rate region is given by \vspace{1mm}
\begin{equation}
	\cR_{\textrm{NEG}}^{(j)} (p_{QUVX}): \, \left\{\begin{array}{rcl}
	R_1 &\leq& I( Q U; Y_j) \ , \\
	R_2 &\leq& I( Q V; Z_j)\ ,  \\
	R_1 + R_2 &\leq& I( U ; Y_j| Q V) + I( Q V; Z_j)\ ,  \\
	R_1 + R_2 &\leq& I( Q U; Y_j) + I( V ; Z_j| Q U) \ . 
		\end{array}\right.
	\end{equation}
for a specific pmf on $p_{QUVX}$.  A simple outer bound on the capacity region of the compound BC is stated in the following theorem. 

\begin{theorem}[Outer bound]\label{theo-outer-bound}
The capacity region of the two-user Compound BC $\cC_{\cJ}$ verifies: 
\begin{equation}
	\cC_{\cJ} \subseteq \,\bigcup_{P_{U}P_V} \,\,\bigcap^N_{j=1} \,\left(\bigcup_{P_{QX|UV}} \, \cR_{\textrm{NEG}}^{(j)} (p_{QUVX}) \right)\ ,
\end{equation}
where the channel input $X$ is a deterministic mapping of $\cQ\times\cU\times\cV$. 
\end{theorem}
It is worth mentioning that when the Compound BC consists in only one BC, the outer bound~\cite{NairELGamal2006}  was not proven to be tight in general. For non-ordered compound setups, the fact of optimizing the common auxiliary RV $Q$ for each channel with index $j$, prevents even further this outer bound from being tight since the encoder is oblivious to the actual channel realization. For instance, it cannot optimize the code for each of the possible channels instances. However, this bound can still be tight in some cases of interest as will be clarified later on.
 
\begin{IEEEproof} 
A sketch of the proof is relegated to Appendix~\ref{ProofOuterBoundGenralCase}.
\end{IEEEproof}

\section{Interference Decoding (ID) in the Compound Broadcast Channel}\label{ID}

We now derive an inner bound on the capacity region of the Compound BC resorting to a class of codes consisting of three RVs, each one encoding a message, generated and mapped via superposition coding and random binning. 

\subsection{Interference Decoding (ID) Inner Bound}\label{IDinnerBound}

The inner bound we derive here shares common ideas with following works~\cite{6691328}. First, the notion of ID used in~\cite{Bacelli} where --roughly speaking-- each receiver is allowed to decode its intended message as well as (non-uniquely) decode or not  the interfering message. Second, the fact that decoding ``non-uniquely" the interfering message alleviates an extra constraint on the information rates yielding  the same result as if the decoder would have to successively decode the interfering and the intended messages which is related 
to~\cite{Nair2009ThreeDegradedMessageSet}.

\begin{theorem}[ID inner bound]\label{theo-ID-CBC}
An inner bound on the capacity region of the Compound BC consists in the set of all rates $(R_0,R_1,R_2)$ included in: 
\begin{equation}
\cR_{\textrm{ID}} \triangleq \bigcup_{p_{QUVX} \in \cP}\, \underbrace{\bigcup_{(T_1,T_2) \in \dsT(p) }}_{\textrm{FME}} \underbrace{\bigcap^N_{j=1}}_{\textrm{(compound)}}   \underbrace{\bigcup^4_{i_j = 1}}_{(\textrm{4 methods)}} \cT^{(j)}_{i_j}  (p,T_1,T_2) \ , 
\end{equation}
where $\cP$ is the set of all input pmfs $p_{QUVX} $ such that $(Q,U,V) \mkv X \mkv (Y_1,\dots, Y_N ,Z_1, \dots ,Z_N)$. \\The rate regions $\cT^{(j)}_{[1:4]}$ and the set $\dsT$ are, respectively, defined as follows: 
	\begin{equation}
	\cT^{(j)}_{1}(p,T_1,T_2): \, \left\{\begin{array}{rcl}
 T_1 & \leq& 	I( U; Y_j| Q)\ , \\
	R_0 + T_1 &\leq& I(Q U; Y_j)\ , \\
	T_2 &\leq& I( V; Z_j |Q)\ , \\
	R_0 + T_2 &\leq& I(Q V; Z_j) \ , 
	\end{array}\right.
	\end{equation}
	\begin{equation}
	\cT^{(j)}_{2}(p,T_1,T_2): \, \left\{\begin{array}{rcl}
	T_1 &\leq& 	I( U; Y_j | Q)\ , \\
	R_0 + T_1 &\leq& I(Q U; Y_j)\ , \\
	T_2	 &\leq& I(V; Z_j U| Q) \ , \\ 
	T_1 + T_2 &\leq& I( U V; Z_j| Q)+I(U;V|Q) \ , \\
	R_0 +T_1 + T_2 &\leq& I(Q U V ; Z_j) +I(U;V|Q) \ , 
	\end{array}\right.
	\end{equation}
\begin{equation}
	\cT^{(j)}_{3}(p,T_1,T_2): \, \left\{\begin{array}{rcl}
	T_1 &\leq& 	I( U; Y_j V| Q)\ ,  \\
	T_1 + T_2 &\leq& I( U V; Y_j | Q) +I(U;V|Q)\ , \\
	R_0 + T_1 + T_2 &\leq& I( Q U V ; Y_j)+I(U;V|Q)\ ,  \\
	T_2 &\leq& I( V; Z_j |Q)\ ,  \\
	R_0 + T_2 &\leq& I(Q V; Z_j)\ , 
		\end{array}\right.
	\end{equation}
	\begin{equation}
	\cT^{(j)}_{4}(p,T_1,T_2): \, \left\{\begin{array}{rcl}
	T_1 &\leq& 	I( U; Y_j V | Q)\ ,  \\
	T_1 + T_2 &\leq& I( U V; Y_j | Q) +I(U;V|Q) \ , \\
	R_0 + T_1 + T_2 &\leq& I( Q U V ; Y_j) +I(U;V|Q)\ ,  \\
	T_2	 &\leq& I(V; Z_j U | Q)\ , \\ 
	T_1 + T_2 &\leq& I( U V; Z_j| Q)+I(U;V|Q)\ ,  \\
	R_0 + T_1 + T_2 &\leq& I(Q U V ; Z_j)+I(U;V|Q)\ , 
		\end{array}\right.
	\end{equation}
	\begin{IEEEeqnarray}{rl}
 \dsT(p) = \Big\{ (T_1, T_2): \quad T_1 & \geq R_1\ ,  \\ T_2 &\geq R_2\ ,  \\
	T_1 + T_2 &> R_1 + R_2+ I(U;V |Q ) \Big\}\ .	
\end{IEEEeqnarray}
\end{theorem}

\begin{IEEEproof} 
The proof is relegated to Appendix~\ref{ProofMainTheorem}. 
\end{IEEEproof}
 
\begin{remark}[Main comments about the proof]
Each user introduces the union of two sets of constraints, corresponding to decoding or not the interference. This  results  --in terms of achievable rates-- in the union of four rate regions: 
\begin{enumerate}
\item The region $\cT^{(j)}_{1}$ is the same rate region as obtained with Marton's inner bound, 
\item The region $\cT^{(j)}_{4} $ is obtained by letting the destinations to decode both the intended and the interfering message, 
\item The regions $\cT^{(j)}_{2}$ and $\cT^{(j)}_{3} $ correspond to each destination  decoding the interfering message at once.  
\end{enumerate}
A slightly similar rate region was also derived in~\cite{Bandemer2012} in a different context, but it does not take advantage of the encoding technique, and thus in our setting it fails at achieving even Marton's inner bound. 
\end{remark}

\begin{remark}[Connection to the standard two-user BC]
Consider the  standard two-user BC where $\cJ=1$. Observe that by allowing both destinations to decode or not the message of the other user --ID scheme-- we found a seemingly larger rate region $\cR_{s,\textrm{ID}}$ than that of Marton~\cite{MartonInnerBound79} which does not use the ID technique. Indeed, these regions are given by 
\vspace{2mm}
\begin{IEEEeqnarray}{rcl}
\cR_{s,\textrm{ID}}\, &\triangleq&\,  \bigcup_{p_{QUVX} \in \cP} \, \bigcup_{(T_1,T_2) \in \dsT(p)} \,\left( \, \bigcup^4_{i=1} \,\,\cT_{i} (p,T_1,T_2)\, \right)\ , \\ 
\cR_{s,\textrm{NID}}\,  &\triangleq& \, \bigcup_{p_{QUVX} \in \cP} \, \bigcup_{(T_1,T_2) \in \dsT(p)} \, \cT_{1}(p,T_1,T_2) \ . \vspace{2mm}
\end{IEEEeqnarray}
It is clear that $\cR_{s,\textrm{NID}} \subseteq \cR_{s,\textrm{ID}}$, but the question is whether or not this inclusion strict. To check this issue, we need to evaluate both regions and thus we resort to FME for $(T_1,T_2)$, and bit recombination between the private rates $(R_1,R_2)$ and the common one $R_0$\footnote{For the interested reader a similar calculation is done in Appendix~\ref{FME}.}. Since the unions commute, we can write that: 
\vspace{1mm}
\begin{IEEEeqnarray}{rcl}
\cR_{s,\textrm{ID}} &=& \bigcup^4_{i=1} \cR_{s,i}\nonumber\\
&=& \cR_{s,\textrm{NID}}  \cup \left(\bigcup^4_{i=2} \cR_{s,i} \right) \ , 
\end{IEEEeqnarray}
\vspace{1mm}
where $\cR_{[2:3]}$ are respectively defined by the following sets of inequalities: 	
	\begin{equation}
	\cR_{s,2}: \, \left\{\begin{array}{rcl}
	R_0 + R_1 &\leq& I(Q U; Y)\ , \\
	R_0 + R_1 + R_2 &\leq& I(V; Z| U Q) + I(Q U; Y)\ ,  \\
	R_0 + R_1 + R_2 &\leq& I(Q U V ; Z) \ , 
	\end{array}\right.
	\end{equation}
	\begin{equation}
	\cR_{s,3}: \, \left\{\begin{array}{rcl}
 R_0 + R_2 &\leq& I(Q V; Z)\ , \\
	R_0 + R_1 + R_2 &\leq& I( U; Y | V Q) + I(Q V; Z)\ , \\
	R_0 + R_1 + R_2 &\leq& I( Q U V ; Y) \ , 
	\end{array}\right.
	\end{equation}
 \begin{equation}
	\cR_{s,4}: \, \left\{\begin{array}{rcl}
 	R_0 + R_1 + R_2 &\leq& I( Q U V ; Y)\ ,  \\
	R_0 + R_1 + R_2 &\leq& I( Q U V ; Z) \ , 
	\end{array}\right. 
	\end{equation}
while $\cR_{s,\textrm{NID}}$ is defined by
 \begin{equation}
	\cR_{s,\textrm{NID}} = \cR_{s,1}: \, \left\{\begin{array}{rcl}	
	R_0 + R_1 &\leq& I(Q U; Y)\ , \\
	R_0 + R_2 &\leq& I(Q V; Z)\ ,  \\
	R_0 + R_1 + R_2 &\leq& I( U; Y | Q) + I(Q V; Z) - I(U;V|Q)\ ,  \\
	R_0 + R_1 + R_2 &\leq& I( Q U; Y ) + I(V; Z|Q) - I(U;V|Q) \ , \\
	2R_0 + R_1 + R_2 &\leq& I(Q U; Y ) + I(Q V; Z) - I(U;V|Q) \ .
	\end{array}\right.
	\end{equation}\vspace{1mm}

From the above rate regions, we observe  that by taking $U= Q$, the region $\cR_{s,\textrm{NID}}$ contains  $\cR_{s,2}$, and similarly, setting $V= Q$ allows $\cR_{s,\textrm{NID}}$ to contain $\cR_{s,3}$ while $U= Q= V$ allows it to contain $\cR_{s,4}$. Hence, using the ID strategy in presence of a single channel per user yields the same rate region as Marton's inner bound. Indeed, the apparently gain provided by choosing to decode  or not the interference is recovered by an optimization of the input distribution. 
\end{remark}

We can observe that by using ID in the compound setting, we get a seemingly larger region than Marton's worst-case inner bound, which is given by:
\begin{equation}
\cR_{\textrm{NID}} \triangleq \bigcup_{p_{QUVX} \in \cP} \, \bigcup_{(T_1,T_2) \in \dsT(p)}  \, \left(\, \bigcap^N_{j=1} \, \cT^{(j)}_{1}(p,T_1,T_2) \, \right)\, .\label{eq-Marton-region}\vspace{1mm}
\end{equation}
It is clear that $\cR_{\textrm{NID}} \subseteq \cR_{\textrm{ID}}$ but yet, no evidence on the strict inclusion has been stated here.  In the sequel, we investigate a Compound BC for which the region based on the usual decoding in Marton's inner bound $\cR_{\textrm{NID}}$ fails at achieving the capacity while $\cR_{\textrm{ID}}$ from Theorem~\ref{theo-ID-CBC} is tight.  The key point in this ``strict inclusion", is that, if the optimizing input pmf varies from one channel to the other (e.g. in terms of superposition ordering of auxiliary RVs), then the joint optimization in the compound setup imposes a stringent limitation on the input pmf. This prevents $\cR_{\textrm{NID}}$ from reaching capacity for some compound models while the ID technique, allowing the choice between two decoding strategies, does not suffer such a loss. 

\subsection{Interference Decoding is Optimal for a Class of Compound Broadcast Channels}\label{Application}
In this section, we will construct a Compound BC model for which Marton's worst-case inner bound, obtained through NID, is strictly sub-optimal compared to ID inner bound where users are allowed to decode or not the interference. 
We first discuss a criterion for the construction of such a compound model and later, prove the optimality of ID. For simplicity, we restrict our analysis to the case $\| \cJ\| = 2$ and private rates only, i.e., $R_0=0$.

\subsubsection{Irrelevant compound models}

The difficult to characterize optimal coding for the Compound BC is inherent to the class of BCs in the set, i.e., the set of channel users over which we define the compound model. We shall refer to as ``irrelevant" models those 
of ordered BCs for which Marton's worst-case inner bound is tight. As a matter of fact, Marton's inner bound achieves the capacity of every BC for which capacity is known. 

Consider the class of broadcast channels: 
\begin{equation}
 \cW= \{ \cW_1,\cW_2\} = \left\{ \cX \mapsto (\cY_j, \cZ_j)\right \}_{j \in \{1,2\}} \ , 
\end{equation} 
where $ Y_2 \preccurlyeq Y_1$ and $Z_1 \preccurlyeq Z_2$. Then, it follows that, whatever the auxiliary RVs $(Q,U) \sim p_{QU}$: 
\begin{equation}
	I(QU;Y_2) \leq I(QU;Y_1) \quad ,\quad I(U;Y_2|Q) \leq I(U;Y_1|Q) \ .
\end{equation}
Thus,  Marton's inner bound based on superposition coding and random binning yields the region 
\begin{equation}
\left\{\begin{array}{rcl}	
	R_1 &\leq& \min \limits_{j=1,2}I(QU;Y_j) \ , \\
R_2 &\leq& \min\limits_{j=1,2} I(QV;Z_j)   \ ,\\
R_1 + R_2 &\leq& \min\limits_{j=1,2}I(U;Y_j|Q)+ \min\limits_{j=1,2} I(QV;Z_j) - I(U;V|Q)  \ ,\\
R_1 + R_2 &\leq& \min\limits_{j=1,2} I(QU;Y_j)+ \min\limits_{j=1,2} I(V;Z_j|Q) - I(U;V|Q) \ ,
	\end{array}\right.
	\end{equation}   \vspace{1mm}
which  reduces to: \vspace{1mm}
\begin{equation}
\left\{\begin{array}{rcl}	
R_1 &\leq& I(QU;Y_2) \ ,  \\
R_2 &\leq& I(QV;Z_1)   \ , \\
R_1 + R_2 &\leq& I(U;Y_2|Q)+ I(QV;Z_1) - I(U;V|Q) \ , \\
R_1 + R_2 &\leq& I(QU;Y_2) + I(V;Z_1|Q) - I(U;V|Q)   \, .
	\end{array}\right.
	\end{equation}    \vspace{1mm}
This is the the rate region obtained by coding for only the pair of users corresponding to the channel $(Y_2,Z_1)$. Furthermore, it is straightforward  to check that if the capacity of this channel is known (e.g. when $Y_2$ and $Z_1$ are ordered in the sense of ``degradedness" or ``less-noisiness"), then Marton's inner bound achieves the capacity region of the Compound BC. Thus, if the marginals seen in set of users $1$, i.e., $(Y_1,Y_2)$ are ordered at least in the known senses of ``less noisiness", and so are those in the set of users $2$, i.e., $(Z_1,Z_2)$, Marton's inner bound for this setup leads to the capacity region of the ``worst" BC formed by the worst pair of users in the set. Hence this class of compound models is irrelevant for our purpose. 

\subsubsection{Compound Binary Erasure and Binary Symmetric BC}\label{ConverseApplication}

In this section, we construct the simplest while relevant Compound BC setting, where: 
\begin{itemize}
\item Set of user 2 contains only one channel instance, i.e., $Z_1 = Z_2 = Z$.
\item Set of user 1 is compound of two possible channel instances denoted by $\{Y_j\}_{j \in \{1,2\}}$.
\end{itemize}
Our aim is to show the desired ``strict" inclusion $\cR_{\textrm{NID}}\subset \cR_{\textrm{ID}}$. 
To this end, we need to find a ``relevant" compound BC where $(Y_1,Y_2)$ are not strongly ordered (e.g. neither degraded nor less-noisy). Otherwise the resulting Compound BC would be formed by $Z$ and the worst channel between $(Y_1,Y_2)$, for which it is straightforward to see that $\cR_{\textrm{NID}}$ achieves the capacity region.

 Besides this argument, if we are to show the strict inclusion of Marton's rate region with respect to the rate region obtained by ID, we need to provide for some inverse orderings in the compound channels formed by all possible pairs of users, so as to impose a tradeoff between two antagonist coding schemes for Marton's coding scheme, i.e.,  two antagonist choices of auxiliary RVs at the encoder. One can then think of a setting where for instance the BC $(Y_1,Z)$ has $Z$ ``better" than $Y_1$ while the BC $(Y_2,Z)$ is ordered in the opposite way, i.e $Y_2$ is better than $Z$.   
 
 \begin{table}
\centering
\caption{\label{TableBECBSC} Different Orderings allowed by the BEC(e)/BSC(p) BC}
 \begin{tabular}{|c|c|c|c|c|}
\hline 
 $0 \leq e \leq 2p$ & $2p < e \leq 4p(1-p)$ & $4p(1-p) < e \leq H_2(p)$ & $H_2(p) < e \leq 1$ \\ 
\hline 
BSC degraded of BEC & BEC Less Noisy BSC & BEC More Capable BSC & BSC Ess. Less Noisy BEC \\
\hline 
\end{tabular}

\end{table}
 
Consider the \emph{Binary Erasure Channel} (BEC) with erasure probability $e$ and the \emph{Binary Symmetric Channel} (BSC) with crossover probability $p$. These have the particularity of allowing for a variety of orderings between the outputs~\cite{NairBECBSC2010}, depending on $(e,p)$, as summarized  in Table~\ref{TableBECBSC}. Define the Compound BC with components: \vspace{1mm}
\begin{equation}
\cW\,:\,\left\{
\begin{array}{lcl}
\cX & \longmapsto & \cZ \,\, \equiv \textrm{BSC(p)}\, , \\
\cX & \longmapsto & \cY_1 \equiv \textrm{BSC($p_1$)}\, , \\
\cX & \longmapsto & \cY_2 \equiv \textrm{BEC($e_2$)}\, . 
\end{array}\right. \vspace{1mm}
\end{equation}
We first start by imposing to  $Y_2$ to be more capable than $Y_1$, which requires: $4 p_1(1-p_1)< e_2 \leq H_2(p_1).$ One possible choice is then to take $Y_1$ as a \emph{physically degraded} version of $Z$, i.e., $p < p_1 < 0.5 $, and $Y_2$ more capable than $Z$, i.e., 
\begin{equation}
 4 p (1-p)< 4p_1(1-p_1)< e_2 \leq H_2(p) \leq H_2(p_1) \ . \label{ParametersInequality}
\end{equation}
This choice fulfills the criteria stated for the construction of a relevant example. For this case, the simple outer bound enunciated in Section~\ref{SimpleConverse} writes as: 
\begin{equation}
\cC_{\cJ} \subset \cC_1  \cap \cC_2  \ , 
\end{equation} 
where: 
\begin{equation}
	\cC_1 \,:\, \left\{\begin{array}{rcl}
 R_1 &\leq& 1-H_2(p_1 \star \alpha) \ , \\ 
 R_2 &\leq& H_2(p \star \alpha) - H_2(p) \ , 
 	\end{array}\right.\label{capacity_1}
\end{equation}
\begin{equation}
	\cC_2 \,:\, \left\{\begin{array}{rcl}
 R_1 &\leq& (1-e_2)\, H_2(\alpha) \ , \\
 R_2 &\leq& 1 - H_2(p\star\alpha)  \ ,\\ 
 R_1+ R_2 &\leq& (1- e_2) \ . 
	\end{array}\right.\label{capacity_2}
\end{equation}
We claim that the capacity region $\cC_1$ is strictly included in $\cC_2$, for which  we can compare: \vspace{1mm}
\begin{equation}
 \max_{(R_1,R_2) \in \cC_1} \frac{R_1}{ (1-H_2(p))- R_2} = \lim_{\alpha \rightarrow 1/2} \frac{1-H_2(p_1\star \alpha) }{1-H_2(p \star \alpha)} \approx \frac{ (1-2 p_1)^2 }{ (1-2 p)^2 } 
\end{equation} 
and 
\begin{equation}
\max_{(R_1,R_2) \in \cC_2} \frac{R_1}{(1-H_2(p))- R_2} \geq \frac{1-e_2}{1-H_2(p)} \ . 
\end{equation}
Our claim simply follows by noticing that from the assumptions on the parameters $e_2$, $p$ and $p_1$, we have that: 
\begin{equation}
 \frac{ (1-2 p_1)^2 }{ (1-2 p)^2 } \leq 1 \leq \frac{1-e_2}{1-H_2(p)} \ , 
\end{equation}
which shows the outer bound reduces to $\cC_1$.

\subsubsection{Evaluation of the ID inner bound of Theorem~\ref{theo-ID-CBC}}
\begin{figure}[th!]
 \centering
 \includegraphics[scale=1]{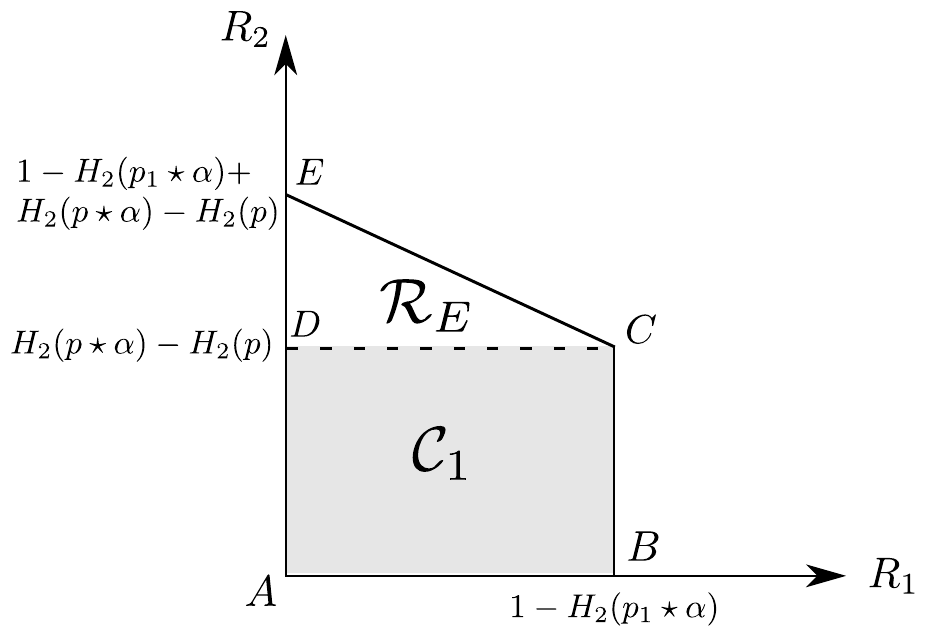}\vspace{-2mm}
 \caption{Comparison between $\cC_1$ and $\cR_{\textrm{ID}}$.}
 \label{InclusionCapacity}\vspace{-2mm}
 \end{figure}

We evaluate the proposed rate region $\cR_{\textrm{ID}}$ of Theorem~\ref{theo-ID-CBC}, which satisfies: 
\begin{equation}
\cR_{\textrm{ID}} \supseteq	\bigcup_{p_{QUVX} \in \cP}\,\bigcup_{(T_1,T_2) \in \dsT(p) } \quad \left( \cT^{(1)}_{3}(p,T_1,T_2) \cap \cT^{(2)}_{4}(p,T_1,T_2)\right) \ ,
\end{equation} 
where $\cT^{(1)}_{3} \cap \cT^{(2)}_{4}$ is defined by the set of inequalities:
\begin{equation}
\left\{\begin{array}{rcl}
T_2 &\leq& I(V;ZU|Q) \ , \\
T_1 + T_2 &\leq& I(UV;Z|Q) + I(U;V|Q) \ , \\
R_0 + T_1 +T_2 &\leq& I(QUV;Z) + I(U;V|Q)\ ,  \\
T_1 &\leq& I(U;Y_2 V|Q)\ ,  \\
T_1 + T_2 &\leq& I(UV;Y_2|Q) + I(U;V|Q)\ ,  \\
R_0 + T_1 +T_2 &\leq& I(QUV;Y_2) + I(U;V|Q)\ ,  \vspace{2mm} \\
T_1 &\leq& I(U;Y_1|Q) \ , \\
R_0 + T_1 &\leq& I(QY;Y_1) \vspace{2mm} \\
T_1 \geq R_1 &,& T_2 \geq R_2 \ , \\
T_1 + T_2 &>& R_1 + R_2 + I(U;V |Q ) \ .
\end{array}\right. 
\end{equation}
This comes to choosing: $i_1 = 3$, i.e., using decoding method~$(3)$ for the BC~$(1)$, while the other channel gets the fourth decoding method: $i_2 = 4 $. These constraints allow $Z$ and $Y_2$ to decode all messages, while forcing $Y_1$ to decode only its own message. In Appendix~\ref{FME}, it is shown after FME on $(T_1,T_2)$, bit recombination, and then setting $R_0=0$, that the previous rate region reduces to the set of rates satisfying: 
\begin{equation}
\left\{\begin{array}{rcl}
R_1 &\leq& I(QU;Y_1)\ , \\
R_1 + R_2 &\leq& I(QU;Y_1) + I(V;Z|QU)\ , \\
R_1 + R_2 &\leq& I(QU;Y_1) + I(UV;Y_2|Q)\ , \\
R_1 + R_2 &\leq& I(QUV;Y_2) \ .
\end{array}\right. 
\end{equation}
Then, letting: $V = X$, $\bar{Q}=(Q,U) $, and using the fact that $Y_2$ is \emph{more capable} than $Z$, yields: 
\begin{equation}
\left\{\begin{array}{rcl}
R_1 &\leq& I(\bar{Q};Y_1)\ ,  \\
R_1 + R_2 &\leq& I(\bar{Q};Y_1) + I(X;Z|\bar{Q}) \ . 
\end{array}\right. 
\end{equation}
This achievable rate region coincides with the outer bound and thus provides the capacity region of the BC $(Y_1,Z)$ for the considered setup. Letting then $\bar{Q} \longmapsto X \equiv \textrm{BSC}(\alpha)$, and $X \sim \textrm{Bern}(1/2)$ we get the following union over all $\alpha \in [0:1]$ of: 
\begin{equation}
	\cR_{\textrm{ID}}\,:\, \left\{\begin{array}{rcl}
 R_1 &\leq& 1-H_2(p_1\star\alpha)\ , \\
 R_1 + R_2 &\leq& 1-H_2(p_1\star\alpha) +H_2(p \star\alpha) - H_2(p)\ .
		\end{array}\right.\label{our-inner-binary}
\end{equation}
In order to check that $\cR_{\textrm{ID}}$  is equal to the outer bound $\cC_1$, we should first start by noticing that it is the exclusive union of two rate regions: $\cC_1$ and $\cR_E$ which are defined by
 \begin{equation}
	\cR_E\,:\, \left\{\begin{array}{rcl}
 R_2 &\geq& H_2(p \star\alpha) - H_2(p) \ ,\\
 R_1 + R_2 &\leq& 1-H_2(p_1\star\alpha) + H_2(p \star\alpha) - H_2(p)\ .
		\end{array}\right. 
\end{equation}
As plot in Fig.~\ref{InclusionCapacity}, this region has four corner points among which $3$ of them are clearly included in $\cC_1$, i.e., $A$, $B$, and $C$. 

 To show that the point $E$ lies in the region $\cC_1$, we first write that: 
\begin{equation}
 E = (0,  1-H_2(p_1\star\alpha) + H_2(p \star\alpha) - H_2(p)) \ . 
\end{equation}
Since $Y_1$ is physically degraded with respect to $Z$, i.e., $p \leq p_1$, and since $\alpha$, $p \star\alpha$ and $p_1 \star\alpha$ are all included in the interval $[0:0.5]$, one can clearly write that: $- H_2(p_1\star\alpha) + H_2(p\star\alpha) \leq 0$. Hence, the point $E$ is dominated by the point $C_2 =( 0; 1 - H_2(p) )$, which is already achievable in $\cC_1$. The line between $C$ and $E$ can is achieved by the convexity of the rate region $\cC_1$. 

\subsubsection{Outer bound on Marton's inner bound}
When restricted to Marton's inner bound, the rate region in expression~\eqref{eq-Marton-region} is included in the union of next constraints:
\begin{equation}
\left\{\begin{array}{rcl}
T_2 &\leq& I(V;Z|Q)\ , \\
R_0 + T_2 &\leq& I(QV;Z)\ , \\
T_1 &\leq& \underset{j=1,2}{\min \, } I(U;Y_j|Q)\ , \\
R_0 + T_1 &\leq& \underset{j=1,2}{\min \, } I(QU;Y_j)\ , \\
T_1 \geq R_1 &,& T_2 \geq R_2\ , \\
T_1 + T_2 &>& R_1 + R_2 + I(U;V |Q ) \ .
\end{array}\right. 
\end{equation}
 
Then, we perform FME on the rates $T_1$ and $T_2$, bit recombination, and we set $R_0 = 0$, which yields the following rate region: 
\begin{equation}
\left\{\begin{array}{rcl}
 R_2 &\leq& I(QV;Z) \\
R_1 &\leq& \underset{j=1,2}{\min} I(QU;Y_j)\ , \\
R_1 + R_2 &\leq& I(V;Z|Q)+ \underset{j=1,2}{\min \, } I(QU;Y_j) - I(U;V|Q)\ , \\
R_1 + R_2 &\leq& I(QV;Z)+ I(U;Y_2|Q) - I(U;V|Q) \  , 
\end{array}\right. 
\end{equation}
where we have used the fact that: $I(Q;Y_1) \leq I(Q;Z)$, i.e., \emph{physical degradedness}. As a matter of fact, the previous rate region is contained in the set of rates verifying: 
\begin{equation}
\left\{\begin{array}{rcl}
R_1 &\leq& \underset{j=1,2}{\min \, } I(QU;Y_j) \ , \\
R_1 + R_2 &\leq& I(X;Z|QU)+ \underset{j=1,2}{\min \, } I(QU;Y_j) \ , 
\end{array}\right. 
\end{equation}
because for each $P_{QUVX}\in \cP$ the next inequalities hold: 
\begin{IEEEeqnarray}{rCl}
 I(QV;Z) &\leq&  I(X;Z) \, , \\
 I(V;Z|Q) + \min_{j=1,2} I(QU;Y_j) - I(U;V|Q)  &\leq& I(X;Z|QU)+ \min_{j=1,2} I(QU;Y_j)  \\
 &\leq&  I(X;Z) \ . 
\end{IEEEeqnarray}
By letting $\bar{Q}=(Q,U)$, we obtain the following constraints:
\begin{equation}
\cR_{\textrm{OuterNID}}\,:\, \left\{\begin{array}{rcl}
R_1 &\leq& \underset{j=1,2}{\min \, } I(\bar{Q};Y_j) \ , \\
R_1 + R_2 &\leq& I(X;Z|\bar{Q})+ \underset{j=1,2}{\min \, } I(\bar{Q};Y_j) \ . 
\end{array}\right.\label{upper bound to Marton } 
\end{equation}
In Appendix~\ref{CardinalityBound}, we show that it suffices to evaluate this bound for all auxiliary RVs $\bar{Q}$ that verify $\|\bar{Q}\|\leq 4$ and $X \sim \textrm{Bern}(1/2)$. 

\begin{figure}[th!]
 \centering
 \includegraphics[scale=1.4]{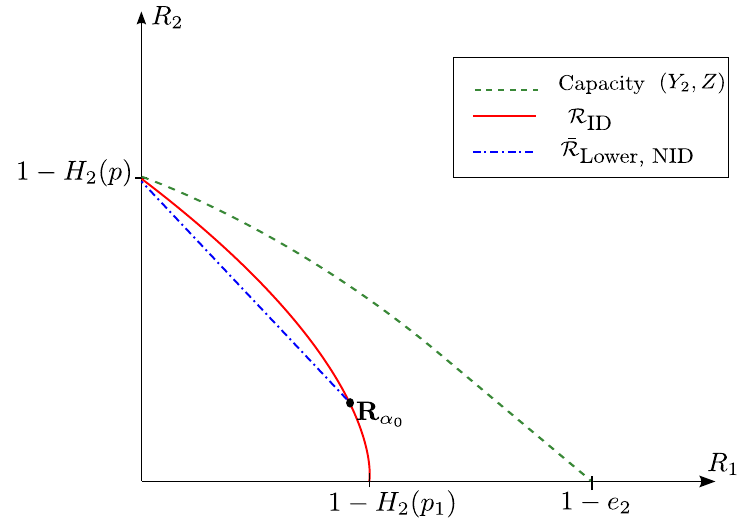}\vspace{-2mm}
 \caption{Comparison between the rate region $\cR_{\textrm{ID}}$ and the convex closure or  $\cR_{\textrm{Lower,NID}}$.}
 \label{FigureRegionsLower}\vspace{-2mm}
 \end{figure}

Though we might state such characteristics about the maximizing distribution, the optimization of this region turns out to be tricky since the usual bounding tools such as ``Mrs. Gerber's Lemma" leads only to the next lower bound: 
\begin{equation}
	\cR_{\textrm{Lower,NID}}\,\subseteq \, 
	\left\{
	\begin{array}{ll}
	R_2 & \leq H_2(p \star\alpha) - H_2(p) \ ,\\
R_1 &\leq \min\{ 1-H_2(p_1\star\alpha), \bar{e}_2 (1-H_2(\alpha)) \} \ . 
		\end{array}\right.\label{outer-Martion-binary}
	\end{equation}
Fig.~\ref{FigureRegionsLower} plots a comparison between these two regions. This lower bound coincides with the capacity region $ \cR_{\textrm{ID}}$ over the interval $R_2 \in [0:H_2(p\star\alpha_0) -H_2(p)]$ or equivalently $R_1 \in [0:1-H_2(p_1\star\alpha_0) ]$ where $\alpha_0$ is given by:
$
 1-H_2(p_1\star\alpha_0) = (1-e_2) (1-H_2(\alpha_0)) \ . 
$
 
In order to derive an upper bound, we study a looser outer bound to $ \cR_{\textrm{Outer,NID}}$, provided that the gap stays strict between the capacity region and this outer bound. Let us define the function $t: [0: 1-H_2(p)]\mapsto \Re_+$ as:
\begin{equation}
	t(x) \triangleq \sup_{p_{XQ} \in \cC(x)} \min \{I(Q;Y_1), I(Q;Y_2)\} \ , 
\end{equation}
where the class $\cC(x)$ is given by 
\begin{IEEEeqnarray}{rCl}
\cC(x) = \big\{ p_{XQ} \in \cP(\cX \times \cQ): \quad & & Q \mkv X \mkv (Z,Y_1,Y_2) 
\IEEEnonumber \\
& & X \sim \textrm{Bern}(1/2)\ , \, I(X;Z|Q ) \geq x  \big\} \label{ConstraintEquality}  \ . 
\end{IEEEeqnarray}
The function $t:x\mapsto t(x)$ characterizes the convex closure of the region $\bar{\cR}_{\textrm{Outer,NID}}$, i.e.,  $(R_1, R_2) \in \bar{\cR}_{\textrm{Outer,NID}}$ thus $R_1 = t(R_2)$. In the same way, define $t_1$ over $[0: 1-H_2(p)]$ by 
\begin{equation}
	t_1(x) \triangleq \sup_{p_{XQ} \in \cC(x)} I(Q;Y_1) \ ,
\end{equation}
where $t_1$ characterizes the convex closure of the region $\bar{\cR}_{\textrm{ID}}$. 

In the sequel, we work towards a closed form evaluation of an upper bound of $t$ that would still be dominated by $t_1$. 

\subsubsection{An upper bound on the function $t(x)$}
We follow the method in~\cite{WitsenhausenEntropyBounds} where 
\begin{IEEEeqnarray}{rCl}
 t(x) &\triangleq& \sup_{p_{XQ} \in\cC(x)} \min \left\{I(Q;Y_1)\,, \,I(Q;Y_2)\right\} \\
&=& \sup_{p_{XQ} \in\cC(x)} \min_{a \in [0:1]} \bigl[ a \, I(Q;Y_1)+ \bar{a}\, I(Q;Y_2) \bigr] \\
& \leq& \min_{a \in [0:1]} \sup_{p_{XQ} \in\cC(x)} \bigl[ a \, I(Q;Y_1)+ \bar{a} \, I(Q;Y_2) \bigr] \ ,
\end{IEEEeqnarray}
for all $x \in [0: 1-H_2(p)]$.  Let define for each $a \in [0:1]$ and $t_a\in [0:1-H_2(p)]$, 
\begin{equation}
	t_a(x) \triangleq \sup_{p_{XQ} \in\cC(x)} \bigl[ a\, I(Q;Y_1)+ \bar{a}\, I(Q;Y_2) \bigr] \ . 
\end{equation} 
Notice that: \begin{itemize}
\item The case $a=1$ was already studied in~\cite{WitsenhausenEntropyBounds} and it was shown that: 
\begin{equation}
	t_1(x) = 1 - H_2(p_1\star p_x) \ , 
\end{equation}
where $H_2(p\star p_x) - H_2(p) = x$.
\item The case $a=0$ can be studied in a very similar fashion as in~\cite{WitsenhausenEntropyBounds} by finding out that: 
\begin{IEEEeqnarray}{rCl}
	t_0(x) &=& \inf_{\lambda \in \cR^+} \left[ F_0(\lambda)- \lambda \,x \right] \\
	&=& (1 - e_2)\, \left(1 - \frac{x}{1-H_2(p)}\right) \ , 
\end{IEEEeqnarray}
where: 
\begin{equation}
F_0(\lambda) = \max \left \{ (1-H_2(p)) \, \lambda , (1-e_2) \right\} \ . 
\end{equation}
\end{itemize}
 
Now, to upper bound $t_a$, we could have written that: 
\begin{IEEEeqnarray}{rCl}
 t_a(x) &\leq& a\, \sup_{\cC(x)} I(Q;Y_1)  + \bar{a}\, \sup_{\cC(x)} I(Q;Y_2)  \label{UpperBoundLoose} \\ 
 &=& a \, t_1(x) + \bar{a}\, t_0(x)\\
 &\geq& t_1(x) \label{UnsuccessfulUpperBound} \ , 
\end{IEEEeqnarray}
where \eqref{UnsuccessfulUpperBound} follows from what we have proved in Section~\ref{ConverseApplication}, i.e., $t_0$ dominates $t_1$ over the interval $[0:1-H_2(p)]$. Thus, we cannot restrict ourselves to the upper bound in \eqref{UpperBoundLoose} on $t_a$ since it is rather loose, and we will hence bound more tightly the function $t_a$. 

 \begin{proposition}
 The function $t_a$ satisfies the following properties:
 \begin{description}
\item{(i)} For all $x \in [0: 1-H_2(p)]$, 
\begin{equation}
t_a(x) = \max_{p_{XQ} \in \cC(x)} \left[ a\, I(Q;Y_1)+ \bar{a}\, I(Q;Y_2) \right] \ ,
\end{equation} 
\item{(ii)}  $t_a$ is concave in $x$,
\item{(iii)}  $t_a$ can be described identically by its supporting lines,
\item{(iv)}  $t_a$ is decreasing in $x$. 
\end{description}
 \end{proposition}
 
\begin{IEEEproof}
The proof is relegated to Appendix~\ref{Proposition3}. 
\end{IEEEproof}

The next result is rather useful since it allows us to transform the optimization of a rate region into optimizing one quantity captured in $F_a(\lambda)$.

\begin{corollary} The following conclusions can be drawn: 
 \begin{description}
\item{(a)}  The constraint in~\eqref{ConstraintEquality} can be transformed into: 
\begin{equation}
I(X;Z|Q) = x \ . 
\end{equation}
\item{(b)}  We have that: 
\begin{IEEEeqnarray}{rCl}
	t_a(x) &=& \inf_{\lambda \in \cR^+} \biggl[ \max_{\cP(\cX \times \cQ)} \bigl[ a\, I(Q;Y_1)+ \bar{a}\, I(Q;Y_2) + \lambda \, I(X;Z|Q ) \bigr] - \lambda \, x \biggr] \\
	&=& \inf_{\lambda \in \cR^+} \biggl[ F_a(\lambda)- \lambda \, x \biggr] \ , 
\end{IEEEeqnarray}
where
\begin{equation}
	F_a(\lambda) \triangleq \max_{p_{XQ} \in\cP(\cX \times \cQ)} \bigl[ a\, I(Q;Y_1) + \bar{a}\, I(Q;Y_2) \bigr] \ . 
\end{equation}
\end{description}
\end{corollary}
\begin{IEEEproof} $(a)$ follows from the non-increasing property of $t_a$ and $(b)$ follows from the concavity of the function $t_a$ since a concave function can be described by its supporting lines~\cite{Eggleston}. 
\end{IEEEproof} 
The analysis of the function $t_a$ for an arbitrary $a$ brings about significant computational complexity, we thus only chose to plot it using stochastic optimization methods. We chose $e_2 = 0.46$, $p =0.1$ and $p_1 =0.13$. It can be readily shown that these parameters verify~\eqref{ParametersInequality}. 

\begin{figure}[ht!]
 \centering
 \includegraphics[scale=0.4]{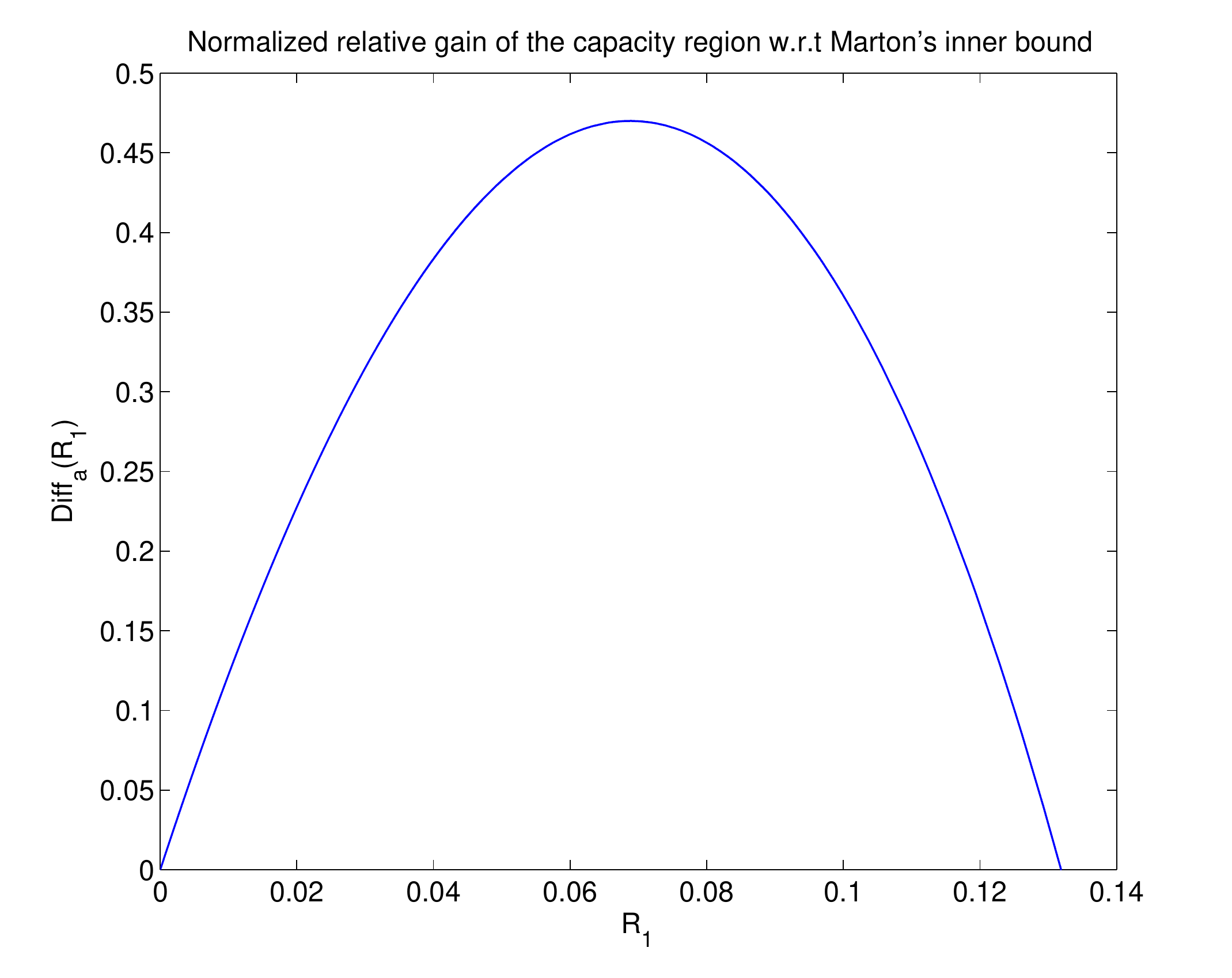} 
 \caption{ $d_a(R_1)$ the normalized relative gain of the capacity region with respect to Marton's inner bound for $a = 0.92$, $e_2 = 0.46$, $ p =0.1 $ and $p_1 =0.13$.}
 \label{FigureDiff} 
 \end{figure}

In Fig.~\ref{FigureDiff}, we chose $a = 0.92 $ and plot the normalized difference function: 
\vspace{1mm}
\begin{equation}
d_a(R_1) = \dfrac{t^{-1}_1(R_1) - t^{-1}_a(R_1)}{\max(\mid t^{-1}_1(R_1) - t^{-1}_a(R_1)\mid )}\ ,
\end{equation}\vspace{1mm}
 over the interval of interest: $[0: 1-H_2(p_1\star \alpha_0)]$ where: $1-H_2(p_1\star \alpha_0) = (1-e_2) (1-H_2(\alpha_0))$. The function $d_a$ being strictly positive, the claim of strict inclusion is thus shown.

We have investigated so far the role that alternative decoding techniques, namely ``Interference Decoding", play in the Compound BC where the users present a given hierarchy unknown at the encoder. The decoding technique takes advantage of the many possible decoding ways to alleviate the constraint of superposition coding at the source which allows the latter to apply a ``symmetric" encoding rule regardless of which channel controls the communication. In the sequel, we analyze a class of non-ordered Compound BC to infer novel strategies when there is no specific order between channels users. In this case, we will not seek to optimize the decoder but rather the encoding technique.  
 
\section{Multiple Description Coding in the Compound Broadcast Channel}\label{MDcompound}
 
In this section,  we investigate a coding technique, referred to as ``Multiple Description (MD) coding", that can enhance the achievable rates in the Compound BC.  The utility of this coding arises especially when no sort of order between the many possible instances of the users channels exists. The main idea behind MD coding is to convey the message intended to the many instances of the same group of users, through a common description as well as a set of dedicated private descriptions which can be easily decoded each at their respective instances. The common description --to be decoded by all users-- will suffer from the compound setup in that the rate has to be enough small to be decodable by all users in the same group whereas the private descriptions suffer no such loss. It is worth mentioning here that the introduction of  private descriptions will also result in a loss tantamount to their ``correlation cost". We aim at exploring the utility of MD coding in the Compound BC setting. 

In the sequel, for a matter of conciseness, we choose to address the Compound BC setting when only one user has  two possible channels, namely $Y_1$ or $Y_2$, whilst the other user suffers from no such uncertainty $Z$.  We first derive two inner bounds on the capacity region to be compared: the Common Description (CD) inner bound that is equivalent to Marton's worst-case inner bound, and the MD inner bound. We then specialize the bounds to the Compound MISO BC and show how MD coding outperforms the standard CD coding. Finally, we analyze the behavior of the obtained rate regions compared to our outer bound. 

\subsection{Multiple Description (MD) Inner Bound} 


\begin{theorem}[MD inner bound]\label{MultipleDescriptions}
An inner bound on the capacity region of $2\times1$ Compound BC is given by the set of rate pairs $(R_1, R_2)$ satisfying:
\begin{IEEEeqnarray}{rCl} \label{MultipleDescriptionsInnerBound}
\IEEEyesnumber\IEEEyessubnumber*
R_1 &\leq& I(U_0 U_1; Y_1 |Q )\ , \\
R_1 &\leq& I(U_0 U_2; Y_2|Q )\ , \\
2 R_1 &\leq& I(U_0 U_1; Y_1|Q ) + I(U_0 U_2; Y_2|Q ) - I(U_1 ; U_2 |Q U_0)\ ,\\
R_2 &\leq& I(V; Z|Q )\ ,\\
R_1 + R_2 &\leq& I(U_0 U_1; Y_1|Q ) + I(V; Z|Q ) - I(U_0 U_1; V|Q ) \ , \\
R_1 + R_2 &\leq& I(U_0 U_1; Y_2|Q ) + I(V; Z|Q ) - I(U_0 U_2; V|Q )\ ,\\
2  R_1 + R_2 &\leq& I(U_0 U_1; Y_1|Q ) + I(U_0 U_2; Y_2|Q ) + I(V; Z|Q )\nonumber \\
&&\qquad \qquad - I(U_0 U_1 U_2; V|Q ) - I(U_1 ; U_2 |Q U_0)\ , \\
2 R_1 + 2 \, R_2 &\leq& I(U_0 U_1; Y_1|Q ) + I(U_0 U_2; Y_2|Q ) + 2 I(V; Z|Q ) \nonumber \\
&&\qquad \qquad - I(U_0 U_1; V|Q ) - I(U_0 U_2; V|Q ) - I(U_1 ; U_2 | Q U_0 V) \ ,
\end{IEEEeqnarray}
for some set of arbitrarily correlated RVs of joint pmf: $P_{Q U_0 U_1 U_2 V X}$ such that the Markov chain $ ( Q,U_0,U_1,U_2,V) \mkv X \mkv ( Y_1, Y_2, Z) $ holds.
\end{theorem}

\begin{IEEEproof}
The proof is given in Appendix~\ref{MDAchievability}.
\end{IEEEproof}

\subsection{Common Description (CD) Inner Bound} 
Inspired by Marton's inner bound, we can derive what we call the ``common description" (CD) coding --worst-case of Marton's inner bound--  that consists of all rate pairs $(R_1, R_2)$ verifying:
\begin{IEEEeqnarray}{rCl}\label{CD}
R_1 &\leq& \underset{j\in\{1,2\}}{\min} I(U; Y_j|Q) \ ,\\
R_2 &\leq& I(V; Z |Q) \ , \\
R_1 + R_2 &\leq& \underset{j\in\{1,2\}}{\min} I(U; Y_j|Q) + I(V; Z|Q) - I(U; V|Q)\ ,
\end{IEEEeqnarray}
where $U$, $V$ and $Q$ are arbitrarily correlated auxiliary RVs. 

Without time-sharing, this inner bound imposes that both users in the compound setting decode the same set of variables and does not allow to treat the two possible outputs differently. However, time-sharing helps enhance the performance of this region since it allows for different signaling strategies across the time slots. The combination of the two techniques is denoted in literature as ``symbol or block expansion"~\cite{WeingartenDoF} and allows CD coding to achieve the optimal DoF for some classes of the compound MISO BC. It is easy to check that MD inner bound \eqref{MultipleDescriptionsInnerBound} recovers the CD inner bound \eqref{CD} by setting both private descriptions equal to: $U_1 \equiv \emptyset$ and $U_2 \equiv \emptyset$. Thus, implying that Marton's inner bound can achieve the optimal DoF for the compound $2\times 1$ Gaussian MISO BC, the question of whether MD inner bound can strictly improve on CD inner bound arises, and will be investigated in this section.  

\subsection{MD Coding over the BC and the Compound Channel} 
In this section, we elaborate on the fact that CD coding performs at least as good as MD coding in both the BC and  the Compound Channel. 

As for the Compound Channel, let us assume that we have a compound model with two possible channel outputs denoted by $Y_1$ and $Y_2$. We want to show that, for all joint pmfs $P_{U_0 U_1 U_2 X}$  there exists a common auxiliary RV $U^\star_0$ that yields a rate greater than the one achieved by using MD coding. Let 
\begin{IEEEeqnarray}{rCl}
R (p_{U_0 U_1 U_2 X}) & \triangleq & \min  \big\{  I(U_0 U_1 ;Y_1) \,, \, I(U_0 U_2;Y_2) \, ,\, \\
&&  \dfrac{1}{2} \big[ I(U_0 U_1 ;Y_1) + I(U_0 U_2;Y_2) - I(U_1;U_2|U_0) \big]  \big\} \ ,  
\end{IEEEeqnarray}
where  we have that: 
\begin{IEEEeqnarray}{rCl}
 I(U_0 U_1 ;Y_1) &\leq& I(U_0 U_1 U_2 ;Y_1)\ , \\
 I(U_0 U_2 ;Y_2) &\leq& I(U_0 U_1 U_2 ;Y_2) \ ,\\
 I(U_0 U_1 ;Y_1) + I(U_0 U_2;Y_2) - I(U_1;U_2|U_0) &\leq& I(U_0 U_1 U_2 ;Y_1) + I(U_0 U_1 U_2 ;Y_2) \ , 
\end{IEEEeqnarray}
and thus, 
\begin{equation}
 R (p_{U_0 U_1 U_2 X}) \leq \min \left\{ I(U_0 U_1 U_2 ;Y_1) \, , \, I(U_0 U_1 U_2;Y_2) \right\} \ . 
\end{equation} 
By letting $U^\star_0 = (U_0 U_1 U_2)$, the desired equality holds\footnote{The inequality in the inverse order is trivial by setting $U_1 = U_2 = \emptyset $.}: 
\begin{equation}
\underset{p_{U_0 U_1 U_2 X}}{\max} R (p_{U_0 U_1 U_2 X}) = \underset{p_{U_0 X}}{\max} \min \{ I(U_0; Y_1) , I(U_0;Y_2) \} \ . 
\end{equation} 

As a matter of fact, for the case of the standard BC it turns out that MD coding do not help much neither. To check this, for $Y_1 \equiv Y_2$, fix a joint pmf $p_{U_0 U_1 U_2|X}$ and let us assume that 
\begin{equation}
I(U_0 U_1 ;Y_1) - I(U_0 U_1; V) \leq I(U_0 U_2 ;Y_1) - I(U_0 U_2; V)\ . 
\end{equation} 
Then, it is easy to see that the choice $U^\star = (U_0 U_2)$ and $U^\star_1 = U^\star_2 = \emptyset$ allows us to get: 
\begin{equation} 
	R (p_{U_0 U_1 U_2 X}) \leq \underset{p_{U_0 X}}{\max} \{ I(U_0;Y_1) - I(U_0;V) \} \ . 
\end{equation} 
Hence, 
\begin{equation} 
\underset{p_{U_0 U_1 U_2 X}}{\max} R (p_{U_0 U_1 U_2 X}) = \underset{p_{U_0 X}}{\max} \,\min \left\{ I(U_0; Y_1) \,, \,I(U_0;Y_2) \right\} \ . 
\end{equation} 
Needless to say that in the compound BC, the previous assertion is not true any longer since it is not known whether the inequalities: 
\begin{IEEEeqnarray}{rCl}
 I(U_0 U_1 ;Y_1) - I(U_0 U_1; V) &\leq& I(U^\star;Y_1) - I(U^\star; V) \ , \\
 I(U_0 U_2 ;Y_2) - I(U_0 U_2; V) &\leq& I(U^\star;Y_2) - I(U^\star; V) \ ,\\
\sum_{j=1}^2\left[ I(U_0 U_j ;Y_j) - I(U_0 U_j; V) \right] - I(U_1;U_2|U_0 V) &\leq& \sum_{j=1}^2 \left[I(U^\star;Y_j) - I(U^\star; V) \right] \ ,
\end{IEEEeqnarray}
still hold for some $U^\star$, and this is the key reason for which MD is useful. However, MD proves to be useless in the cases of  the BC and the Compound Channel  while no evidence on its role in the Compound BC was stated. This motivates the following comparison between the CD and MD coding techniques for the $2\times 1$ Compound MISO BC.

\section{The Real Compound MISO BC and MD Based DPC}\label{MDinnerBounds}

The optimal transmit strategy for the non-ordered Gaussian MISO BC is to apply Dirty-Paper Coding (DPC)~\cite{caire-2003, 1056659}, which is a non-linear coding technique that allows the decoder to suppress the interference. In the sequel, we derive  the inner bounds resulting from an adequate use of the DPC scheme with the MD coding technique, referred to as MD-DPC, and later study a specific class of Compound MISO BC for which MDs are of consequent utility compared to the basic CD coding, referred to as CD-DPC. 

Consider the Compound MISO BC which consists of a source equipped with $2$ antennas and $2$ single antenna receivers. Receiver $1$ has two possible outputs, namely, $Y_1$ and $Y_2$, and let $Z$ be the channel output of the receiver $2$, where these outputs at time $t=[1,\dots,n]$ are given by
\begin{equation}\label{Setting}
\left\{\begin{array}{lll}
y_{j,t} &=& \bh_j^t \, \bx_t + n_{j,t} \ , \\
 z_{t} &=& \bg^t \, \bx_t + w_{t} \ , 
\end{array}\right.
\end{equation}
for $j \in \{1, 2\}$, where: $\bh_j$ and $\bg$ are $2 \times 1$ generic real channel vectors that are assumed to be constant throughout the transmission. Moreover, it is assumed that any subset of $2$ channels among them are linearly independent; $\bx$ is the $2\times 1$ power limited channel input vector so that $\dsE [ \bx^t \bx ] \leq P $ and last, the noise sequences $\{n_{j,t}\}$ and $ \{w_{t}\}$ are assumed to be i.i.d. draws according to a standard Gaussian distribution $\cN(0,N)$.

In this section, we will compare the CD to the MD inner bound under two different coding techniques depending  on the correlation between the private auxiliary RVs. We first start with the case where the private descriptions are \emph{uncorrelated} in the way that the encoder communicates part of the time a private description $U_1$ to help user $Y_1$ to decode the intended message, and a private description $U_2$ during the remaining part of the time to help user $Y_2$. Later, we consider \emph{arbitrary correlation} between the private descriptions in that both are transmitted all along time, resulting in a non-zero correlation cost. 


\subsection{Preliminaries and Useful Definitions}\label{preliminaries}

In the sequel, we resort to DPC~\cite{Costa1983a} in its vector formulation, thus some basic definitions and analytic formulas will be introduced herein to lighten the notation afterwards.

Let us consider the following coding scheme:
\begin{equation}\label{CommonDPC}
\left\{\begin{array}{rcl}
 U_0 &=& X_u + \alpha X_v \ , \\
 V &=& X_v \ , \\
	\bX &=& X_u \, \bB_u + X_v \, \bB_v \ ,
\end{array}\right.
\end{equation}
where $X_u \sim \cN(0, P_u)$ and $X_v \sim \cN(0, P_v)$ are independent RVs such that $P_u + P_v \leq P$. 
It is then easy to check that: 
\begin{equation}
I(U_0;Y_j) - I(U_0;V)= \log_2\left( \dfrac{ h_{j,u}^2 P_u + N }{I_j (\alpha - \beta_j)^2 + N } \right) \ , 
\end{equation} 
where: 
\begin{equation}\label{alpha0P0} 
\beta_j = \dfrac{P_u \, h_{j,u} \,h_{j,v} }{h_{j,u}^2\,P_u + N} \  \textrm{ and } \ I_j = \left(\dfrac{P_v}{P_u}\right) \dfrac{\left( h_{j,u}^2 P_u + N \right)^2 }{ h_{j,u}^2 P_u + h_{j,v}^2 P_v + N } \ . 
\end{equation}

We now choose to transmit an additive private description $X_p \sim \cN(0,x)$ while keeping the total useful power equal to $P_u$, i.e., $0 \leq x \leq P_u$. Then, with the following coding scheme: 
\begin{equation} \label{DPCPrivate}
\left\{\begin{array}{rcl}
 U_0 &=& X_{u} + \alpha X_v \ , \\
 U_j &= &X_{p} + \alpha_j X_v \ , \\
 \bX &=& ( X_{u}+ X_{p}) \bB_u + X_v \bB_v \ ,
\end{array}\right.
\end{equation}
we can optimize the value of the private DPC parameter $\alpha_j$ to state the following result. 

\begin{lemma}[Optimizing the private descriptions]\label{LemmaPrivate}
\begin{equation} 
 \max_{\alpha_j \in \mathds{R}} \left[ I_{\alpha_j}(U_0U_j;Y_j) - I_{\alpha_j}(U_0U_j;V) \right]= \frac{1}{2}\log_2\left( \dfrac{ h_{j,u}^2 P_u + N }{\dfrac{I^x_j N(\alpha - \beta^x_j)^2  }{ h_{j,u}^2 x+N } + N } \right) \ , 
\end{equation} 
 and where, for $ j \in \{1,2\} $, we have: 
\begin{equation}\label{alphaxPx}
\beta^x_j = \dfrac{(P_u -x) \, h_{j,u} \,h_{j,v} }{h_{j,u}^2 P_u + N } 
 \ \text{ and } \ 
I^x_j = \left(\dfrac{P_v}{P_u-x}\right) \dfrac{\left( h_{j,u}^2 P_u+ N \right)^2 }{ h_{j,u}^2 P_u + h_{j,v}^2 P_v + N } \ . 
\end{equation}
\end{lemma}
\vspace{1mm}
 \begin{IEEEproof} 
 The key point of the proof is that the private description, when optimized, yields an interference free link: 
 \begin{IEEEeqnarray}{rcl}
	 \max_{\alpha_j \in \mathds{R}}\left[ I_{\alpha_j }(U_j;Y_j|U_0) - I_{\alpha_j }( U_j;V|U_0) \right] &=& I(X_{p} ; Y_j| X_{u} X_v ) \\
	 &=& \frac{1}{2}\log_2\left( \dfrac{ h_{j,u}^2 x + N }{N} \right) \ .
\end{IEEEeqnarray}
The rest of the proof is relegated to Appendix~\ref{ProofLemmaDescriptions}. 
\end{IEEEproof}

 \subsection{Common Description DPC (CD-DPC)}
Consider the channel model defined by \eqref{Setting} and let us define the two following rate regions resulting from two antagonist DPC schemes:  
\begin{equation} 
\cR_{1}: \, \left\{\begin{array}{rcl} 
 R_1 &\leq& \underset{\alpha}{\max} \, \underset{j\in \{1,2\}}{\min}\,\dfrac{1}{2}  \log_2\left( \dfrac{ h_{j,u}^2 P_u + N }{I_j (\alpha - \beta_j)^2 + N } \right) \ , \\ 
R_2 &\leq& \dfrac{1}{2} \log_2 \left( \dfrac{ g_u^2 P_u + g_v^2 P_v +N }{ g_u^2 P_u + N } \right) \ ,
\end{array}\right.
\end{equation}
where $\beta_j$ and $I_j$ are given similarly by: 
\begin{equation} 
\beta_j = \dfrac{P_u \, h_{j,u} \,h_{j,v} }{h_{j,u}^2\,P_u + N} \ \text{ and } \ I_j =\left( \dfrac{P_v}{P_u} \right)\dfrac{\left( h_{j,u}^2 P_u + N \right)^2 }{ h_{j,u}^2 P_u + h_{j,v}^2 P_v + N } \, . \vspace{1mm}
\end{equation} 
The second rate region is given by the set of rate pairs satisfying: 
\begin{equation} 
\cR_2 \,: \left\{\begin{array}{rcl} 
R_2 &\leq&\dfrac{1}{2} \log_2 \left( \dfrac{ g_v^2 P_v +N }{N} \right) \ , \\
R_1 &\leq& \underset{j= 1,2}{\min}\,\dfrac{1}{2}  \log_2 \left( \dfrac{ h_{j,u}^2 P_u + h_{j,v}^2 P_v +N }{ h_{j,v}^2 P_v +N } \right) \ .
\end{array}\right. 
\end{equation}
 
\begin{proposition}[CD inner bound]
An inner bound on the capacity region of the Compound MISO BC defined in \eqref{Setting} is given by the set of rates satisfying: 
\begin{equation} 
\cR_{\textrm{CD-MISO BC}} = \bigcup_{\substack{(P_u, P_v) \\ P_u + P_v \leq P}}\,\bigcup_{\substack{\bB_u,\bB_v \\ \|\bB_u \|= 1 \\ \|\bB_v \| = 1}} \left[ \cR_1(\bB_u,\bB_v,P_u,  P_v)  \cup \cR_2(\bB_u,\bB_v,P_u,  P_v)\right]  \ . 
\end{equation}
\end{proposition}
\begin{IEEEproof}
First, note that the rate regions $\cR_1$ and $\cR_2$ are nothing but the two corner points of the CD rate region given in~\eqref{CD}. The rate region $\cR_1$ is obtained by evaluating the corner point: 

\begin{equation} 
\left\{
\begin{array}{rcl}
R_1 &\leq& \underset{j\in\{1,2\}}{\min} I(U; Y_j|Q) - I(U;V |Q) \\
R_2 &=& I(V; Z |Q) 
\end{array}\right. \ , 
\end{equation}
using the following coding scheme: 
\begin{equation} 
\left\{\begin{array}{rcl} 
\bX &=& X_u \bB_{u} + X_v\bB_v \ ,\\ 
U &=& X_u + \alpha X_v = X_u + \alpha V \ , 
\end{array}\right. 
\end{equation}
where $X_u \sim \cN(0, P_u)$ and $X_v \sim \cN(0, P_u)$ are independent RVs such that $P_u + P_v \leq P$. 

As for the second rate region $\cR_2$, it results from the evaluation of the second corner point of CD under the antagonist coding scheme, where $V$ dirty-paper codes the codewords $U$; the calculations follow in a similar manner. 
\end{IEEEproof}
 
\subsection{MD-DPC with Uncorrelated Private Descriptions}
 
In the sequel, we will evaluate the MD inner bound given in Theorem~\ref{MultipleDescriptions}. To this end, we explore two different approaches for MD-DPC depending on the existing correlation between the private descriptions, for which it will be enough to study  the specific corner points:
\begin{equation} \label{MDPrivate}
 \left\{
\begin{array}{rcl}
R_1 &\leq& \underset{j\in\{1,2\}}{\min}\big[ I(U_0 U_j; Y_j|Q) - I(U_0 U_j;V |Q) \big] \\
2 R_1 &\leq& \underset{j\in\{1,2\}}{\sum} \big[ I(U_0 U_j; Y_j|Q) - I(U_0 U_j;V |Q) \big] - I(U_1; U_2 | U_0 V Q) \\
R_2 &=& I(V; Z |Q) \ . 
\end{array}\right.
\end{equation}
The MD inner bound we derive here is based on the evaluation of \eqref{MDPrivate} via a time-sharing argument~\cite{WeingartenDoF}, where, unlike the common description, each private description is transmitted only part of the time. Both common and private descriptions apply a DPC scheme, but with difference parameters and signallings as will be clarified later. Let $Q$ be a binary valued time-sharing RV such that:
\begin{equation}
 \dsP(Q=1) = 1-\dsP(Q=2) \triangleq t \ .
\end{equation} 
Let us define the following rate region $\cR_{u}$ as: 
\begin{equation*}
\cR_{u}: \left\{\begin{array}{rcl} 
R_1 &\leq& \underset{ \alpha }{\max} \underset{ j \in \{1,2\} }{ \min} \Biggl\{ \dfrac{1}{2} \, p_Q (j) \, \log_2 \left( \dfrac{ h_{j,u}^2\, x +N}{N}\right) \\
&& \qquad \qquad \quad  + \dfrac{1}{2} \log_2 \left( \dfrac{ h_{j,u}^2 P_u + N } {I^x_{j} \, \left( \alpha - \beta^x_j \right)^2 + N + h_{j,u}^2 x } \right) \Biggr\} \ ,\\
R_2 &\leq& \dfrac{1}{2} \log_2 \left( \dfrac{g^2_{u}\, P_u+g_{v}^2\, P_v +N}{ g_{u}^2\, P_u + N} \right) \ , 
\end{array}\right.
\end{equation*}
where $\beta^x_j$ and $I^x_j$ are chosen as follows: 
 \begin{equation} 
\beta^x_j = \dfrac{(P_u -x) \, h_{j,u} \,h_{j,v} }{h_{j,u}^2 P_u + N } 
 \ \text{ and } \ 
I^x_j =\left( \dfrac{P_v}{P_u-x}\right) \dfrac{\left( h_{j,u}^2 P_u+ N \right)^2 }{ h_{j,u}^2 P_u + h_{j,v}^2 P_v + N } \ . 
\end{equation}

\begin{proposition}[MD-DPC inner bound with uncorrelated private descriptions]
An inner bound on the capacity region of the Compound MISO BC defined in~\eqref{Setting} is given by: 
 \begin{equation} 
\cR_{\textrm{MDindep-MISO BC}} = \bigcup_{ t \in [0:1] }\, \bigcup_{\substack{(P_u, , P_v) \\ P_u + P_v \leq P \\ 0\leq x \leq P_u}}\,  \bigcup_{\substack{\bB_u,\bB_v \\ \|\bB_u \|= 1 \\ \| \bB_v \| = 1}} \cR_{u}(\bB_u,\bB_v ,x,t,P_u,  P_v)  \ . 
\end{equation}
\end{proposition}
\begin{IEEEproof}
For $Q=1$, we let: 
\begin{equation}
\left\{ \begin{array}{rcl}
\bX &=& ( X_{u} + X_{p})\, \bB_u + X_v \bB_v \ ,\\
 U_0 &=& X_{u} + \alpha \, X_v \ ,\\
 U_2 &=& \emptyset \ ,\\
 U_1 &=& X_{p} + \alpha_1 \, X_v\ .
\end{array}\right.
\end{equation}
And alternately for $Q=2$, let: 
\begin{equation}
\left\{ \begin{array}{rcl}
\bX &=& ( X_{u} + X_{p})\, \bB_u + X_v \bB_v \ ,\\ 
U_0 &=& X_{u} + \alpha \, X_v \ ,\\
U_1 &=& \emptyset  \ , \\
U_2 &=& X_{p} + \alpha_2\, X_v \ .
\end{array}\right.
\end{equation}
In this case, the correlation term becomes null since $U_1$ and $U_2$ are never activated in the same time slot.
Hence, \eqref{MDPrivate} becomes equal to: 
 \begin{IEEEeqnarray}{rCl}
R_1 &\leq& I(U_0;Y_1|Q) - I(U_0;V|Q) + t \, \Bigl[ I(U_1 ;Y_1| U_0 ,Q\!=\!1) - I(U_1 ;V| U_0, Q=1) \Bigr]\ , \\ 
R_1 &\leq& I(U_0;Y_2|Q) - I(U_0;V|Q) + \bar{t} \, \Bigl[ I(U_2;Y_2| U_0 ,Q\!=\!2) - I(U_2;V| U_0, Q=2) \Bigr]\ ,\\ 
R_2 &\leq& I(V;Z|Q) \ .
\end{IEEEeqnarray}
The key point is then to note that, for $j\in\{1,2\}$:
\begin{equation}
I(U_0;Y_j|Q) - I(U_0;V|Q) \overset{(a)}{=} \dfrac{1}{2} \log_2 \left( \dfrac{ h_{j,u}^2 P_u + N } {I^x_{j} \, \left( \alpha - \beta^x_j \right)^2 + N + h_{j,u}^2 x } \right) , 
\end{equation}
where $(a)$ is a result of that the CD suffers from the interference of the private description power $h_{j,u}^2 x $ over both time slots in the exact same manner, be it from the private description $U_1$ or from $U_2$. 
Finally, the  result follows by using Lemma~\ref{LemmaPrivate} to maximize the private DPC parameters $\alpha_1$ and $\alpha_2$. 
\end{IEEEproof}

 \subsection{MD-DPC with Correlated Private Descriptions}
 In this section, we allow the private descriptions $U_1$ and $U_2$ in~\eqref{MDPrivate} to be arbitrarily  correlated. Let the set of rate pairs $\cR_{c}$ defined by: 
 
\begin{equation*} 
\cR_{c}: \left\{ \begin{array}{rcl} 
R_1 &\leq& \min \{ f_1(\alpha,x) , f_2(\alpha,x) \} \ , \\ 
R_1 &\leq& \dfrac{1}{2} \left[ f_1(\alpha,x) + f_2(\alpha,x) - \dfrac{1}{2} \log_2 ( 2 \pi e x) \right] \ , \\ 
R_2 &\leq& \dfrac{1}{2} \log_2 \left( \dfrac{ g_u^2 P_u+ g_v^2 P_v +N }{ g_u^2 P_u+ N } \right) \ , 
\end{array}\right. 
\end{equation*}
where: 
\begin{equation} 
f_j(\alpha,x) \triangleq \dfrac{1}{2} \log_2 \left( \dfrac{ h_{j,u}^2\, P_u + N }{ \dfrac{ I^x_jN(\alpha - \beta^x_{j})^2 }{ h_{j,u}^2 x+N }  + N } \right) \ , 
\end{equation}
and $\beta^x_j$ and $I^x_j$ are given similarly to~\eqref{alphaxPx} by:
\begin{equation} 
\beta^x_j = \dfrac{(P_u -x) \, h_{j,u} \,h_{j,v} }{h_{j,u}^2 P_u + N } 
 \ \text{ and } \ 
I^x_j = \left(\dfrac{P_v}{P_u-x}\right) \dfrac{\left( h_{j,u}^2 P_u+ N \right)^2 }{ h_{j,u}^2 P_u + h_{j,v}^2 P_v + N } \, . 
\end{equation}

\begin{proposition}[MD inner bound with correlated private descriptions]
An inner bound on the capacity region of the Compound MISO BC is given by: 
\begin{equation} 
\cR_{\textrm{MDcorr-MISO BC}} = \bigcup_{\substack{\bB_u,\bB_v \\ \|\bB_u \|= 1 \\ \| \bB_v \| = 1}} \, \bigcup_{\substack{(P_u,  P_v) \\ P_u + P_v \leq P \\ 0\leq x \leq P_u}} \,\bigcup_{\alpha \in \mathds{R}} \,\cR_{c}(\bB_u,\bB_v ,\alpha,x,P_u,  P_v) \ . 
\end{equation}
\end{proposition}

\begin{IEEEproof}
To prove our claim, we resort to the MD coding inner bound letting, in the single letter, the two ARV be $U_1$ and $U_2$ equal given $Q$, $U_0$, and $V$. The correlation term becomes thus: 
\begin{equation}
 I(U_1;U_2|Q U_0 V) = H(U_1 | Q U_0 V) = H(U_2 | Q U_0 V) \ . 
\end{equation}

Let us use the following coding scheme: 
\begin{equation} 
\left\{\begin{array}{rcl}
 \bX &=& ( X_{u}+ X_{p}) \bB_u + X_v \bB_v \ , \\
 U_0 &=& X_{u} + \alpha X_v \ , \\
 U_1 &=& X_{p} + \alpha_1 X_v \ , \\
 U_2 &=& X_{p} + \alpha_2 X_v \ , \\
 V &=& X_v \ .
\end{array}\right.
\end{equation}
It is then straightforward with the result of Lemma~\ref{LemmaPrivate}, that the achievable rates are those given in the proposition. 
\end{IEEEproof}

\subsection{MD-DPC strictly outperforms CD-DPC}

Let us now be in the presence of the most stringent compound model where $\bh_1$ and $\bh_2$ are unit-norm orthogonal channels. Assume also that the other user's channel is quite accommodating such that $\bg$ is orthogonal to the ``mean channel" of user $1$, 
\begin{equation}
 \bg \,\bot\, \dfrac{1}{\sqrt{2}}(\bh_1 + \bh_2) = \bh_{1,2}\  .
\end{equation} 
In order to show that MD-DPC strictly outperforms CD-DPC for this setting, we need to evaluate CD-DPC inner bound based on the corresponding channel models. Then, we show that the MD-DPC inner bound strictly outperforms it. 
\subsubsection{CD-DPC inner bound}
We start by characterizing CD-DPC inner bound in a closed form. 

\begin{proposition}[CD-DPC inner bound]
The CD-DPC inner bound writes as the set of rate pairs satisfying: 
\begin{equation} 
\left\{ \begin{array}{rcl}
 R_1 &\leq& \dfrac{1}{2} \log_2 \Biggl( \dfrac{P_u + 2 N}{P(\eta) + 2 N } \Biggr) \ , \\ 
R_2 &\leq& \dfrac{1}{2} \log_2 \left( \dfrac{ (1-\eta) P_v + 2 N }{2 N} \right) \ , 
\end{array}\right.
\end{equation}
for some $\eta \in [-1:1]$, where
\begin{equation}
 P(\eta) \triangleq \dfrac{ (1-\eta )P_v P_u}{P + 2 N + \sqrt{(P+ 2N)^2+ (\eta^2-1) P_v^2} } \ . 
\end{equation} 
\end{proposition} 

\begin{IEEEproof} 
The proof is relegated to Appendix~\ref{CDCornerPoints}.
\end{IEEEproof}

\begin{remark}
In order to derive the optimal value of $\eta$ for the overall rate region, we look at the resulting weighted sum-rate. If we let $\mu \in \Re_+$, then the optimization of $R_1 + \mu R_2$ over $\eta$ depends on the value of $\mu$. For $\mu = 0 $, the optimal choice is $\eta = 1$ that is we have to transmit in a direction that is collinear with the mean channel $\bh_{1,2}$, as for the case $\mu \rightarrow \infty $, the optimal choice is to let $\eta = -1$, which means to transmit the information for the second user in a direction that is collinear to its channel. For intermediate values of $\mu$, the weighted sum-rate is not necessarily maximized with either choices of $\eta$. 
\end{remark}

We evaluate the two MD-DPC inner bounds as a function of $x$, the power dedicated to private descriptions, and compare them to the case $x=0$, i.e., the CD-DPC inner bound. We let $\bB_u = \bh_{1,2}$ and thus, by transmitting information to user $1$ orthogonal to the channel of user $2$. \vspace{2mm}

\subsubsection{MD-DPC with correlated private descriptions outperforms CD-DPC}~\\
To evaluate the gain of MD-DPC inner bound with \emph{arbitrarily correlated} private descriptions, note that if at least $0 \leq x \leq (2\pi e)^{-1} $, then the bound on $R_1$ can be written as follows: 
\begin{IEEEeqnarray}{rCl}
R_1 &\leq& \dfrac{1}{2} \underset{ \alpha\in \mathds{R} }{\max} \min \left\{ \, f_1(\alpha ,x)\ , \ f_2(\alpha ,x) \right\} \\ 
 &=&\dfrac{1}{2} \log_2 \left( \dfrac{P_u + 2 N}{ \dfrac{ (P_u- x)}{ x + 2 N } \dfrac{2N}{P_u} P(\eta) + 2 N } \right) \\ 
 & \overset{(a)}{\geq} & \dfrac{1}{2} \log_2 \left( \dfrac{P_u + 2 N}{ P(\eta) + 2 N } \right) \ , 
\end{IEEEeqnarray} 
where $(a)$ follows from the fact that the function $ f: x \mapsto \dfrac{ (P_u- x)}{ x + 2 N } $ is strictly decreasing in $x$. Indeed, the inequality in $(a)$ is strict for non-degenerate power parameters $P_v \neq 0$ and $\eta \neq 1 $, which corresponds to $R_2 \neq 0$ and yields the proof of the claim.\vspace{2mm}

\subsubsection{MD-DPC with uncorrelated private descriptions outperforms CD-DPC}
As for MD-DPC inner bound with \emph{uncorrelated private descriptions}, the constraint on the rate $R_1$ writes as:
 \begin{equation} 
 R_1 \leq \dfrac{1}{2} \log_2 \left( \dfrac{P_u + 2 N}{ \dfrac{ (P_u- x)}{ \sqrt{x + 2 N} }\dfrac{\sqrt{2N}}{P_u}P(\eta) + \sqrt{2N} \, \sqrt{x + 2N} } \right) \ , 
\end{equation}
for which we have considered a time-sharing $t = \bar{t} = 0.5$. Now, the function given by
\begin{equation}
g(x) \triangleq \dfrac{ (P_u - x)}{ \sqrt{x + 2 N} }\dfrac{ P(\eta) }{P_u} + \sqrt{x + 2N} \ , 
\end{equation}
 is not compulsorily strictly decreasing in $x$ for all values of $\eta$. However, it is clear that: 
 
\begin{equation}
g^\prime(x) = \dfrac{ (x+2N)P(\eta) + (P_u+2N)(P_u-P(\eta) ) }{2P_u(x+2N)^{3/2}} \ , 
\end{equation}
and since $ 0 \leq x \leq P_u $, then:
\begin{equation}
g' (x) \leq \dfrac{1}{4 P_u N \sqrt{2 N} }(P_u+2N)\bigl(P_u- 2 P(\eta)\bigr) \ . 
\end{equation}

Thus, $P(\eta) > \frac{P_u}{2} $ suffices to have the function $g$ strictly decreasing in $x$, and thus, the claim of strict optimality would be proved. Note that if e.g. $P \geq 4N$, then for values of $\eta$ close to $-1$, i.e., $R_2$ close to second user's capacity, the gain is strictly positive and more significant. \vspace{2mm}

\subsubsection{Comparison of the MD-DPC inner bounds} 
\begin{figure}[ht]
 \centering
 \includegraphics[scale=0.6]{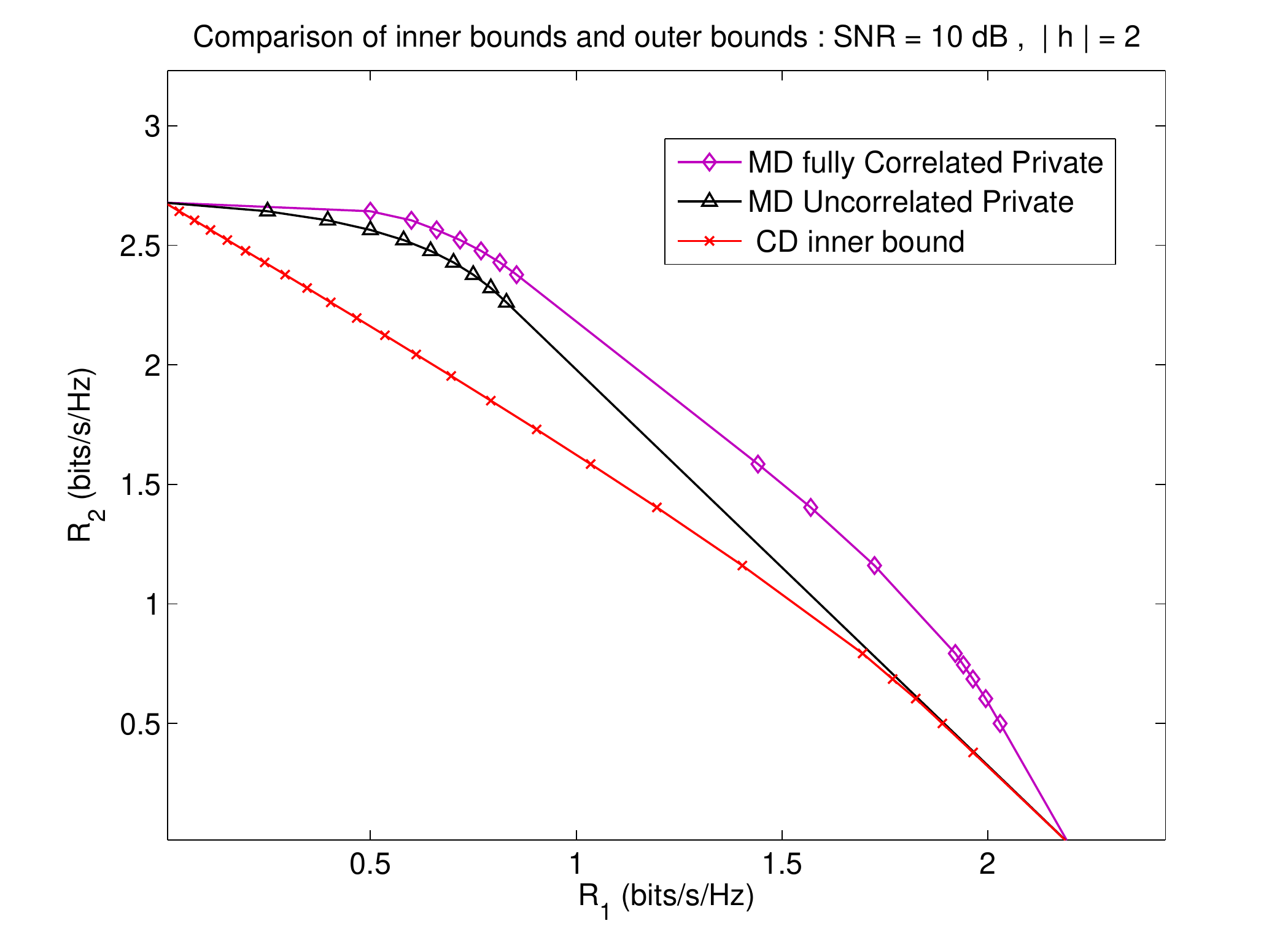} 
 \caption{Comparison of the CD-DPC and the MD-DPC inner bounds with uncorrelated and correlated private descriptions: $\textrm{SNR} = 10$~dB, $\|\bh_1\| = \|\bh_2\| = 2 $.}
 \label{FigureInnerBounds} 
 \end{figure}

An interesting question to investigate is whether the MD inner bound with correlated descriptions outperforms or not the same with uncorrelated descriptions. These two bounds compare differently following the values of the channel gains. The MD with uncorrelated private descriptions makes each user loose:  
\begin{equation}
\dfrac{1}{4} \log_2 \left( \dfrac{\|\bh\|^2 x + 2 N}{ 2 N} \right)
\end{equation}
compared to the single rates of the MD-DPC with correlated descriptions. Whereas the latter, through the correlation coefficient, engenders a loss of 
\begin{equation}
\dfrac{1}{4} \log_2 \left( 2 \pi e x \right) \ .
\end{equation}
Thus, the relative behavior of these two bounds depends on the specific values of $N$, $P_u $ and $\|\bh\|^2 $. In Fig.~\ref{FigureInnerBounds}, we plot the corresponding rate regions for $\textrm{SNR} = 10$~dB, $\|\bh_1\| = \|\bh_2\| = 2 $ and the assumptions made on the channels' structure.

\subsection{Block Expansion}
\begin{figure}[ht]
 \centering
 \includegraphics[scale=0.6]{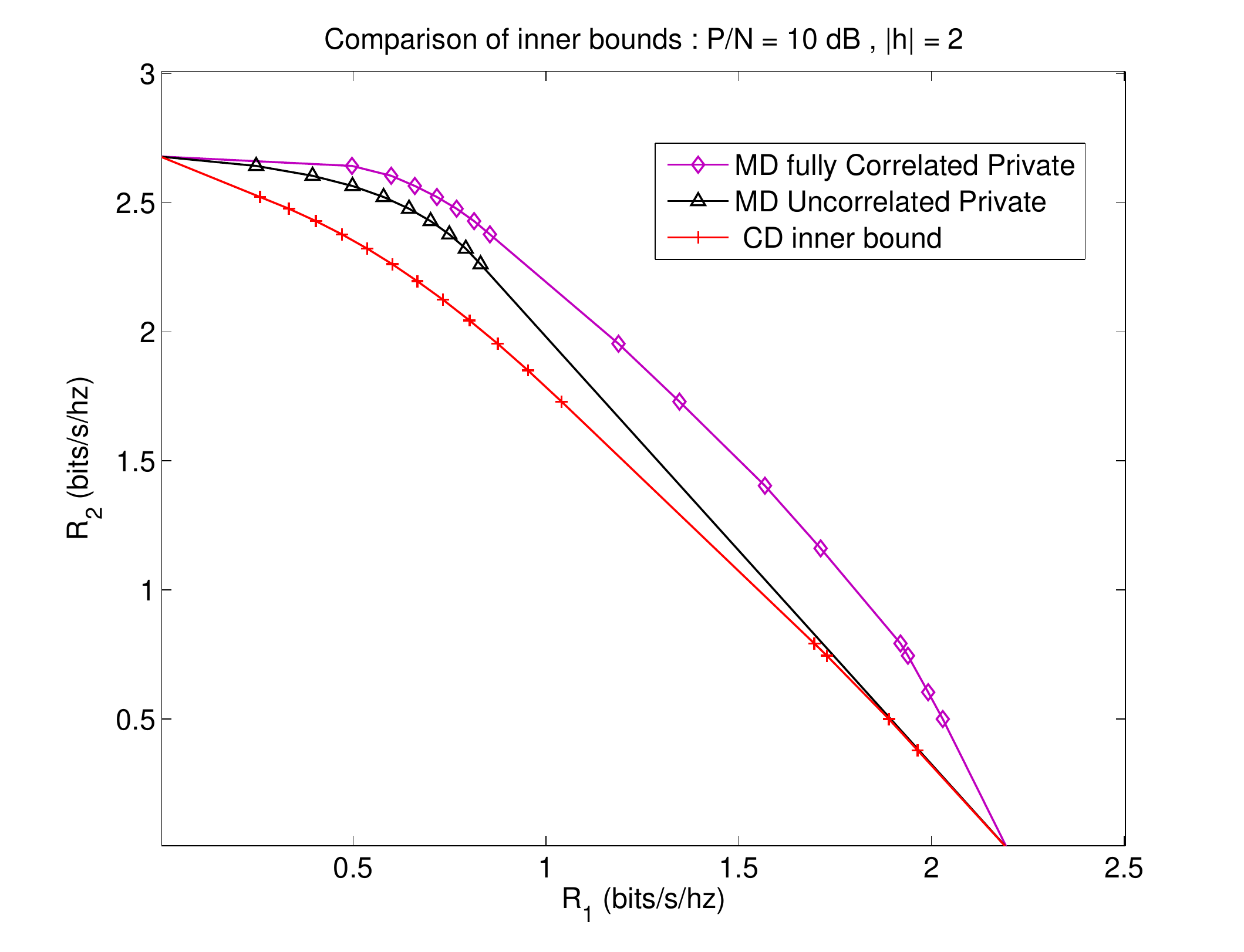} 
 \caption{Comparison of the CD-DPC and MD-DPC with uncorrelated and correlated private descriptions inner bounds with a time-sharing argument: $\textrm{SNR} = 10$~dB, $\|\bh_1\| = \|\bh_2\| = 2 $. }
 \label{FigureInnerBoundsTimeSharing} 
 \end{figure}

Last, the bounds we have studied so far did not allow for different encoding parameters across time slots. The reason is that the question we were exploring is one of the utility of private descriptions in the Compound MISO BC. Now, if we combine CD inner bound and MD inner bound with correlated private descriptions both with a time-sharing argument where in each time slot, a new coding scheme is used (in terms of beams, power allocations and DPC parameters), then one could expect that the behavior is still captured by the obtained bounds. Fig.~\ref{FigureInnerBoundsTimeSharing} corroborates the previous statement. 

 Yet, Block Expansion does not enhance much the performance of MD-DPC coding scheme, the reason being these schemes allow already for good coding schemes, however, CD-DPC is much more enhanced by Block Expansion. Indeed, in the DoF analysis, Time Sharing is crucial for CD-DPC to be DoF optimal \cite{WeingartenDoF}.
 
 \subsection{Outer Bound on the Capacity of the Compound MISO BC}\label{OuterBoundMD}
In this section, we present an outer bound on the capacity region of the Compound MISO BC which consists in the intersection of some rate regions. 

Let us introduce the following channel matrices: 
\begin{IEEEeqnarray}{rCl}
\bg_{1,2} &\triangleq& \left[ \bg \ \bh_1 \ \bh_2 \right]\ , \\
\bh_{1,z} &\triangleq & \left[ \bh_1\ \bg \right] \ ,\\
\bh_{2,z} &\triangleq& \left[ \bh_2\ \bg \right] \ .
\end{IEEEeqnarray}
We then define the corresponding channel outputs to the channel $\bg_{1,2}$, that has the same marginal as the output formed by the concatenation of $[Z \ Y_1 \ Y_2 ] $, as $Z_{1,2}$, and we define similarly the two outputs $Y_{1,z}$ and $Y_{2,z}$. The following theorem gives the resulting outer bound. 

\begin{theorem}[Outer bound on the capacity of the Compound MISO BC]\label{outer-MISO-BC}
An outer bound on the capacity region of the Compound MISO BC is given by the set of rate pairs: 
\begin{equation}
\cO = \cC_1 \cap \cC_2 \cap \cC_{1,2} \cap \cC_{z} \ , 
\end{equation}
where $\cC_j$ is the capacity region of the BC with outputs $(Y_j , Z)$, for $j \in \{1,2\}$,  
\begin{IEEEeqnarray}{rCl}
\cC_j = \underset{{\substack{(\bK_u , \bK_v) \\ tr(\bK_u + \bK_v)\leq P}}}{ \bigcup }&&\biggl\{ (R_1,R_2)\in  \mathds{R}^2_+ \, , \nonumber\\
R_1 &\leq& \dfrac{1}{2} \log_2 \left( \dfrac{ \bh_{j}^t \bK_u \bh_{j}+ N }{N} \right) \\
R_2 &\leq& \dfrac{1}{2} \log_2 \left( \dfrac{ \bg^t ( \bK_u + \bK_v) \bg +N }{\bg^t \bK_u \bg+N } \right) \\
& \text{or}&\nonumber  \\
R_1 &\leq& \dfrac{1}{2} \log_2 \left( \dfrac{ \bh_{j}^t ( \bK_u + \bK_v)\bh_{j}+N }{ \bh_{j}^t \bK_v \bh_{j}+N } \right) \\
R_2 &\leq& \dfrac{1}{2} \log_2 \left( \dfrac{ \bg^t \bK_v \bg +N }{ N } \right) \biggr\} \ , 
\end{IEEEeqnarray}
$\cC_{1,2}$ is the capacity region of the Compound MISO BC with outputs $(Y_1 , Z_{1,2})$ and $(Y_2 , Z_{1,2})$, 
\begin{IEEEeqnarray}{rCl}
\cC_{1,2} = \underset{{\substack{(\bK_u , \bK_v) \\ tr(\bK_u + \bK_v)\leq P}}}{ \bigcup } &&\biggl\{ (R_1,R_2) \in  \mathds{R}^2_+ \, , \nonumber\\
R_1 &\leq& \underset{j \in \{1,2\}}{\min} \,\dfrac{1}{2}\log_2 \left( \dfrac{ \bh_{j}^t (\bK_u + \bK_v )\bh_{j}+N}{\bh_{j}^t \bK_v \bh_{j}+ N} \right)\ , \\
R_2 &\leq& \dfrac{1}{2} \log_2 \left( \dfrac{\left| \bg_{1,2}^t \bK_v \bg_{1,2}+N \mathbf{I}_3\right|}{N^3} \right) \biggr\}
\end{IEEEeqnarray}
and finally, $\cC_{z}$ is the capacity region of the Compound BC with outputs $(Y_{1,z},Z)$ and $(Y_{2,z},Z)$, 
\begin{IEEEeqnarray}{rCl}
\cC_{z} = \underset{{\substack{(\bK_u , \bK_v) \\ tr(\bK_u + \bK_v)\leq P}}}{ \bigcup }&&\biggl\{ (R_1,R_2) \in  \mathds{R}^2_+ \, ,\nonumber \\
R_1 &\leq&\underset{j \in \{1,2\}}{\min}\,  \dfrac{1}{2}\log_2 \left( \dfrac{ \left| \bh_{j,z}^t \bK_u \bh_{j,z}+N \mathbf{I}_2 \right|}{ N^2} \right) \\
R_2 &\leq& \dfrac{1}{2} \log_2 \left( \dfrac{ \bg^t (\bK_u + \bK_v ) \bg +N } { \bg^t \bK_u \bg + N } \right) \biggr\}\ .
\end{IEEEeqnarray}
\end{theorem}\vspace{2mm}

\begin{IEEEproof}
The proof is straightforward  from the following observations. The fact that the capacity of the considered  compound model is always included in the intersection of the capacities of the BCs $\cC_1$ and $\cC_2$, and that this setting is a degraded version of the setups where there is a least one user with an extra receive antenna, whose capacities are given in references~\cite{WeingartenDegradedMIMO},~\cite{HonFaChong2013}. 
\end{IEEEproof} 

 \begin{figure}[ht]
 \centering
 \includegraphics[scale=0.6]{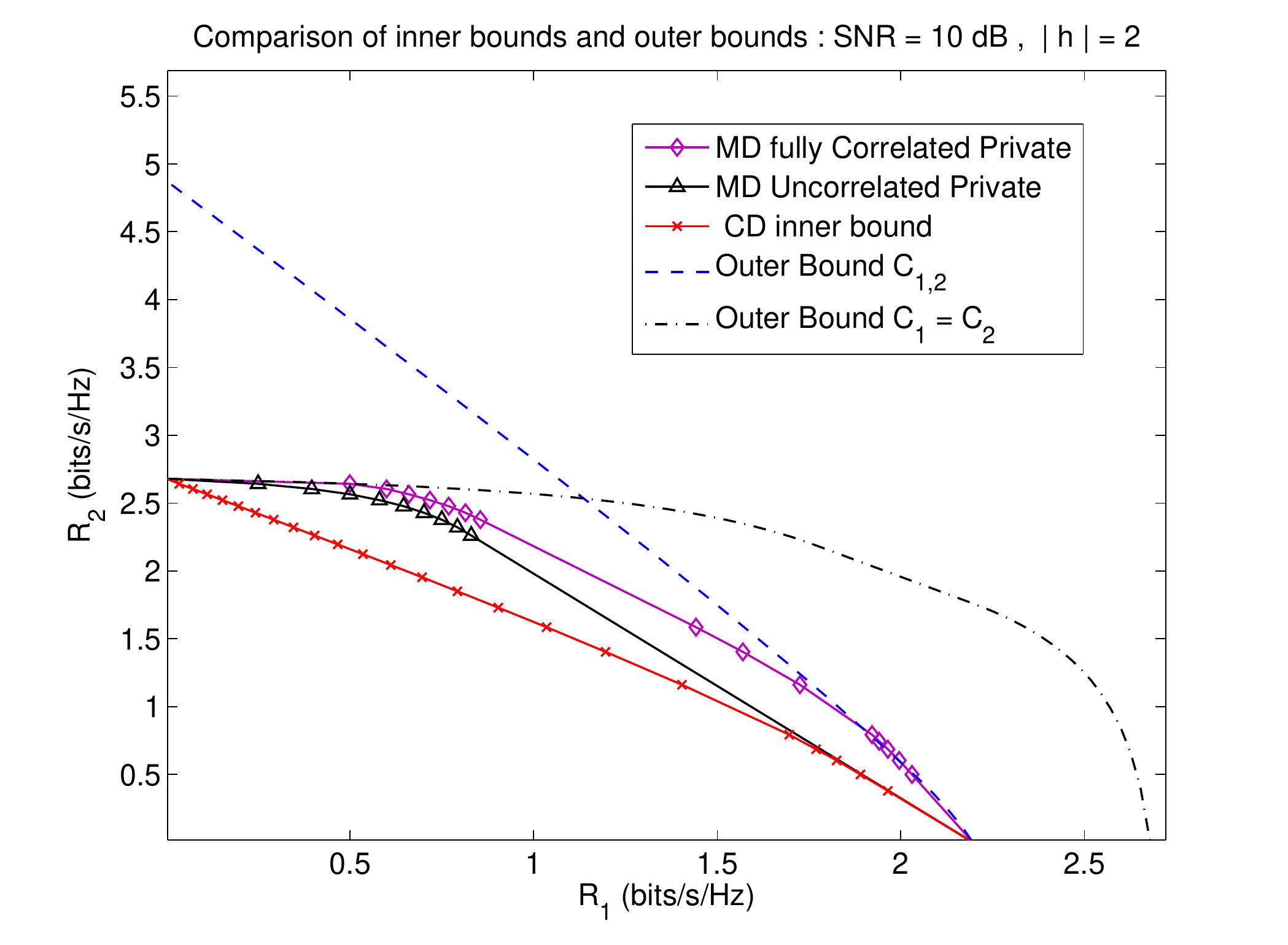} 
 \caption{Comparison of the inner bounds and the intersection of the outer bounds: $\textrm{SNR} = 10$~dB, $\|\bh_1\| = \|\bh_2\| = 2 $. }
 \label{InnerOuterComparison} 
 \end{figure}

\begin{remark}
The outer bound stated in Theorem~\ref{outer-MISO-BC} is tight in the high SNR regime and thus is DoF optimal. To check this, notice that the bounds $\cC_1$, $\cC_2$ and $\cC_z$ attain each the points $(d_1\leq 1 \ , \ d_2\leq 1)$ by letting $\bK_u = \bg^\perp \times (\bg^{\perp})^t $. As for the bound $\cC_{1,2}$, it achieves all the points $( 2 d_1 + d_2 \leq 2)$, thus the intersection of these two regions leads to the optimal DoF. 
\end{remark}

In Fig.~\ref{InnerOuterComparison}, we plot the inner and outer bound for  intermediate SNR values.  Although the gap with the outer bound suggests that the inner and outer regions do not meet,  it is our belief that the inner bound is tight while our outer remains rather loose.

 \section{Summary and Discussion}\label{Conclusion}
We start our conclusions with the analysis of the relative behavior of the MD and the ID inner bounds, to understand if there is any mutual inclusion between the two bounds. The question we want to answer is whether introducing multiple descriptions, one for each instance in the compound setting, allows to recover the ID inner bound. We also would like to understand to what extent decoding interference is crucial for Marton's worst case inner bound. 
 
\subsection{Can Multiple Descriptions or Interference Decoding techniques recover each other?}\label{MDVsID}
For this sake, we evaluate the MD inner bound in the case of the discrete example studied in Section~\ref{Application} and try to identify a set of auxiliary RVs yielding the capacity region. For the discrete Compound BC we studied earlier, we assumed that user $1$ could observe one of two possible channel instances, namely, $Y_1$ and $Y_2$, such that $Y_2$ is more capable than both $Y_1$ and $Z$, and $Y_1$ be a degraded version of $Z$. The maximizing choice of auxiliary RVs led to $Z$ and $Y_2$ decoding all the signal and $Y_1$ decoding only its intended information. 

The capacity region is of the form: 
\begin{equation} 
\left\{
\begin{array}{rcl}
R_1 &\leq& I(Q ; Y_1 ) \ ,\\ 
R_1 + R_2 &\leq& I(Q ; Y_1 ) + I(X; Z|Q) \ .
\end{array} \right.\vspace{2mm}
\end{equation}
We next discuss a formulation of the MD inner bound that captures the intuition of the capacity achieving choice of auxiliary RV for ID inner bound. Indeed, the encoder does not transmit a common description to the two users interested in the same message, but communicate only private descriptions to them. However, in the present case the common auxiliary RV $Q$ is no longer a time-sharing variable as it was the case in Section~\ref{MDcompound}, it can carry common information to all receivers as well. With this, we can achieve the set of rate pairs satisfying: 

\begin{equation}\label{R_3ARV}
\cR_{\textrm{3-ARV}} = \bigcup_{p_{Q U_1U_2VX}}\left(  \bigcup_{( T_{1,1}, T_{1,2}, T_2) \in \dsT(p)} \cM( p,T_{1,1}, T_{1,2}, T_2) \right) \ ,
\end{equation}
where $\cM$ and $\dsT$ are respectively defined by the following: 
\begin{equation}
	\cM: \, \left\{\begin{array}{rcl}
 T_2 & \leq& I( V; Z | Q)\,,\\
	R_0 + T_2 &\leq& I(Q V; Z )\,, \\
	T_{1,1} &\leq& I( U_1; Y_1 |Q)\,, \\
	R_0 + T_{1,1} &\leq& I(Q U_1; Y_1)\,, \\
	T_{1,2} &\leq& I( U_2;Y_2 |Q)\,, \\
	R_0 + T_{1,2} &\leq& I(Q U_2; Y_2)\,, \\
	\end{array}\right.
\end{equation}
\begin{IEEEeqnarray}{rcl}
 \dsT = \biggl\{ (T_{1,1}, T_{1,2} ,T_2 ): \quad T_2 \geq R_2 \, &,&\, \min \{ T_{1,1},T_{1,2} \} \geq R_1\,,\\
	T_{1,1} - R_1 + T_2 - R_2 &>& I(U_1;V|Q)\,, \\
 T_{1,2} - R_1 + T_2 - R_2 &>& I(U_2;V|Q) \,,\\
 T_{1,1} - R_1+ T_{1,2} - R_1 &>& I(U_1;U_2|Q)\,, \\
T_{1,1} + T_{1,2} - 2 \, R_1 +T_2 - R_2 &>& I(U_1;U_2|Q) + I(U_1 U_2;V|Q) \biggr\} 	\ .
\end{IEEEeqnarray}

\begin{IEEEproof} 
The proof is relegated to Appendix~\ref{Proof3ARV}. 
\end{IEEEproof}

We know that an optimal transmission scheme to achieve the capacity region of the considered BEC/BSC requires both users $Z$ and $Y_2$ to decode all messages while restricting the weaker user $Y_1$ to decode only the common message. Hence, we rely on this argument to build the straightforward extension of Marton's coding scheme, i.e., $V = U_2 = X$ and $U_1 = Q$, which along with rate splitting leads to the following achievable rate region: 
\begin{equation}
\left\{\begin{array}{rcl}
 R_1 &\leq& I(Q;Y_1)\,, \\
 R_1 + R_2 &\leq& I(X;Z|Q) + I(X;Y_2|Q) + \min\{I(Q;Y_1), I(Q;Y_2)\} - H(X|Q)\, . 
\end{array} \right.
\end{equation}
In the general case, there is strong evidence that the above rate region induced by MD is strictly included in the capacity region given by: 
\begin{equation}
\left\{\begin{array}{rcl}
 R_1 &\leq& I(Q;Y_1)\,, \\
 R_1 + R_2 &\leq& I(X;Z|Q) + I(Q;Y_1)\, ,
\end{array} \right.
\end{equation}
that is achieved by using ID, which yields: 
\begin{equation}
\left\{\begin{array}{rcl}
 R_1 &\leq& I(Q;Y_1)\,, \\
 R_1 + R_2 &\leq& \min \{ I(X;Z|Q) , I(X;Y_2|Q)  \} + I(Q;Y_1)\ , \\
 R_1 + R_2 &\leq& \min \{ I(X;Z ) , I(X;Y_2 )  \} \ , 
\end{array} \right.
\end{equation}
where $Y_1$ is degraded with respect to $Z$ and $Y_2$ is more capable than $Z$. The inclusion results from the fact that there exist $P_{X|Q}$ for which
\begin{equation}
 I(X;Y_2|Q) - H(X|Q) < 0 \ . 
\end{equation}
Thus, MD does not seem to be enough to achieve the capacity region of the compound model investigated in Section~\ref{Application}. This is due to the fact that the cost engendered by precoding against interference prevents from decoding it which results in a loss proportional  to its entropy. Therefore, it appears that ID outperforms MD in some cases.  

On the other hand, in the MISO case, imposing users to decode interference is sub-optimal, at least from a DoF perspective, since ID introduces sum-rates constraints of the form 
 \begin{equation}
 R_1 + R_2 \leq I(X;Y_1) \ ,
 \end{equation}
and thus, prevents the sum-DoF from reaching values greater than $1$ which we already know is sub-optimal. Therefore, it is crucial to precode against interference. 
 
Summarizing, since neither MD coding or ID seem to generalize all the results obtained herein one can benefit from the combination of both techniques and thus, from the optimization of both encoding and decoding schemes.
 
\subsection{Summary}\label{Summary}
In this work, we explored a decoding and a encoding technique for the two-user memoryless Compound Broadcast Channel (BC). We first studied the role of Interference Decoding (ID) where an achievable rate region is derived by using ``single per-message description" codes constructed via \emph{superposition coding} and \emph{random binning}. At the decoders, the constraint of decoding only the intended message is alleviated to allow each of the users to decode or not the other user's (interference) message. Unlike for the standard two-user BC, this strategy proves to be useful in compound setups, where channel uncertainty prevents the encoder from coding optimally for each possible BC formed by all pairs of channels in the set. A simple outer bound is also derived based on the best outer bound hitherto known on the capacity region of the two-user BC. This outer bound captures one of the most stringent effects of simultaneity of users over the random codes constructed: antagonist coding strategies. 

Surprisingly enough, ID not only outperforms Non-Interference Decoding (NID) technique, i.e., Marton's worst-case rate region, but also allows to achieve the capacity of a class of non-trivial BC while NID stays strictly suboptimal. Thus, though the coding scheme is simple (in terms of the number of auxiliary variables involved and of the complexity of the encoding operation) the decoders' optimization allows to palliate the uncertainty at the source. 

Later, we studied an encoding technique with a more evolved coding strategy, namely Multiple Description (MD) coding. The source transmits to the group of users, interested in the same message, common and private descriptions. For the specific case of the Compound MISO BC, resorting to MD is essential since a common description, i.e., applying DPC with a single description  cannot accommodate the interference seen by each instance of the users channels in the set, unless combining it with a time-sharing argument. The key point in the MISO BC setting is that using a fraction of power to transmit the private descriptions is useful for all SNR ranges while turns out to be DoF optimal. Indeed, each private description creates an interference free link and thus each user can recover a part of its rate interference free. 

Finally, we addressed the question of whether MD or ID may generalize each other. It appears that none of these schemes can perform well for ordered and non-ordered class of Compound BCs at once, mainly because the two strategies strongly rely on two different interference mitigation techniques: precoding against interference and decoding interference. The first results in a rate loss tantamount to a correlation cost while the latter results in an extra sum-rate constraint.

 As a conclusion, it would be worth mentioning the benefits of combining these two schemes to yield a larger inner bound, and thus, full advantage would be taken from the joint optimization of the encoding technique (MD coding) and the decoding technique (ID). 


\newpage

\appendices
\section{Useful Notions and Auxiliary Results}

In this appendix we provide basic notions on some concepts used in this paper.

\label{sec:typical}

Following~\cite{csiszar1982information}, we use in this paper \emph{strongly typical sets} and the so-called \emph{Delta-Convention}. 
Some useful facts are recalled here.
Let $X$ and $Y$ be RVs on some finite sets $\cX$ and $\cY$, respectively. We denote by $p_{XY}$ (resp. $p_{Y|X}$, and $p_X$) the joint pmf of $(X,Y)$ (resp. conditional distribution of $Y$ given $X$, and marginal distribution of $X$). 

\begin{definition}
For any sequence $x^n\in\cX^n$ and any symbol $a\in\cX$, notation $N(a|x^n)$ stands for the number of occurrences of $a$ in $x^n$.
\end{definition}

\begin{definition}
A sequence $x^n\in\cX^n$ is called \emph{(strongly) $\delta$-typical} w.r.t. $X$ (or simply \emph{typical} if the context is clear) if
\[
\abs{\frac1n N(a|x^n) - P_X(a)} \leq \delta \ \text{ for each } a\in\cX \ ,
\]
and $N(a|x^n)=0$ for each $a\in\cX$ such that $P_X(a)=0$.
The set of all such sequences is denoted by $\typ{X}$.
\end{definition}

\begin{definition}
Let $x^n\in\cX^n$.
A sequence $y^n\in\cY^n$ is called \emph{(strongly) $\delta$-typical (w.r.t. $Y$) given $x^n$} if
\[
\abs{\frac1n N(a,b|x^n,y^n) - \frac1n N(a|x^n)P_{Y|X}(b|a)} \leq \delta \ \text{ for each } a\in\cX, b\in\cY \ ,
\]
and, $N(a,b|x^n,y^n)=0$ for each $a\in\cX$, $b\in\cY$ such that $P_{Y|X}(b|a)=0$.
The set of all such sequences is denoted by $\typ{Y|x^n}$.
\end{definition}

\emph{Delta-Convention~\cite{csiszar1982information}:\ }
For any sets $\cX$, $\cY$, there exists a sequence $\{\delta_n\}_{n\in\bN^*}$ such that lemmas below hold.\footnote{As a matter of fact, $\delta_n\to0$ and $\sqrt{n}\,\delta_n\to\infty$ as $n\to\infty$.}
From now on, typical sequences are understood with $\delta=\delta_n$. 
Typical sets are still denoted by $\typ{\cdot}$.

\begin{lemma}[{\cite[Lemma~1.2.12]{csiszar1982information}}]\label{lem-concentration}
There exists a sequence $\eta_n\toas{n\to\infty}0$ such that
\[
P_X^n(\typ{X}) \geq 1 - \eta_n \ .
\]
\end{lemma}

\begin{lemma}[{\cite[Lemma~1.2.13]{csiszar1982information}}]
\label{lem:cardTyp}
There exists a sequence $\eta_n\toas{n\to\infty}0$ such that, for each $x^n\in\typ{X}$,
\begin{IEEEeqnarray*}{c}
\abs{\frac1n \log \norm{\typ{X}} - H(X)} \leq \eta_n 		\ ,\\
\abs{\frac1n \log \norm{\typ{Y|x^n}} - H(Y|X)} \leq \eta_n	\ .
\end{IEEEeqnarray*}
\end{lemma}

\begin{lemma}[Asymptotic equipartition property]
\label{lem:AEP}
There exists a sequence $\eta_n\toas{n\to\infty}0$ such that, for each $x^n\in\typ{X}$ and each $y^n\in\typ{Y|x^n}$,
\begin{IEEEeqnarray*}{c}
\abs{-\frac1n \log P^n_X(x^n) - H(X)} \leq \eta_n 				\ ,\\
\abs{-\frac1n \log P^n_{Y|X}(y^n|x^n) - H(Y|X)} \leq \eta_n		\ .
\end{IEEEeqnarray*}
\end{lemma}

\begin{lemma}[Joint typicality lemma~\cite{cover2006elements}]
\label{lem:jointTypicality}
There exists a sequence $\eta_n\toas{n\to\infty}0$ such that
\[
\abs{-\frac1n \log P^n_Y(\typ{Y|x^n}) - I(X;Y)} \leq \eta_n \
\textrm{ for each }x^n\in\typ{X}\ .
\]
\end{lemma}

\section{Sketch of the Proof of Theorem~\ref{theo-ID-CBC}}\label{ProofMainTheorem}
Let $j \in \cJ$ be the index of an arbitrary pair of users in the compound set. We first show the achievability of the union of the four regions for this channel $\bigcup_{i \in [1:4]} \cT_{i}$. For convenience of notations we drop the index $j$. 
\subsection{Outline of Proof}
The coding scheme we use is as follows: 
\begin{itemize}
	\item We use three auxiliary RVs, one for each message, 
	\item We perform binning on the two auxiliary RV that code for the private messages, superposing them over the auxiliary RV coding for the common message, 
	\item The decoding will introduce the principle of list decoding, which will allow us to combine two decoding techniques, 
	\item The error probability will be shown to be directly related to the list size, and thus, bounding the list size will allow us to have a tight bound on the average probability of error, 
	\item The intersection of the union of the regions comes from the fact that we use two different decoding functions at the two users. 
	\end{itemize}
	
\subsection{Detailed Proof}
	\underline{\textit{Codebook generation:}}
	The encoding is similar to that of Marton's coding with a common message.
	
	Fix, $P_{Q}, P_{U|Q}, P_{V|Q}$ and let $T_1 \geq R_1$ and $T_2\geq R_2$ be four positive rates. 
	
	Generate $2^{nR_0}$ $n$-sequences $q^n(w_0) , w_0 \in \cM_0$ following each the probability distribution: 
	\begin{equation}
	P^n_{Q} (w_0)= \prod^n_{i=1} P_{Q} (q_i (w_0) ) \ , 
 \end{equation} and set them all in $\cC_0 $. 
 
	For each $q^n(w_0)$, generate $2^{nT_1}$ $n$-sequences $u^n(l_1,w_0)$, $l_1 \in [1:2^{nT_1}]$ each following 
	\begin{equation}
	P^n_{U|Q}( u^n(l_1,w_0) )= \prod^n_{i=1} P_{U|Q} (u_i (l_1,w_0)|q_i (w_0)) \ , 
 \end{equation} and map all these sequences in $2^{nR_1}$ bins, each indexed with $w_1 \in [1:2^{nR_1}]$: $\cC(w_1,w_0)$.
 
	Generate similarly $2^{nT_2}$ n-sequences $v^n(l_2,w_0)$, $l_2 \in [1:2^{nT_2}]$ each following $P^n_{V|Q} (v^n(l_2,w_0))$ and map them into $2^{nR_2}$ bins: $\cC(w_2,w_0)$.
	
\underline{\textit{Encoding:}}

To send a message vector: $(W_0, W_1, W_2)$, the encoder first finds a pair of sequences $(u^n(L_1,W_0),v^n(L_2,W_0))$ in the product bins $\cC(W_j,W_0)$ such that: 
 \begin{equation}
 \big( q^n(W_0),u^n(L_1,W_0),v^n(L_2,W_0) \big) \in \typ{QUV} \ , 
 \end{equation}
and then transmits: $x^n\bigl(q^n(W_0) ,u^n(L_1,W_0),v^n(L_2,W_0) \bigr) $ which is generated via a random mapping.

\underline{\textit{Decoding:}} 

First, assume that no `` encoding error: $\epsilon_0$" has occurred, and note: $(L_1, L_2)$ the chosen indices. For a matter of conciseness, we consider only Decoder 1. 
	
Given a received sequence $y^n$, define the two lists: 
	\begin{IEEEeqnarray}{rCl}
	 \cL_1 (y^n) &\triangleq& \Bigl\{ (w_0,w_1)\,\big |\, (q^n(w_0),u^n(l_1,w_0),y^n) \in \typ{QUY} \text{ for } u^n(l_1,w_0) \in \cC(w_1,w_0) \, \Big\} \\
	 	\cL_2 (y^n) &\triangleq& \Bigl\{ (w_0,w_1)\,\big |\,(q^n(w_0),u^n(l_1,w_0),v^n(l_2,w_0),y^n) \in \typ{QUVY} \IEEEnonumber \\
	 	&& \text{for some } w_2 \, , \, v^n(l_2,w_0) \in\cC(w_2,w_0) \, , \, \text{ and } u^n(l_1,w_0) \in\cC(w_1,w_0) ) \, \Big\}. 
	\end{IEEEeqnarray} 
	These lists correspond to two different decoding functions: ``non-unique" decoding of the other user's message, and ``not" decoding it.
	Denote the intersection of these two lists by 
	\begin{equation}\label{ListIntersectionDefinition}
	\cL^{(n)} \triangleq \cL_1 (y^n) \cap \cL_2 (y^n).
 	\end{equation}	
	\underline{\textit{Analysis of the probability of error:}} To analyze the probability of error at user 1, we need to control the expected cardinality of the intersection of the above lists. The next lemma (shown in Appendix \ref{Lemma1And2} ) states this result. 
	\begin{lemma}\label{LemmaList}
	For every $\epsilon_1>0$, the average probability of error is linked to the list size as follows: 
	\begin{equation}
	 	P^{(n)}_e \leq \dsP \{\| \cL^{(n)} \| \geq 2 \} + \epsilon_1 
	\end{equation}
	 for $n> \exists\, N_1$ large enough.
	 \end{lemma}
	Now, bounding the probability of error will mainly consist in bounding the decoding list size.
	
	\underline{\textit{Bounding the list size:}} 
	
	On one hand, the list size being an integer valued RV, we can write: 
	\begin{equation}
		\dsP\{\| \cL^{(n)} \| \geq 2\} \leq \dsE [\| \cL^{(n)} \| ] - \dsP\{\| \cL^{(n)} \| \geq 1\}.
		\label{EqProof1}
	\end{equation}
	On the other hand: 
	\begin{IEEEeqnarray}{rCl}
		\dsE \| \cL^{(n)} \| &=& \dsP\{ (W_0,W_1) \in \cL^{(n)}\} + \sum_{\mathclap{(w_0,w_1) \neq (W_0,W_1)}} \dsP\{(w_0,w_1) \in \cL^{(n)}\}.
		\label{EqProof2}
	\end{IEEEeqnarray}	
The next lemma provides a bound on the expected list size from the RHS of \eqref{EqProof2}. The proof is relegated to Appendix \ref{Lemma1And2} .\vspace{1mm}

	\begin{lemma}[Bounding the probability of undetected errors]\label{LemmaProbError}
	The probability of decoding $(w_0,w_1) \neq (W_0,W_1)$, can be upper-bounded as follows: 
	\begin{equation}
		\sum_{(w_0,w_1) \neq (W_0,W_1)} \dsP\{(w_0,w_1) \in \cL^{(n)}\} \leq \min \{I^{(n)}_1, I^{(n)}_2\} \ ,
		\label{EqProof3}
	\end{equation}
	for $n$ large enough, i.e. $n> N_2$, for some $N_2$, where: 
	\begin{IEEEeqnarray}{ll}
		I^{(n)}_1 &\triangleq \exp_2\bigl( n\, [T_1 - I(U; Y|Q) +\epsilon_2 ]\bigr) + \exp_2\bigl(n \, [R_0 + T_1 - I(Q U; Y )+\epsilon_2 ] \bigr) \ , \\
 I^{(n)}_2 &\triangleq \exp_2\bigl(n \, [T_1 - I(U; YV| Q)+ \epsilon_3] \bigr) + \exp_2\bigl(n \, [T_1 + T_2 - I(U V; Y|Q) - I(U;V|Q) + \epsilon_3] \bigr) \IEEEnonumber \\
 & + \exp_2\bigl(n\, [R_0 + T_1 + T_2 - I(Q U V; Y) - I(U;V|Q) + \epsilon_3] \bigr).
	\end{IEEEeqnarray}	
	\end{lemma}
Hence, from \eqref{EqProof1}, \eqref{EqProof2} and \eqref{EqProof3} we can write that: 
\begin{IEEEeqnarray}{rCl}
 \dsP\{\| \cL^{(n)}\| \geq 2\} &\leq& \min\{I^{(n)}_1,I^{(n)}_2\} \ . 
\label{EqProof4}
\end{IEEEeqnarray}	
Then Lemma 1 and \eqref{EqProof4}, imply that for n large enough: 
\begin{IEEEeqnarray}{rCl}
	P^{(n)}_e &\leq& \dsP\{\| \cL^{(n)} \| \geq 2\} + \epsilon_1 \leq \min\{I^{(n)}_1;I^{(n)}_2\} +\epsilon_1 \ . 
\end{IEEEeqnarray}
Thus, provided that: 
\begin{equation}		
\limsup_{n \rightarrow \infty} \min \{I^{(n)}_1, I^{(n)}_2\} = 0 \ , 
\end{equation}
the probability of error at user 1, knowing that no encoding error occurred, will tend to $0$ as $n \rightarrow \infty $.

Following the proof of the Covering lemma \cite{cover2006elements}, the probability of encoding error can be upper bounded as $n$ grows large enough as follows:
\begin{equation}
	\dsP(\epsilon_0) \leq \exp_2\bigl( n \, [I(U;V|Q) - (T_1 -R_1 + T_2 - R_2) + \epsilon^\prime] \bigr) .
\end{equation}
The condition for no such error does not depend on the users pair index, and thus, it intersects the union of all regions, which concludes the proof.
\section{The probability of error is linked to list size}\label{Lemma1And2}
\subsubsection{Proof of Lemma \ref{LemmaList}}
Let us start by recalling: 
\begin{equation} 
\cL_1(Y^n) \cap \cL_2(Y^n) = \cL^{(n)}.
\end{equation}
Let $(\hat{W_0},\hat{W_1})$ be the estimated messages at decoder 1, where 
\begin{IEEEeqnarray}{ll}
&\dsP\{ (\hat{W_0},\hat{W_1}) \neq (W_0, W_1) \} \IEEEnonumber \\
&= \delta \dsP\{ \exists (\hat{w_0},\hat{w_1}) \neq (W_0, W_1)\,:\,(\hat{w_0},\hat{w_1}) \in \cL^{(n)} | (W_0,W_1) \in \cL^{(n)} \} + \IEEEnonumber \\
& (1-\delta) \dsP\{ \exists (\hat{w_0},\hat{w_1}) \neq (W_0, W_1) \,: \,(\hat{w_0},\hat{w_1}) \in \cL^{(n)} | (W_0,W_1) \notin \cL^{(n)} \} \\
&\leq \dsP\{ \|\cL^{(n)} \| > 1\} + (1-\delta)\ , 
\end{IEEEeqnarray} 
with $	(1-\delta) \triangleq \dsP\{ (W_0, W_1) \notin \cL^{(n)}\}$. 

Then, following standard arguments, by the LLN and independence of codebooks, we can easily show that, for all $\epsilon_1>0$, $\exists \,N_1$ such that for $n \geq N_1$, we have $(1-\delta) \leq \epsilon_1$.	 

This ends the proof of the statement: 
\begin{equation} 
 P^{(n)}_e \leq \dsP \{\| \cL^{(n)} \| \geq 2 \} + \epsilon_1 \ . 
\end{equation}
 
\subsubsection{Proof of Lemma \ref{LemmaProbError}}
	Let $(w_0,w_1) \neq (W_0,W_1)$ be the supposedly decoded pair of messages. We have, recalling \eqref{ListIntersectionDefinition}, that:
	\begin{equation}
	\dsP\{(w_0,w_1) \in \cL^{(n)}\} \leq \min_{j=1,2} \, \dsP\{(w_0,w_1) \in \cL_j(Y^n)\} \ . 
	\end{equation}
For the first list, we have, following similar arguments of Lemma \ref{lem:jointTypicality}, that: 
	 \begin{IEEEeqnarray}{rCl} 
	 \dsP\{(W_0,w_1) \in \cL_1(Y^n)\} &=& \dsP\{ (q^n(w_0),u^n(l_1,w_0),y^n) \in \typ{QUY} \text{ for } l_1 \in [1:2^{n(T_1- R_1)}]  \} \qquad \\
	 &\leq& \sum_{l_1 \in [1:2^{n(T_1- R_1)}]} \dsP\{ (q^n(w_0),u^n(l_1,w_0),y^n) \in \typ{QUY} \}\\	 
	 &\leq& \exp_2 \bigl(n \,[T_1 - R_1 - I( U; Y | Q)+ \epsilon_2 ] \bigr) \ , 
	 \end{IEEEeqnarray}
and similarly, if moreover $w_0 \neq W_0$, 
	\begin{IEEEeqnarray}{rCl} 
	\dsP\{(w_0,w_1) \in \cL_1(Y^n)\} \leq \exp_2 \bigl(n \,[T_1 - R_1 - I(Q U; Y )+ \epsilon_2 ] \bigr) \ . 
	\end{IEEEeqnarray}
Now, for the second list, i.e, decoding method, we know that: \\
1) If $w_0 = W_0 $, $ w_1 \neq W_1$ and $l_2 = L_2$ which implies $w_2= W_2$: 
	\begin{IEEEeqnarray}{rCl}
&&\dsP\bigl\{ (Q^n(W_0),U^n(l_1,W_0),V^n(L_2,W_0),Y^n) \in \typ{QUVY} \text{ for } l_1 \in [1:2^{n(T_1- R_1)}] \bigr\} \IEEEnonumber \\
&\leq& \exp_2 \bigl( n \,[ T_1 - R_1 + H(QUVY) - H(Q) - H(U|Q) - H(YV|Q) + \epsilon_3] \bigr)\\
&=& \exp_2 \bigl( n \,[ T_1 - R_1 -I(U;YV|Q)+\epsilon_3] \bigr) \ , 
 	\end{IEEEeqnarray} 
where we used the fact that, since $w_1 \neq W_1$, then $U^n(l_1,W_0)$ and $V^n(L_2,W_0)$ are independent conditionally on $Q^n(W_0)$.\\
2) If $w_0 = W_0$, $w_1 \neq W_1$, and $l_2 \neq L_2 $ then: 
	\begin{IEEEeqnarray}{rCl}
&&\dsP\bigl\{ (Q^n(W_0),U^n(l_1,W_0),V^n(l_2,W_0),Y^n) \in \typ{QUVY} \text{ for } l_1 \in [1:2^{n(T_1- R_1)}] \bigr\} \IEEEnonumber \\
&\leq& \exp_2 \bigl(n \,[  T_1 - R_1 +H(QUVY) - H(Q) - H(U|Q) - H(V|Q) - H(Y|Q) + \epsilon_3] \bigr) \\
&=& \exp_2 \bigl(n \,[ T_1 - R_1 -I(UV;Y|Q) - I(U;V|Q)+\epsilon_3] \bigr) \ . 
 	\end{IEEEeqnarray} 
3) Finally, if $w_0 \neq W_0 $, then whatever $l_1$ and $l_2$: 
	\begin{IEEEeqnarray}{rCl}
&&\dsP\bigl\{ (Q^n(w_0),U^n(l_1,w_0),V^n(l_2,w_0),Y^n) \in \typ{QUVY} \bigr\} \IEEEnonumber \\
&\leq& \exp_2 \bigl(n \,[  H(QUVY) - H(Q) - H(U|Q) - H(V|Q) - H(Y) + \epsilon_3] \bigr) \\
&=& \exp_2 \bigl(n \,[  -I(QUV;Y) - I(U;V|Q)+\epsilon_3] \bigr) \ . 
 	\end{IEEEeqnarray}
This ends the proof of Lemma 2. 
 
 \section{Outer Bound Derivation for the Compound BC}\label{ProofOuterBoundGenralCase}
 We need to recall that the proof in~\cite{NairELGamal2006} of the outer bound for users' pair $(k)$, uses the specific choice of auxiliary RV: 
\begin{equation}
 \left\{\begin{array}{rcl}
		U_i &=& W_1 \ , \\
		V_i &=& W_2 \ , \\
		Q^{(k)}_i &=& (Y^{i-1,(k)} , Z^{n, (k)}_{i+1}) \ . 
		\end{array}\right.
	\end{equation}
Here, we notice that the auxiliary RV $(U_i, V_i)$ do not the depend on the users' pair index. Thus, we can show that for all channel indices $(k)$ with the specific choice: $U_i = W_1 \, , \, V_i = W_2$, 
\begin{equation}
	\cR_{\textrm{NEG}}^{(k)} (p_{QUVX}): \, \left\{\begin{array}{rcl}
	n R_1 &\leq& \displaystyle \sum_{i=1}^n I( Q_{k,i} U_i; Y_{k,i})\ , \\
	n R_2 &\leq& \displaystyle \sum_{i=1}^n I( Q_{k,i} V_i; Z_{k,i}) \ , \\
	n (R_1 + R_2) &\leq& \displaystyle \sum_{i=1}^n \big [ I( U_i ; Y_{k,i}| Q_{k,i} V_i) + I( Q_{k,i} V_i; Z_{k,i})\big ] \ ,\\
	n (R_1 +  R_2) &\leq& \displaystyle \sum_{i=1}^n \big [ I( Q_{k,i} U_i; Y_{k,i}) + I( V_i ; Z_{k,i}| Q_{k,i} U_i)\big ] \ ,
		\end{array}\right.
	\end{equation}
where $Q_{k,i} = (Y^{i-1}_{k,1} , Z^{n}_{k,i+1}) $. Thus, we could possibly factor the resulting joint pmf on $(U_i, V_i)$ over all compound channel indices, and let only the common variable choice vary from one channel to another. Moreover, we can show in the same fashion as in~\cite[Lemma 3.2]{NairELGamal2006}, that the maximizing distribution of the input $p_{X|QUV}$ is a deterministic mapping.


\section{Proof of Achievability of the Capacity}\label{FME}

From Theorem~\ref{theo-ID-CBC}, we can see that the region $\cR_{\textrm{SNID}}$ verifies: 
\begin{equation}
\cR_{\textrm{SNID}} \supseteq	\bigcup_{p_{QUVX} \in \cP}\,\bigcup_{(T_1,T_2) \in \dsT(p) } \,\left( \cT^{(1)}_{3} (p,T_1,T_2)\cap \cT^{(2)}_{4}(p,T_1,T_2)\right)\, ,
\end{equation}
In this section, we evaluate the region thus obtained by: 
\begin{equation}
\cR^\star_{\textrm{SNID}} \triangleq \bigcup_{p_{QUVX} \in \cP}\, \bigcup_{(T_1,T_2) \in \dsT(p) } \,\left( \cT^{(1)}_{3} (p,T_1,T_2)\cap \cT^{(2)}_{4}(p,T_1,T_2)\right) \, , 
\end{equation}
where $\cT^{(1)}_{3} \cap \cT^{(2)}_{4}$ is the subset of $\Re_{+}^2$ defined by the inequalities:
\begin{equation}
\left\{\begin{array}{rcl}
T_2 &\leq& I(V;ZU|Q) \\
T_1 + T_2 &\leq& I(UV;Z|Q) + I(U;V|Q) \\
R_0 + T_1 +T_2 &\leq& I(QUV;Z) + I(U;V|Q) \\
T_1 &\leq& I(U;Y_2 V|Q) \\
T_1 + T_2 &\leq& I(UV;Y_2|Q) + I(U;V|Q) \\
R_0 + T_1 +T_2 &\leq& I(QUV;Y_2) + I(U;V|Q) \vspace{2mm}\\
T_1 &\leq& I(U;Y_1|Q) \\
R_0 + T_1 &\leq& I(QU;Y_1) \\
T_1 \geq R_1 &,& T_2 \geq R_2 \, , \\
T_1 + T_2 &>& R_1 + R_2 + I(U;V |Q ) \,
\end{array}\right. 
\end{equation} 
Recalling here that $Y_1$ is physically degraded towards $Z$, we can first rewrite the decoding constraints as the following: 
\begin{equation}
\left\{\begin{array}{rcl}
T_2 &\leq& I(V;ZU|Q) \\
T_1 &\leq& \min\{I(U;Y_1|Q),I(U;Y_2 V|Q) \} \\
R_0 + T_1 &\leq& I(QU;Y_1) \\
T_1 + T_2 &\leq& I(UV;Y_2|Q) + I(U;V|Q) \\
R_0 + T_1 + T_2 &\leq& I(QUV;Y_2) + I(U;V|Q) \ . 
\end{array}\right. 
\end{equation}
 
The, we can run FME over the binning rate pair $(T_1, T2)$ to get the following region: 
 \begin{equation}
\left\{\begin{array}{rcl}
R_2 &\leq& I(V;ZU|Q) \\
R_1 &\leq& \min\{I(U;Y_1|Q),I(U;Y_2V|Q) \} \\
R_0 + R_1 &\leq& I(QU;Y_1) \\
R_1 + R_2 &\leq& I(UV;Y_2|Q) \\
R_1 + R_2 &\leq& I(V;Z|UQ) + \min\{I(U;Y_1|Q),I(U;Y_2 V|Q) \}\\
R_0 + R_1 +R_2 &\leq& I(QUV;Y_2) \\
R_0 + R_1 +R_2 &\leq& I(V;Z|UQ) + I(QU;Y_1) \ . 
\end{array}\right. 
\end{equation}
Later, we chose to apply \emph{bit recombination} on the admissible rates $(R_0,R_1,R_2)$ as follows: 
\begin{equation}
\left\{\begin{array}{rcl}
 R_0 &=& R^\star_{0} + R^\star_{01} + R^\star_{02} \ ,\\ 
R_1 &=& R^\star_1 - R^\star_{01} \geq 0 \ ,\\
R_2 &=& R^\star_2 - R^\star_{02} \geq 0 \ ,\\
R^\star_{01} &\geq 0& \ , \ R^\star_{02} \geq 0 \ . 
\end{array}\right. 
\end{equation}
 
It is straightforward that this bit recombination fits the decoding logic of the terminals, i.e., part of the private messages is mapped into the common message, enabling each terminal to still recover the totality of its intended message. The region writes thus as: 
\begin{equation}
\left\{\begin{array}{rcl}
R^\star_2 - R^\star_{02} &\leq& I(V;ZU|Q) \\
R^\star_1 - R^\star_{01} &\leq& \min\{I(U;Y_1|Q),I(U;Y_2 V|Q) \} \\
R^\star_0 + R^\star_1 + R^\star_{02} &\leq& I(QU;Y_1) \\
R^\star_1 - R^\star_{01} + R^\star_2 - R^\star_{02}&\leq& I(UV;Y_2|Q) \\
R^\star_1 - R^\star_{01} + R^\star_2 - R^\star_{02} &\leq& I(V;Z|UQ) + \min\{I(U;Y_1|Q),I(U;Y_2 V|Q) \} \\
R^\star_0 + R^\star_1 + R^\star_2 &\leq& I(QUV;Y_2) \\
R^\star_0 + R^\star_1 + R^\star_2 &\leq& I(V;Z|UQ) + I(QU;Y_1) \\
R^\star_1 \geq R^\star_{01} \ , \ R^\star_2 \geq R^\star_{02} \ & ,& \ R^\star_{01} \geq0 \ , \ R^\star_{02} \geq 0 
\end{array}\right. 
\end{equation}
 
Performing again FME over the splitting rate pair $( R^\star_{01}, R^\star_{02})$, we get the following region: 
\begin{IEEEeqnarray}{rCl}
R^\star_0 + R^\star_1 &\leq& I(QU;Y_1) \IEEEnonumber \\
R^\star_0 + R^\star_1 + R^\star_2 &\leq& I(QU;Y_1) + I(UV;Y_2|Q) \IEEEnonumber \\
R^\star_0 + R^\star_1 + R^\star_2 &\leq& I(QU;Y_1) + I(V;ZU|Q) \label{RS_1} \\ 
R^\star_0 + R^\star_1 + R^\star_2 &\leq& I(QU;Y_1) + I(V;Z|UQ) + \min\{I(U;Y_1|Q),I(U;Y_2 V|Q) \} \label{RS_2} \\
R^\star_0 + R^\star_1 + R^\star_2 &\leq& I(QUV;Y_2)\IEEEnonumber \\
R^\star_0 + R^\star_1 + R^\star_2 &\leq& I(U;Y|VQ) + I(QU;Y_1) \label{RS_3} \ . 
\end{IEEEeqnarray}
We clearly notice that the constraints: \eqref{RS_1} and \eqref{RS_2} are implied by \eqref{RS_3}, thus, the resulting region $\cR^\star_{\textrm{SNID}}$ is defined by the following constraints: 
\begin{equation}
\left\{\begin{array}{rcl}
 R^\star_0 + R^\star_1 &\leq& I(QU;Y_1) \\
R^\star_0 + R^\star_1 + R^\star_2 &\leq& I(QU;Y_1) + I(UV;Y_2|Q) \\
R^\star_0 + R^\star_1 + R^\star_2 &\leq& I(QUV;Y_2) \\
R^\star_0 + R^\star_1 + R^\star_2 &\leq& I(V;Z|UQ) + I(QU;Y_1) \ . 
\end{array}\right. 
\end{equation} 
Thus, letting $R^\star_0 = 0$, and noting the rate pairs as $(R_1,R_2)$, one gets the desired rate region. 
 
\section{Cardinality bounds}\label{CardinalityBound}
Consider a pair of RVs $(Q,X)$ following the joint probability distribution $p_Q p_{X|Q}$.
Since the input is binary, let the four continuous functions on $P_{X|Q}$: 
\begin{IEEEeqnarray}{rcl}
f_1\bigl(P_{X|Q}(0|q)\bigr) &=& P_{X|Q}(0|q) \ , \\ 
f_2\bigl(P_{X|Q}(0|q)\bigr) &=& H(Z|Q=q) = H_2(p\star  P_{X|Q}(0|q)) \ , \\
f_3\bigl(P_{X|Q}(0|q)\bigr) &=& H(Y_1|Q=q) = H_2(p_1 \star P_{X|Q}(0|q)) \ , \\
f_4\bigl(P_{X|Q}(0|q)\bigr) &=& H(X|Q=q) = H_2(P_{X|Q}(0|q)) \ . 
\end{IEEEeqnarray}
By the usual consequence of Fenchel-Eggleston-Caratheodory theorem~\cite{csiszar1982information}, we can construct an auxiliary RV $Q'$ such that: 
 \begin{IEEEeqnarray}{rcl}
\sum_q P_Q(q) P_{X|Q}(0 |q) &=& \sum_{q^\prime} P_{Q^\prime}(q^\prime) P_{X|Q}(0|q^\prime) = P_X(0) \ , \\
H(Z | Q) &=& H(Z | Q^\prime) \ , \\
H(Y_1| Q) &=& H(Y_1| Q^\prime) \ , \\
H(X | Q) &=& H(X | Q^\prime) \ , \\
\|Q'\| &\leq& 4 \ . 
\end{IEEEeqnarray}
Thus, we conclude that with this new auxiliary RV $Q'$, the region is unchanged: 
 \begin{IEEEeqnarray}{rcl}
I(X;Z|Q) &=& H(Z|Q) - H(Z|X) = H(Z|Q^\prime) - H_2(p) = I(X;Z|Q^\prime) \ , \\
I(Q;Y_1) &=& H(Y_1) - H(Y_1| Q)= H_2\left(p_1 \star P_X(0)\right) - H(Y_1| Q^\prime ) = I(Q^\prime;Y_1) \ , \\
I(Q;Y_2) &=& (1-e) \left(H(X) - H(X | Q)\right) = \bar{e} \left(H_2(P_X(0)\right) - H(X | Q^\prime)) = I(Q^\prime;Y_2) \ . 
\end{IEEEeqnarray}

\subsubsection{Input uniformity}
In \cite{NairBECBSC2010} lies a definition of the ``c-symmetric broadcast channel" as being the BC formed by $2$ c-symmetric channels. Following this same idea, and considering equivalently the Compound BC or the Compound Channel, we can say that the BC resulting from the simultaneity of two c-symmetric BC is c-symmetric.
 
As it is shown in~\cite[Lemma 2]{NairBECBSC2010} that uniform input distribution is optimal for such a channel, we conclude that $X \sim \textrm{Bern}(1/2)$ is optimal for the Compound BC. 
 
\section{Proof of Proposition 3}\label{Proposition3}
Recall that: 
\begin{equation}
\begin{array}{rcl}
 t_a(x) \triangleq \sup\limits_{p_{QX}\in\cC(x)} \bigl[ a\, I(Q;Y_1)+ \bar{a}\,I(Q;Y_2) \bigr] \ . 
\end{array} 
\end{equation} 
We want to show that: 
 \begin{enumerate}[i)]
\item For all $x \in [0: 1-H_2(p)]$, 
\begin{equation} 
 t_a(x) = \max_{p_{QX}\in\cC(x)} \bigl[ a\,I(Q;Y_1)+ \bar{a}\,I(Q;Y_2) \bigr]
\end{equation} 
\item $t_a$ is concave in $x$.
\item $t_a$ can be described identically by its supporting lines. 
\item $t_a$ is decreasing in $x$. 
\end{enumerate}
\begin{IEEEproof} 
i) We have that: 
\begin{IEEEeqnarray}{rCl}
\cC(x) = \biggl\{ p_{XQ} \in \cP(\cX \times \cQ): \ & & Q \mkv X \mkv  (Y,Z_1,Z_2) ,\\
& & X \sim \textrm{Bern}(1/2) , \ I(X;Z|Q ) = x  \biggr\} \ . 
\end{IEEEeqnarray} 
Since, we have proved that the optimizing probabilities have a finite cardinality, the conditional mutual information being continuous, $\cC(x)$ is thus compact. As the probability space $ \cP(\cX \times \cQ)$ has a finite dimension, the set $\cC(x)$ is thus closed. Thus, the supremum is achieved. 

ii) Concavity: 

Let $x_1 ,x_2 \in [0: 1-H_2(p)]$ and let $ \alpha \in [0:1]$. 
Denote $x = \alpha\, x_1 + (1-\alpha)\,x_2$. 
We need to show that: $ t_a(x) \geq \alpha\,t_a( x_1) + (1-\alpha)\,t_a(x_2)$. 

Let for $i \in \{1,2\}$, 
\begin{equation} 
 P_{X_i, Q_i} = \argmax{p_{QX}\in\cC(x)} \bigl[ a\,I(Q;Y_1)+ \bar{a}\,I(Q;Y_2) \bigr] \ . 
\end{equation}
Define moreover: $T \sim \textrm{Bern}(t)$ independent of all other RVs. Define
\begin{equation}
(X,Q_T) = \left\{\begin{array}{rcl}
(X_1,Q_1) &\textrm{if}& \quad T = 0\ ,\\
(X_2,Q_2)  &\textrm{if}& \quad T = 1 \ , 
\end{array}\right.
\end{equation}
and by letting $Q=(Q_T,T)$, we have: 
\begin{itemize}
\item $X \sim \textrm{Bern}(1/2)$.
\item $Q \mkv X \mkv (Y,Z_1,Z_2)$ is a valid Markov chain. 
\item And the following equalities hold: 
\begin{IEEEeqnarray}{rCl}
I(X;Z|Q) &=& \alpha\,I(X_1;Z|Q_1) + (1-\alpha)\, I(X_2;Z|Q_2)\\
&=& \alpha\, x_1 + (1-\alpha)\,x_2 = x \ . 
\end{IEEEeqnarray} 
\end{itemize}
We thus have that: $p_{XQ} \in \cC(x) $. Thus, 
\begin{IEEEeqnarray}{rCl}
 \alpha\,t_a(x_1) + (1-\alpha)\,t_a(x_2) &=& \alpha\, \bigl( a\,I(Q_1;Y_1)+ \bar{a}\,I(Q_1;Y_2) \bigr) \IEEEnonumber\\
 && \quad \quad \quad + (1-\alpha)\, \bigl( a\,I(Q_2;Y_1)+ \bar{a}\,I(Q_2;Y_2) \bigr) \\
 &=& a\,I(Q_T;Y_1 | T) + (1-a)\, I(Q_T;Y_2 | T) \\
 &\leq& a\, I(T Q_T;Y_1 ) + (1-a)\, I(T Q_T;Y_2 ) \\
 &=& a\, I(Q;Y_1 ) + (1-a)\, I(Q;Y_2 )\\
 &\leq& \max_{p_{QX}\in\cC(x)} \bigl[a\, I(Q;Y_1 ) + (1-a)\, I(Q;Y_2 ) \bigr] \\
 &=& t_a(x) \ , 
\end{IEEEeqnarray} 
which concludes the proof of concavity. \\
iii) This property follows from the concavity of $t_a$. \\
iv) Monotony: \\
Since $t_a$ is concave, we have that:
\begin{equation}
 	 t^\prime_a(x) \leq t^\prime_a(0) = \lim_{x \rightarrow 0^+} \frac{t_a(x) - t_a(0)}{x} \ . 
\end{equation}
Since, 
\begin{equation} 
t_a(0) = a\,(1-H_2(p_1)) + (1-a)\, (1-e_2) > t_a(x) \ , 
\end{equation} 
for all $x \in [0:1-H_2(p)] $, we have that:
\begin{equation}
 	 t^\prime_a(x) \leq t^\prime_a(0) \leq 0 \ ,
\end{equation} 
$t_a$ is thus decreasing in $x$. 
\end{IEEEproof}

\section{Proof of achievability of Multiple Description inner bound}\label{MDAchievability}
In this section, we establish the achievability of the MD inner bound \eqref{MultipleDescriptions}. 
Let $W_1$ be the message decoded by user $1$, and let $W_2$ be the message decoded by user $2$, plus let $R_1$ and $R_2$ denote their respective rates. Let $T_1$ and $T_2$ denote the corresponding binning rates. We construct the following code.

\underline{\textit{Codebook generation: }}

 	Generate $2^{n\, T_1}$ $n$-sequences $u_0^n(l_1)$, $l_1 \in [1:2^{nT_1}]$ each following: $$ P^n_{U_0}( u_0^n(l_1) )= \prod^n_{i=1} P_{U_0} (u_{0,i} (l_1)) \ , $$ and map all these sequences in $2^{nR_1}$ bins, each indexed with $w_1 \in [1:2^{nR_1}]$: $\cC_0(w_1)$.
	
	Generate similarly $2^{nT_2}$ n-sequences $v^n(l_2)$, $l_2 \in [1:2^{nT_2}]$ each following $P^n_{V} (v^n(l_2))$ and map them into $2^{nR_2}$ bins: $\cC_v(w_2)$.
	
	For each $u_0^n(l_1)$, $l_1 \in [1:2^{nT_1}]$, generate $2^{n\, \hat{R}_j}$ $n$-sequences $u_j^n(s_j,l_1)$, $s_j \in [1:2^{n \hat{R}_1}]$ following each: $$ P^n_{U_j|U_0}( u_j^n(s_j,l_1) )= \prod^n_{i=1} P_{U_j|U_0} (u_{j,i} (s_j,l_j)|u_{0,i} (l_1)) \ .$$ 
	
	\underline{\textit{Encoding:}} 

 	To send a message pair $( W_1, W_2) $, the encoder finds a couple of sequences $u_0^n (l_1)$ and $v^n(l_2)$ in the product bin $\cC_0(W_1) \times \cC_v(W_2)$ and a couple of indices $ (s_1, s_2) $ such that 	\begin{equation}
 \big( u_0^n(l_1),u_1^n(s_1,l_1),u_1^n(s_2,l_1),v^n(l_2 ) \big) \in \typ{U_0U_1U_2V} \ . 
 \end{equation}
 It then transmits an n-sequence $x^n\left(u_0^n(l_1),u_1^n(s_1,l_1),u_1^n(s_2,l_1),v^n(l_2 ) \right) $ which is generated via a random mapping. 
 
 Using the well known second order moment method, one can make the probability of the encoding error event arbitrarily close to $0$ if: 
 \begin{equation}
 \left\{
 \begin{array}{rcl}
 T_1 - R_1 + \hat{R}_1 + \hat{R}_2 &\geq& I(U_1; U_2 |U_0)\ ,\\
 T_1 - R_1 + T_2 - R_2 &\geq& I(U_0;V)\ , \\
 T_1 - R_1 + \hat{R}_1 + T_2 - R_2 &\geq& I(U_0 U_1 ; V)\ ,\\
 T_1 - R_1 + \hat{R}_2 + T_2 - R_2 &\geq& I(U_0 U_2 ; V)\ , \\
 T_1 - R_1 + \hat{R}_1 + \hat{R}_2 + T_2 - R_2 &\geq& I(U_0 U_1 U_2 ; V) + I( U_1 ; U_2|U_0)\ . 
 \end{array} \right. 
 \end{equation}
 
 	\underline{\textit{Decoding:}} 
 	
The second user, upon receiving the sequence $z^n$, looks for the unique index $w_2$ such that for some $v^n(l_2) \in \cC_v(w_2) $, the following holds: 
 \begin{equation}
 \big( v^n(l_2 ) , z^n \big) \in \typ{VZ} \ . 
 \end{equation}
The probability of error in such a decoding rule is arbitrarily small provided that: 
\begin{equation}
T_2 \leq I(V;Z) \ . 
\end{equation}

Concerning the two instances of the first user $Y_1$ and $Y_2$ let us assimilate each of them to a decoder. Decoder $j$ finds the unique index $l_1$ such that for some $s_j$ where, the following joint typicality holds: 
 \begin{equation}
 \big( u_0^n(l_1),u_j^n(s_j,l_1) , y_j^n \big) \in \typ{U_0U_jY_j} \ . 
 \end{equation}
The probability that the decoded $l_1$ does not fall into the bin specified by $ w_1$ is made arbitrarily provided that: 
\begin{equation}
T_1 + \hat{R}_j \leq I (U_0 U_j;Y_j) \ . 
\end{equation}

Then the overall decoding error events occur with arbitrary small probability provided that: 
 \begin{equation}
 \left\{
 \begin{array}{rcl}
T_1 + \hat{R}_1 & \leq& I(U_0 U_1;Y_j) \ ,\\
T_1 + \hat{R}_2 & \leq& I(U_0 U_2;Y_j)\ , \\
T_2 &\leq& I(V;Z) \ .
 \end{array} \right. 
 \end{equation}
After running FME on the system of inequalities bearing in mind the natural encoding constraints: 
 \begin{equation}
 \left\{
 \begin{array}{rcl}
 \hat{R}_1 & \geq& 0 \ ,\\
 \hat{R}_2 & \geq& 0 \ ,\\
 T_1 &\geq& R_1 \ ,\\
 T_2 &\geq& R_2 \ , 
 \end{array} \right.
 \end{equation}
 the region given in \eqref{MultipleDescriptions} follows immediately. 

\section{Proof of Lemma \ref{LemmaPrivate}}\label{ProofLemmaDescriptions}
We derive the optimal rate obtained when the following coding scheme is used: 
\begin{align}
\bX &= ( X_{u} + X_{p} ) \bB_u + X_v \bB_v\ , & U_0 &= X_{u} + \alpha X_v\ ,  \\
 V &= X_v\ ,  & U_1 &= X_{p} + \alpha_1 X_v \ , 
\end{align}
where $X_{p} \sim \cN(0, x)$, $X_{u} \sim \cN(0, P_u - x)$ and $X_{v} \sim \cN(0, P_{v})$ are pairwise independent RV and such that: $P_{u} \leq P - P_v$. \\
This means that we transmit two descriptions intended for user 1 making these two descriptions compensate ``jointly" the interference, hence, we are interested in computing the rate: $R_{0,1} = I(U_0 U_1;Y) - I(U_0 U_1;V)$. Some algebraic manipulations lead us to the following result: 
\begin{equation}
R_{0,1} = \frac{1}{2}\, \log_2 \left( \dfrac{ h_u^2\, P_u +N }{\dfrac{P_v \left( h_u^2\, P_u + N \right)}{ h_u^2\, P_u + h_v^2\, P_v +N } P(\alpha, \alpha_1) + N} \right) \ , 
\end{equation}
where the quadratic polynomial $P(\alpha,\alpha_1)$ is given by: 
\begin{equation} 
 P(\alpha ,\alpha_1) = h_u^2 (\alpha_1 - \beta^x_1 + \alpha - \beta^x )^2 + \dfrac{ N}{x} (\alpha_1 - \beta^x_1 )^2 + \dfrac{ N}{P_{u}-x} (\alpha - \beta^x )^2 \ , 
\end{equation}
and, $ \beta^x= \dfrac{( P_{u} -x) \, h_u \,h_v }{h_u^2\,P_u + N} $ and $ \beta^x_1 = \dfrac{x \, h_u \,h_v }{h_u^2\,P_u + N} $. 

An interesting insight brought by this expression is that to achieve the optimal DoF, we need only have $ \alpha_1 + \alpha = \beta^o_1 + \alpha^o $ rather than pairwise equality as might be suggested by the previous section. This translates perfectly the ``joint" interference management of both decoded descriptions $U_0$ and $U_1$, recovering trivially the optimal interference free rate as both descriptions cancel the interference fully each on their own $\alpha_1 - \alpha^\star_1 = \alpha_0 - \alpha^\star_0 = 0$. 

Upon optimizing the polynomial $P(\alpha,\alpha_1)$ over $\alpha_1$, the resulting rate is given by the rather simple expression: 
\begin{equation} \label{2aRV}
R_{0,1}= \frac{1}{2}\, \log_2 \left( \dfrac{ h_u^2\, P_u +N }{\dfrac{P_v}{(P_{u}-x)} \dfrac{\left( h_u^2\ P_u +N \right)^2}{\left( h_u^2 P_u + h_v^2 P_v +N \right)} \dfrac{N}{ h_u^2 x +N} (\alpha - \beta^x)^2 + N} \right) \ , 
\end{equation} 
It can be readily checked that this expression corresponds to the following formulation of the rate: 
\begin{equation}
R_{0,1} = I(U_0;Y) - I(U_0;V) + I(X_{p} ; Y | X_{u} X_v ) \ , 
\end{equation}
where 
\begin{equation}
	 I(X_{p} ; Y| X_{u} X_v ) = \dfrac{1}{2} \log_2 \left( \dfrac{h_u^2 x +N}{N}\right) \ , 
\end{equation}
 and where $I(U_0;Y) - I(U_0;V)$ corresponds to the case where $X_u$ dirty-paper codes $X_v$ under the noise component variance: $h_u^2 x +N $. 
 
This means that the optimal choice of the variable $U_1$ is the one that maximizes the DPC term $I(U_1; Y |U_0) - I(U_1 ;V|U_0)$. 

\section{Optimization of Common Description inner bound:}\label{CDCornerPoints} 
 Let us first optimize the second corner point of the CD inner bound. We have that 
\begin{equation} 
\begin{array}{rcl}
\cR_2 = \Biggl\{ (R_1,R_2)\in \mathds{R}^2_+ \, , \, \quad R_2 &\leq&\dfrac{1}{2} \log_2 \left( \dfrac{ g_v^2 P_v +N }{N} \right)\ , \\
R_1 &\leq&  \underset{j= 1,2}{\min} \,\dfrac{1}{2}\log_2 \left( \dfrac{ h_{j,u}^2 P_u + h_{j,v}^2 P_v +N }{ h_{j,v}^2 P_v +N } \right) \Biggr\} \ .
\end{array}
\end{equation}
We have that what maximizes $\bh_1$ and $\bh_2$ are orthogonal and of unit norm, thus, we can write that: $h_{1,u}^2 = 1 - h_{2,u}^2$ and $h_{1,v}^2 = 1 - h_{2,v}^2$. The rate $R_2$ does not depend on the beam $\bB_u$, thus, we start by optimizing the rate $R_1$ over it. The two min operands are monotonic in inverse directions and have the same minimum value $0$, thus, the maxmin point corresponds to the equality point. Which by simple algebraic calculations leads to the condition:
\begin{equation} 
 h^2_{1,u} = \dfrac{h^2_{1,v} P_v + N }{P_v + 2 N} \ , 
\end{equation}
and yields then a rate (independent of the beam $\bB_v$) equal to: 
\begin{equation} \label{CornerPoint2optimization} 
R_1 \leq \dfrac{1}{2} \log_2 \left( \dfrac{ P_u + P_v + 2 N }{ P_v + 2 N } \right) \ . 
\end{equation}
Note then that the maximizing beam direction $\bB_v = \bg $, thus one can easily check that this verifies: $h_{1,v} = - 1/\sqrt{2}$ and thus, from \eqref{CornerPoint2optimization}, that $ | h_{1,u} | = 1/\sqrt{2}$. Thus transmitting the first user's signal in the mean channel direction is an admissible optimizing solution. Later in the proof, we show that this corner point is dominated by the first corner point of the CD inner bound. In the sequel, we will perform the optimization under the choice of $ h_{1,u} = 1/\sqrt{2}$ and $g_u = 0$, i.e., we transmit the signal intended to user $1$ in the mean channel direction, which makes it orthogonal to the second user's channel; the optimality of which is given in Appendix~\ref{CDoptimization}. 

We can rewrite the first corner point of the CD inner bound as follows:
\begin{equation} 
\begin{array}{rcl}
\cR_1 = \underset{a \in [0:1]}{\bigcup}\Biggl\{ (R_1 ,R_2) &\in& \Re^2_+ \, , \, \\
 R_1 &\leq&  \underset{\alpha \in \mathds{R}}{\max} \,\underset{j \in \{1,2\}}{\min} \,\dfrac{1}{2}\log_2\left( \dfrac{ P_u + 2 N }{\dfrac{ P_v}{P_u} \dfrac{\left( P_u + 2 N \right)^2 }{ P_u + 2 N + 2 h_{j,v}^2 P_v } (\alpha - \alpha_j)^2 + 2 N } \right) \\ 
R_2 &\leq& \dfrac{1}{2} \log_2 \left( \dfrac{ g^2_v P_v + N }{N} \right)  \Biggr\} 
\end{array}
\end{equation}
where $\alpha_j = \dfrac{\sqrt{2}P_u}{P_u + 2 N} h_{j,v}$. Since $\|\bh_j\| =\|\bB_v\| = 1$ and, $\bh_1$ and $\bh_2$ are orthogonal, we can let $h_{1,v} = \cos (\theta_v) $ and $h_{2,v} = \sin (\theta_v)$. 

The key point in the optimization is to solve the equation: 
\begin{equation}
\dfrac{ (\alpha - \alpha_1)^2}{ P_u + 2 N + 2 \cos(\theta_v)^2 P_v } = \dfrac{( \alpha - \alpha_2)^2}{ P_u +2 N + 2 \sin(\theta_v)^2 P_v } \ . 
\end{equation}
The optimization of the rate of the first user $R_1$ yields the following:  

(i) If $\cos^2(\theta_v) = \dfrac{1}{2}$ and $\cos( \theta_v ) = - \sin(\theta_v)$, then the optimal rate is given by: 

\begin{IEEEeqnarray}{rCl}
	R_1 &\leq&  \underset{\alpha}{\max} \, \underset{j\in \{1,2\}}{\min} \,\dfrac{1}{2}\log_2\left( \dfrac{P_u + 2 N }{ \dfrac{P_v }{P_u } \dfrac{\left(P_u + 2 N \right)^2}{\left( P + 2 N \right)} (\alpha - \alpha_j)^2 + 2 N } \right)\\
	&=&  \underset{\alpha}{\max} \, \dfrac{1}{2}\log_2\left( \dfrac{ P_u + 2 N }{ \dfrac{P_v }{P_u} \dfrac{\left(P_u + 2 N \right)^2}{\left( P + 2 N \right)} (|\alpha| + |\alpha_j|)^2 + 2 N } \right)\\
	&=& \dfrac{1}{2} \log_2\left( \dfrac{ P_u + 2 N }{2 N + P_v \dfrac{P_u }{ P + 2N} }\right) \\
	&=& \dfrac{1}{2} \log_2\left( \dfrac{ P_u + P_v +2 N }{P_v + 2 N}\right)
 \end{IEEEeqnarray} 
where $ \alpha_1 = - \alpha_2 = \dfrac{ P_u }{P_u + 2 N} $ .
It turns out then, that the optimization over the DPC parameter $\alpha$ yields $\alpha = 0$, i.e. the dilemma at the transmitter is so strong that the optimal choice is not to cancel interference and send in a direction that does not yield privilege to any of the channel instances $\bh$. A very important remark, is that this yields exactly the first corner point of the region. 

(ii) If $\cos^2(\theta_v) = \dfrac{1}{2}$ and $\cos( \theta_v ) = \sin(\theta_v)$, then the optimal rate is given by: 
\begin{IEEEeqnarray}{rCl}
	R_1 &\leq& \dfrac{1}{2} \underset{\alpha}{\max} \, \underset{j\in \{1,2\}}{\min} \log_2\left( \dfrac{P_u + 2 N }{ \dfrac{P_v }{P_u } \dfrac{\left(P_u + 2 N \right)^2}{\left( P + 2 N \right)} (\alpha - \alpha_j)^2 + 2 N } \right)\\
	&=& \dfrac{1}{2} \log_2\left( \dfrac{ P_u + 2 N }{ 2 N } \right) \ , 
 \end{IEEEeqnarray}
which corresponds to the point where $h_{1,v} = h_{2,v}$ i.e. $\alpha_1 = \alpha_2 $. Thus, we would have $ \bh_1 -\bh_2 $ orthogonal to $\bB_v$, but since $ \bh_1 -\bh_2 $ is collinear to the second user's channel, then it means that no information is transmitted to it with the beam $ \bB_v $ . The power optimization of this point corresponds to the corner point $(C_1, 0)$. 

(iii) If $\cos^2 (\theta_v)\neq \dfrac{1}{2}$, then there are two optimizing solutions $\alpha^\star_1 $ and $\alpha^\star_2$ such that: \vspace{1mm}
\begin{IEEEeqnarray}{rCl}
 \alpha^\star_1 - \alpha_1 &=& \dfrac{P_u}{P_u + 2 N} \, \dfrac{ \left(- \cos(\theta_v)+ \sin(\theta_v) \right) \sqrt{P_u/2 + N + \cos(\theta_v)^2 P_v} } { \sqrt{ P_u + 2 N + 2 \sin^2(\theta_v) P_v} + \sqrt{ P_u + 2 N + 2 \cos^2(\theta_v) P_v} } \ ,\\ 
 \alpha^\star_2 - \alpha_1 &=& \dfrac{P_u}{P_u + 2 N} \, \dfrac{\left( \cos(\theta_v) - \sin(\theta_v) \right) \sqrt{P_u/2 + N + \cos(\theta_v)^2 P_v}}{ \sqrt{ P_u + 2 N + 2\sin^2(\theta_v)P_v } - \sqrt{ P_u + 2 N + 2 \cos^2(\theta_v) P_v} } \ . \vspace{1mm}
 \end{IEEEeqnarray}
The root that yields the greater rate is $\alpha^\star_1 $. Then, we can rewrite with the following transformation $ y = \sin(2 \theta_v)$ that: 
\begin{IEEEeqnarray}{rCl}
 (\alpha^\star_1 - \alpha_1)^2 &=& \dfrac{2 P_u^2}{( P_u + 2 N)^2}\, \dfrac{\cos^2 (\theta_v + \pi /4 )\quad (P_u/2 + N + \cos(\theta_v)^2 P_v)}{ \left( \sqrt{ P_u + 2 N + 2 \sin^2(\theta_v)P_v } + \sqrt{ P_u + 2 N + 2 \cos^2(\theta_v) P_v} \right)^2 }\,\,\,\,\,\, \,\,\, \\ 
&=& \dfrac{ P_u^2}{2( P_u + 2 N)^2} \, \dfrac{(1-y)\quad(P_u/2 + N + \cos(\theta_v)^2 P_v)}{P + 2 N + \sqrt{(P_u + 2N)(P + 2N + P_v)+ y^2 P_v^2}}\\
&=& \dfrac{ P_u^2}{2( P_u + 2 N)^2} \, \dfrac{(1-y)\quad (P_u/2 + N + \cos(\theta_v)^2 P_v)}{P + 2 N + \sqrt{(P+ 2N)^2+ (y^2-1) P_v^2}}\ .
\end{IEEEeqnarray} 
Note that the value of $y=-1$, i.e., $\theta_v = -\pi/4$, is included in this expression. Thus we drop the case distinctions $\cos^2(\theta_v) = 1/2$ and $\cos^2(\theta_v) \neq 1/2$.

As a conclusion, CD inner bound writes as: 
\begin{equation} 
\begin{array}{rcl}
\cR_{CD} = \underset{y \in [-1:1]}{\bigcup}\Biggl\{ (R_1 ,R_2) &\in& \Re^2_+ \, , \, \\
 R_1 &\leq& \dfrac{1}{2} \log_2 \left( \dfrac{P_u + 2 N}{ \dfrac{P_v P_u (1-y )}{P + 2 N + \sqrt{(P+ 2N)^2+ (y^2-1) P_v^2} } + 2 N } \right) \\ 
R_2 &\leq& \dfrac{1}{2} \log_2 \left( \dfrac{ (1-y) P_v + 2 N }{2 N} \right)  \Biggr\} \ .
\end{array}
\end{equation}

\section{Beamforming optimization for the CD-DPC inner bound}\label{CDoptimization}
In this section, we show, that with a strong uncertainty over the channel instances of user $1$, i.e., $\bh_1$ and $\bh_2$ being orthogonal, when resorting to a CD-DPC, the source has no choice but to send over the mean channel $ \bh_{1,2}$. The proof of this claim is quite evolved and requires the use of many analytical manipulations when solving optimization problems.  

Let us use the following notations. We previously introduced $\theta_v$ such that $h_{1,v} = \cos (\theta_v) $ and $h_{2,v} = \sin (\theta_v)$. Since $\|\bh_j\| =\|\bB_u\| = 1$ and, $\bh_1$ and $\bh_2$ are orthogonal, we can similarly define $\theta_u$ such that $h_{1,u} = \cos (\theta_u) $ and $h_{2,u} = \sin (\theta_u)$. Let us define: 
 \begin{equation}
s_{u} \triangleq \dfrac{\sin(2 \theta_u)}{|\sin(2 \theta_u) |} \ \text{and } \ s_v \triangleq \dfrac{\sin(2 \theta_v)}{|\sin(2 \theta_v)|} \ , 
\end{equation}
when $\sin(2 \theta_u) \neq 0$ and  $\sin(2 \theta_v) \neq 0$.

In this section, we prove that it is optimal to transmit the signal in the directions given by: $ \cB_u = \bh_{1,2}$, which comes to having $h_{1,u} = h_{1,v} = \dfrac{1}{\sqrt{2}}$. 

Thus, we need to solve the optimization problem given by: 
\begin{equation}\label{OptimizatinProblem}
 \bigcup_{\cB_u, \cB_v} \left\{ \begin{array}{rcl} R_1 &\leq&  \displaystyle\max_{\alpha} \displaystyle\min_{j = 1,2} \log_2 \left( \dfrac{1}{A_j(\alpha - \alpha_j)^2 + c_j  } \right)  \ ,  \\
 R_2 &\leq&\log_2 \left( 1 + \dfrac{g_v^2}{g_u^2 P_u + N } \right)  \ ,
 \end{array}\right.
\end{equation}
where: 
\begin{IEEEeqnarray}{rCl} 
A_j &\triangleq& \dfrac{P_v}{P_u} \dfrac{h_{j,u}^2 P_u + N }{h_{j,u}^2 P_u + h_{j,v}^2 P_v + N} \ ,  \\
c_j &\triangleq& \dfrac{N}{h_{j,u}^2 P_u + N} \ , \\
\text{and} \qquad \alpha_j &\triangleq& P_u \dfrac{h_{j,u}h_{j,v}}{h_{j,u}^2 P_u + N } \ .
\end{IEEEeqnarray}
This, in part, requires solving the following optimization problem: 
\begin{equation}\label{MinOptimization} 
\max_{\alpha} \min_{j = 1,2} \log_2 \left( \dfrac{1}{A_j \alpha^2 -2 B_j \alpha + D_j } \right)\ , 
\end{equation}
where: 
\begin{IEEEeqnarray}{rCl} 
B_j &\triangleq& P_v \dfrac{h_{j,u}h_{j,v}}{h_{j,u}^2 P_u + h_{j,v}^2 P_v + N} \ ,  \\
 \text{and} \qquad D_j &\triangleq& \dfrac{h_{j,v}^2 P_v + N }{h_{j,u}^2 P_u + h_{j,v}^2 P_v + N} \ .
\end{IEEEeqnarray}
Finding the optimal DPC parameter $\alpha$ to use requires solving the equation: 
\begin{equation}\label{EquationProblem}
 (A_1 - A_2) \alpha^2 - 2 ( B_1 - B_2) \alpha + (D_1 - D_2) = 0 \ , 
\end{equation}
which yields: 
\begin{itemize}
\item a) If $A_1 = A_2 $ and $B_1 = B_2$ while $D_1 \neq D_2$, no solution exists, 
\item b) If $A_1 = A_2 $ and $B_1 = B_2$ and $D_1 = D_2$, every $\alpha$ is a solution, 
\item c) If $A_1 = A_2 $ and $B_1 \neq B_2$, then there exists only one solution: $\alpha_{opt} = \dfrac{D_1 - D_2}{2(B_1 - B_2)}$,
\item d) If $A_1 \neq A_2 $ and $ ( B_1 - B_2 ) ^2 = (A_1 - A_2 )(D_1 - D_2) $, then there exists only one solution: 
 $\alpha_{opt} = \dfrac{B_1 - B_2}{A_1 - A_2}$,
\item e) If $A_1 \neq A_2 $ and $ ( B_1 - B_2 ) ^2 < (A_1 - A_2 )(D_1 - D_2) $ no solution exists, 
\item f) $A_1 \neq A_2 $ and $ ( B_1 - B_2 ) ^2 > (A_1 - A_2 )(D_1 - D_2) $ then there exist two solutions.
\end{itemize}
Next, we can deduce the optimal values of \eqref{MinOptimization} to be used by the source as given in table \ref{TableCases}. 
\begin{table}
\caption{Optimal DPC parameter for the CD-DPC}
\begin{center}
 \begin{tabular}{| >{\centering\arraybackslash}m{2cm}| >{\centering\arraybackslash}m{5cm}|>{\centering\arraybackslash}m{4cm}|>{\centering\arraybackslash}m{3cm}|}
 \hline 
 $ $ & $B_1 = B_2$ & \includegraphics[scale=.5]{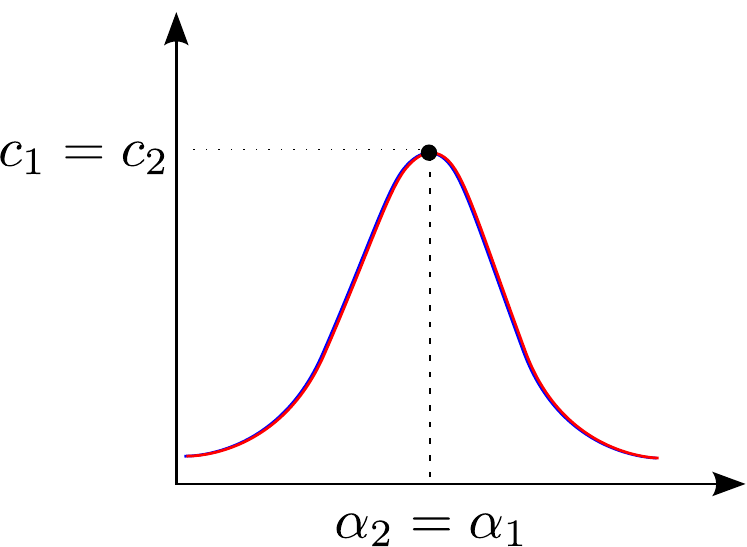} &   $ \alpha_1$  \\ 
 \cline{2-4}  
   $A_1= A_2 $&  $B_1 \neq B_2$  \newline{$ |c_1 - c_2| > A (\alpha_1 - \alpha_2)^2$} &  \includegraphics[scale=.5]{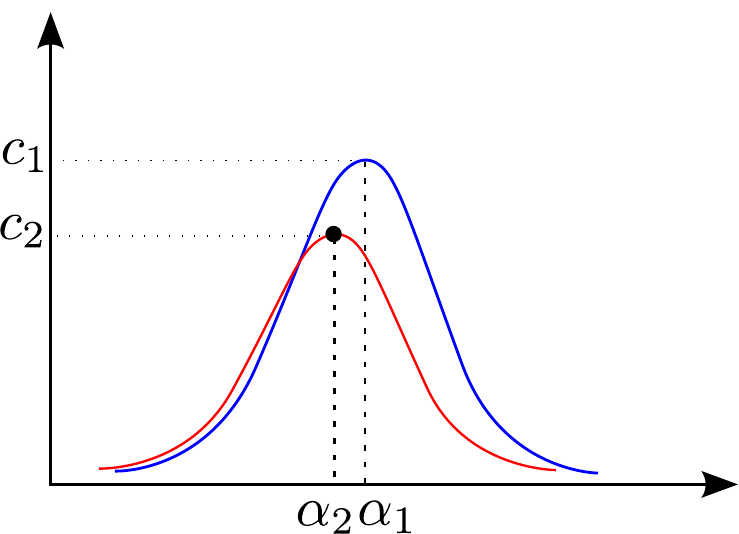} & $\alpha_2$ \\ 
 \cline{2-4} 
  $ $ & $ B_1 \neq B_2$  \newline{$ |c_1 - c_2| \leq A(\alpha_1 - \alpha_2)^2$} &   \includegraphics[scale=.5]{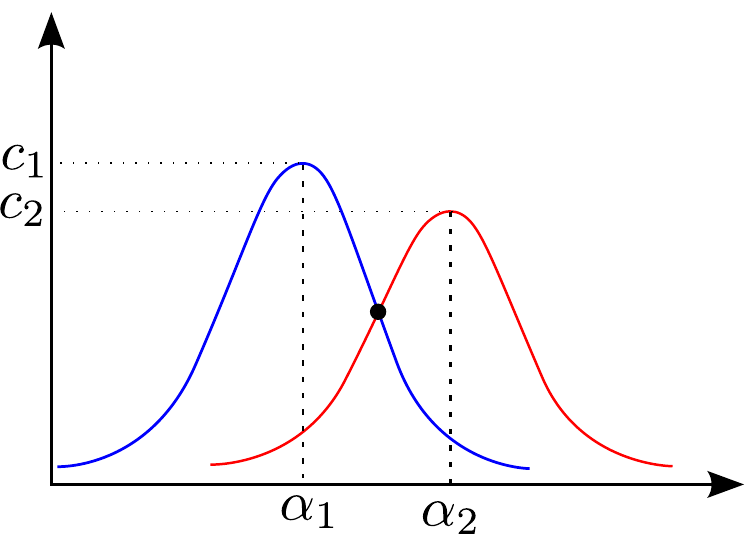} &  $\dfrac{D_1 - D_2}{2(B_1 - B_2)}$  \\ 
 \hline 
 $ $ & $\Delta < 0  $ &  \includegraphics[scale=.5]{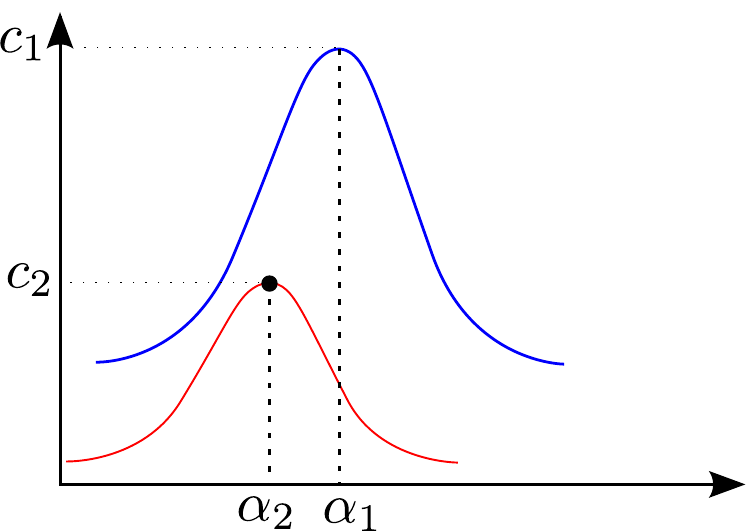}&  $ \alpha_2$  \\ 
 \cline{2-4}  
   &  $\Delta = 0  $ & \includegraphics[scale=.5]{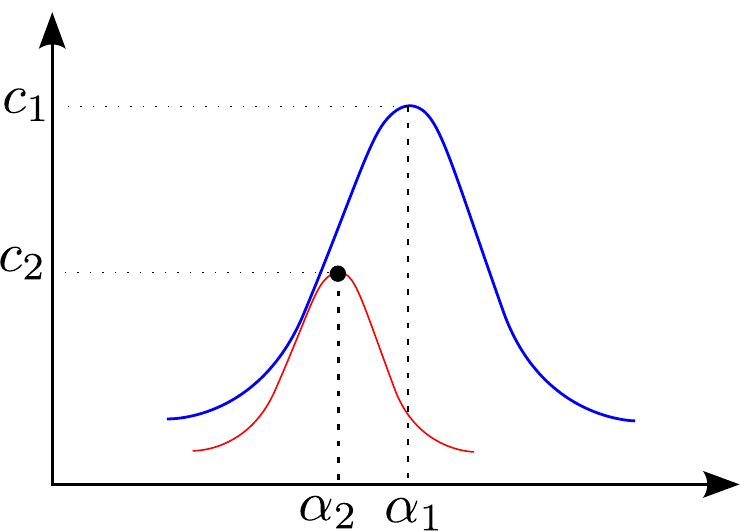}& $ \alpha_2$ \\ 
 \cline{2-4}  
$ A_1\neq A_2$ &  $\Delta > 0$  \newline{$|c_1 - c_2|\leq \min(A_1, A_2)(\alpha_1 - \alpha_2)^2$} & \includegraphics[scale=.5]{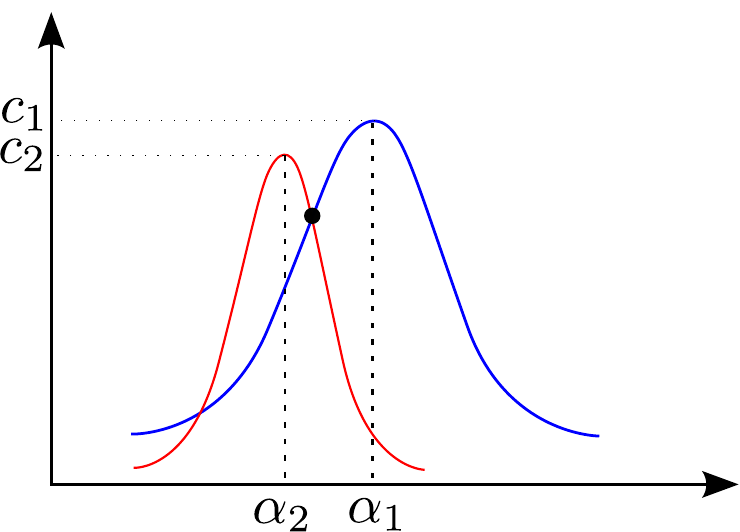}&  $\alpha_1 + \dfrac{A_2|\alpha_1- \alpha_2| - \sqrt{\Delta}}{A_1 - A_2}$  \\ 
 \cline{2-4}   
   &  $\Delta > 0$ \newline{$|c_1 - c_2| > \min(A_1, A_2)(\alpha_1 - \alpha_2)^2 $} & \includegraphics[scale=.5]{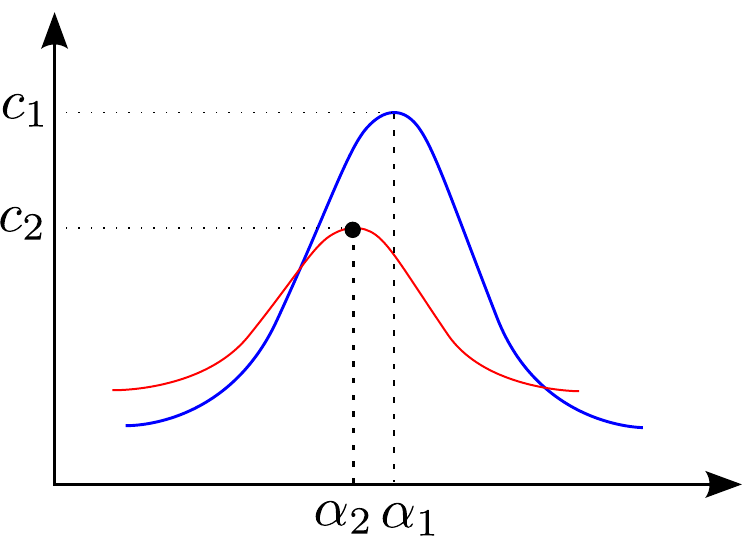}& $\alpha_2$ \\ 
 \hline 
 \end{tabular} \label{TableCases}
 \end{center}
\end{table}
Our aim is to show that, over all these cases, the optimal beamforming strategy is to let $\bB_u = \bh_{1,2}$. 

\subsection{Case of $A_1 = A_2 $ and $B_1 = B_2$}
We show in the following that if $A_1 = A_2 $ and $B_1 = B_2$, then it follows that $D_1= D_2$ and that $h_{j,u} = - h_{j,v} =\dfrac{1}{\sqrt{2}}$.
Thus letting $ \bB_u = \bh_{1,2}$ would yield the optimal solution given by: 
\begin{equation}
 \min_{\alpha} \max_{j = 1,2} A_j(\alpha - \alpha_j)^2 + c_j = c_1 = c_2 = \dfrac{2 N}{P_u + 2N} \ .
\end{equation} 
Hereafter, the details of the proof. 

First, note that since $A_1 = A_2$, then we can write that: 
\begin{equation}
h_{j,v}^2 = \dfrac{h_{j,v}^2 P_u + N}{P_u +2 N}  \ .
\end{equation}
As such, we can say that: $h_{1,v} h_{2,v} \neq 0$.
More over, all quantities write then as: 
 \begin{IEEEeqnarray}{rCl} 
A_1 &=& A_2 = A \triangleq \dfrac{P_v}{P_u} \dfrac{P_u + 2N}{P +2N} \ , \\
\alpha_j &=& \dfrac{P_u }{P_u +2N} \dfrac{h_{j,u}}{h_{j,v}} \ ,  \\
B_j &=& P_v \dfrac{h_{j,u}h_{j,v}}{h_{j,u}^2 P_u + h_{j,v}^2 P_v + N} = \dfrac{P_v}{P+2N} \dfrac{h_{j,u}}{h_{j,v}} \ , \\
D_j &=& \dfrac{h_{j,v}^2 P_v + N }{h_{j,u}^2 P_u + h_{j,v}^2 P_v + N} = \dfrac{h_{j,v}^2 P_v + N }{h_{j,v}^2 (P + 2N}   \ .
\end{IEEEeqnarray}
Now, we have that: 
\begin{IEEEeqnarray}{rCl}
B_1 = B_2 &\Leftrightarrow& P_v \left( \dfrac{h_{1,u}}{h_{1,v}} - \dfrac{h_{2,u}}{h_{2,v}} \right) = 0\ , \\
&\Leftrightarrow& P_v = 0 \quad \text{or} \quad \cos(\theta_u)\sin(\theta_v) - \cos(\theta_v)\sin(\theta_u) = 0 \ , \\ 
&\Leftrightarrow& P_v = 0 \quad \text{or} \quad \sin(\theta_u - \theta_v) = 0 \ ,\\ 
&\Leftrightarrow& P_v = 0 \quad \text{or} \quad \theta_u = \theta_v [\pi ] \ ,
\end{IEEEeqnarray}
but since $ \cos^2(\theta_v) = \dfrac{\cos^2(\theta_u)P_u +N}{P_u + 2N}$, then, one can write that: 
\begin{equation}
B_1 = B_2 \ \text{and} \ A_1 = A_2 \qquad \Rightarrow \qquad P_v = 0 \ \text{or} \ \cos^2(\theta_u) = \dfrac{1}{2} \ .
\end{equation} 
This implies then that: 
\begin{IEEEeqnarray}{rCl}
c_1 &=& c_2 = \dfrac{P_v + 2N}{P + 2N} \ , \\
 \alpha_1 &=& \alpha_2 = \pm \dfrac{P_u}{P_u + 2N}\ .
\end{IEEEeqnarray}
Thus in both cases of $P_v = 0 $ and $P_v \neq 0$, the optimal solution is given by: 
\begin{equation}
 \min_{\alpha} \max_{j = 1,2} A_j(\alpha - \alpha_j)^2 + c_j = c_1 = c_2 = \dfrac{2 N}{P_u + 2 N} \ .
\end{equation} 
Note that, since $\theta_u = \theta_v [\pi ]$, then $2 \theta_u = 2\theta_v [2\pi ]$, thus $s_u = s_v$.\\
Thus, as for the rate of user $2$, two cases unfold:
\begin{itemize}
\item Case of $s_u = s_v = 1$, which corresponds to $B_u = \bh_{1,2}$, and in this case $B_v$ is co-linear to $\bh_{1,2}$ and thus orthogonal to user 2's channel $\bg$ leading to a zero achievable rate:
 \begin{equation}
 R_2 = 0 \ .
 \end{equation}
 The power optimization of this point will yields the single capacity point $(C_1,0)$.
 \item Case of $s_u = s_v = -1$, which corresponds to $B_u \bot \bh_{1,2}$, and in this case $B_v $ is co-linear to user 2's channel $\bg$ leading to the achievability of all rate pairs satisfying:
 \begin{equation} \label{TimeSharingSuboptimal}
 \left\{\begin{array}{rcl}
 R_1 &\leq&  \dfrac{1}{2} \log_2 \left(  \dfrac{P_u + 2N}{2N}\right) \ , \\
 R_2 &\leq&  \dfrac{1}{2} \log_2 \left(  \dfrac{P + N}{P_u + N}\right) \ .
 \end{array}\right.
 \end{equation}
 The set of rate pairs obtained can be shown to perform worse than time sharing as is explained hereafter. 
 To show this, let $\alpha \in [0:1]$ such that: 
 \begin{equation}
 R_1 = \dfrac{1}{2} \log_2 \left(  \dfrac{P_u +2N}{2N}\right) =  \dfrac{\alpha }{2} \log_2 \left(  \dfrac{P +2N}{2N}\right) \ .
 \end{equation}
 We need to show that: 
 \begin{equation}
 R_2 = \dfrac{1}{2} \log_2 \left(  \dfrac{P + N}{P_u + N}\right) \leq \dfrac{ (1-\alpha)}{2}  \log_2\left(  \dfrac{P + N}{ N}\right)\ .
 \end{equation}
 To see this, note that: 
  \begin{IEEEeqnarray}{rCl} 
  \dfrac{P_u+2N}{2N}  = \dfrac{(P+2N)^ \alpha }{(2N)^ \alpha} &\Rightarrow & \dfrac{P_u}{N} + 1  = 2 \left( \dfrac{P}{2N} +1 \right)^\alpha - 1 \\
  &\Rightarrow & R_2 = \dfrac{1}{2} \log_2 \left( \dfrac{\frac{P}{N}+1}{2 \left( \frac{P}{2N} +1 \right)^\alpha - 1 } \right) \\
  &\overset{(a)}{\Rightarrow} & R_2 \leq \dfrac{1}{2}  (1- \alpha) \log_2\left(  \dfrac{P + N}{ N}\right) \ ,
 \end{IEEEeqnarray}
 where $(a)$ is a result of that the function: 
 \begin{equation}
 \begin{array}{rcl}
  [1:\infty[ &\mapsto & \mathds{R}\\
  x &\mapsto & 2 (x+1)^\alpha -1 - (2 x + 1)^\alpha
 \end{array}
 \end{equation}
by a quick function study, is positive. And thus: 
\begin{equation}
 2 \left( \frac{P}{2N} +1 \right)^\alpha - 1  \geq \left(\frac{P}{N} + 1 \right)^\alpha  \ , 
\end{equation}
which proves our claim.
\end{itemize}
To end the discussion of this case, it turns out that the optimal points obtained are the two single capacity points $(C_1,0)$ and $(0,C_2)$. 

\subsection{Case of $A_1 = A_2 = A$ and $B_1 \neq B_2$ and $|c_1 - c_2| \leq A (\alpha_1 - \alpha_2)^2$}
In this case, the optimal solution of the problem \eqref{EquationProblem} is given by: 
\begin{equation}
\alpha_{opt} = \dfrac{D_2-D_1}{2(B_2 - B_1)} = \dfrac{c_2 - c_1}{2(\alpha_2 - \alpha_1)} + \frac{1}{2} (\alpha_2 + \alpha_1) \ .
\end{equation}
Thus, the minimum value of the function to optimize in \eqref{OptimizatinProblem} is given by: 
\begin{equation}
 F_{opt} \triangleq \dfrac{(c_2-c_1)^2}{4 A (\alpha_2 - \alpha_1)^2} + \frac{1}{2} ( c_2 + c_1) + \frac{A}{4} (\alpha_2 - \alpha_1)^2 \ ,
\end{equation}
where as for previously: 
\begin{IEEEeqnarray}{rCl}
A &=& \dfrac{P_v}{P_u} \dfrac{P_u + 2N}{P+2N} \ ,\\
c_j &=& \dfrac{N}{P_u + 2N} \dfrac{1}{h_{j,v}^2} \ ,\\
\alpha_j &=& \dfrac{P_u}{P_u+2N}\dfrac{h_{j,u}}{h_{j,v}} \ . 
\end{IEEEeqnarray}
After some analytic manipulations we end up with the following expression of the optimal solution: 
\begin{equation}\label{Fopt}
 F_{opt} = \dfrac{1}{(P_u + 2N) \sin^2(2 \theta_v)} \left[ \dfrac{N^2 (P+ 2N)}{P_uP_v} \dfrac{\cos^2(2 \theta_v)}{\sin^2(\theta_u - \theta_v)} + \dfrac{P_u P_v}{P+ 2N} \sin^2(\theta_u - \theta_v) + 2N \right] \ ,
\end{equation}
Now, using the fact that: 
\begin{equation}
\cos^2(\theta_v) = \dfrac{\cos^2(\theta_u) P_u + N}{P_u + 2N}\ ,
\end{equation} 
we can write that: 
\begin{equation}\label{BoundX}
 \cos(2 \theta_v) = \dfrac{P_u}{P_u + 2N} \cos(2 \theta_u)\ ,
\end{equation}
which implies  
\begin{equation}
 \sin(2 \theta_u) = s_u \sqrt{1 - \cos^2(2 \theta_u) } =  s_u \sqrt{ 1 - \dfrac{(P_u + 2 N)^2}{P_u^2} \cos^2(2 \theta_v)} \ ,
\end{equation}
where we recall that:
\begin{equation}
s_{u} = \dfrac{\sin(2 \theta_u)}{|\sin(2 \theta_u) |} \ \text{and } \ s_v = \dfrac{\sin(2 \theta_v)}{|\sin(2 \theta_v)|} \ .
\end{equation}
In the sequel, we define the two variables: 
\begin{IEEEeqnarray}{rCl}
x &\triangleq& \cos^2(2 \theta_v) \ ,\\
a &\triangleq& \dfrac{(P_u+ 2N)^2}{P_u^2} \ . 
\end{IEEEeqnarray}
As defined, and recalling \eqref{BoundX}, we can conclude that $x$ lies in the set $\left[0:  \dfrac{1}{a}\right]$.
To further simplify \eqref{Fopt}, we need to express the following quantity: 
\begin{equation}
\sin^2(\theta_u - \theta_v) = \dfrac{1}{2 P_u} \left[ P_u \left( 1 - x - s_{u}s_v \sqrt{\left( 1- x\right) \left(1 - \dfrac{(P_u + 2N)^2}{P_u^2} x \right) }\right) - 2N x \right] \ .
\end{equation}
Letting then:
\begin{equation}
\begin{array}{rcl}
 g(x,s_u,s_v) &\triangleq& 2 P_u \sin^2(\theta_u - \theta_v) \\
 &=& P_u \left( 1 - x - s_{u}s_v \sqrt{\left( 1- x\right) \left(1 - \dfrac{(P_u + 2N)^2}{P_u^2} x \right) }\right) - 2N x \ ,
\end{array}
\end{equation}
one ends up with the following optimal function expressed in $x$, $s_u$ and $s_v$:
 \begin{equation}
 F_{opt} (x) = \dfrac{1}{ (1-x)} \left[ \dfrac{2 N^2 (P+ 2N)}{ P_v } \dfrac{x}{g(x,s_u,s_v)} + \dfrac{ P_v}{2(P+ 2N)} g(x,s_u,s_v) + 2N \right] \ .
\end{equation}  
Now, the rate of the second user is given by: 
\begin{equation}
 R_2 = \log_2 \dfrac{1}{2}  \left(1+ \dfrac{g_v^2 P_v}{g_u^2 P_u + 2N}\right)  \ .
 \end{equation}
 Then, we express: 
\begin{IEEEeqnarray}{rCl}
g_v^2 = \cos^2 \left(\theta_v + \frac{\pi}{4} \right)&=& \dfrac{1 - \sin\left( 2 \theta_v \right) }{2}    \\
&=& \dfrac{1 - s_v \sqrt{1-x}}{2} \ .
\end{IEEEeqnarray}
Similarly, we can show that: 
\begin{equation}
 g_u^2  =  \dfrac{1}{2}  - \dfrac{s_u}{2} \sqrt{1-\dfrac{(P_u+ 2N)^2}{P_u^2}x} \ ,
\end{equation}
Thus, the overall optimization problem is given as: 
\begin{equation} 
 \bigcup_{ (x,s_u,s_v) \in \cS} \left\{\begin{array}{rcl}
  R_1 &\leq&  \dfrac{1}{2}\log_2 \left( \dfrac{1}{F_{opt}(x,s_u,s_v) } \right) \ ,\\
  R_2 &\leq&  \dfrac{1}{2}\log_2 \left( 1 + \dfrac{ P_v(1 - s_v \sqrt{1-x} )}{ P_u \left(1 -s_u \sqrt{1- ax}\right)  + N } \right)  \ ,
\end{array}\right.
\end{equation}
where we define the optimization $\cS$ as: 
\begin{equation}
 \cS \triangleq \biggl\{  x \in \left[0:\frac{1}{a}\right] \ , \ (s_u , s_v)  \in \{-1,1\}^2 \ ,\   \text{s.t} \quad 
  s_u s_v = 1 \Rightarrow x\neq 0
 \biggr\}\  , 
\end{equation}
Hereafter, we study two distinct cases: $s_u s_v =-1 $ and $s_u s_v =1$. 


 \subsubsection{ $s_u s_v = + 1 $ } ~\\
 we show in the following section that this case is impossible because $s_u s_v = + 1 $ contradicts the existence of $x$ such that 
   $$|c_1 - c_2| \leq A (\alpha_1 - \alpha_2)^2 \ ,$$
%
%
\subsubsection{ $s_u s_v = - 1 $ } ~\\
 When $s_u s_v = -1$, we express the first derivative of the function $F_{opt}$ and show that it is always positive, leading us to the claim that $F_{opt}$ is strictly increasing. Thus, the rate of user $1$, $R_1$ is decreasing in $x$. 
 
 If $s_{u } = 1$ and $s_{v} =  -1$, then $R_2$ is easily shown to be decreasing in $x$, and thus, the optimal rate pair that is achievable is given by $x = 0$ leading thus to:
 \begin{equation}\label{OptimalCOrner}
\left\{ \begin{array}{rcl}
  R_1 &\leq& \dfrac{1}{2} \log_2\left( \dfrac{P + 2N}{P_v + 2N} \right) \ ,\\
  R_2 &\leq& \dfrac{1}{2} \log_2\left( \dfrac{P_v +  N}{ N} \right) \ .  
 \end{array} \right.
 \end{equation}
If $s_{u} = 1$ and $s_{v} = -1$, then we can show that $R_2$ can not be greater than: $\log_2\left( \dfrac{P_v + N}{N} \right)$, and thus, the achievable rate region is dominated by \eqref{OptimalCOrner}.
 
 To see this, note that: $R_2$ is strictly increasing in $x$ and thus, its maximum value is attained for $x = \frac{1}{a}$. Then, one can easily check that: 
 \begin{IEEEeqnarray}{rCl}
 R_2 &=& \dfrac{1}{2}  \log_2 \left( 1 + P_v \dfrac{1 - \sqrt{\frac{4N(P_u + N)}{P_u^2}}}{P_u + 2N }\right)\\
 &\leq&  \dfrac{1}{2}  \log_2 \left( 1 + P_v \dfrac{1}{P_u + 2N }\right) \\
& \leq&  \dfrac{1}{2}  \log_2 \left( 1 +  \dfrac{P_v}{N}\right) \ ,
 \end{IEEEeqnarray}
which proves our claim.

Thus, the overall obtained rate region for this case, does not outperform the second corner point of CD-DPC, which is already included in the first corner point. 

 \subsection{Case of $A_1 = A_2 = A$ and $B_1 \neq B_2$ and $|c_1 - c_2| > A (\alpha_1 - \alpha_2)^2$}
 In this case, we show that the obtained rate region does not outperform time sharing. 
 
\begin{figure}[h!]
 \centering
 \includegraphics[scale=0.8]{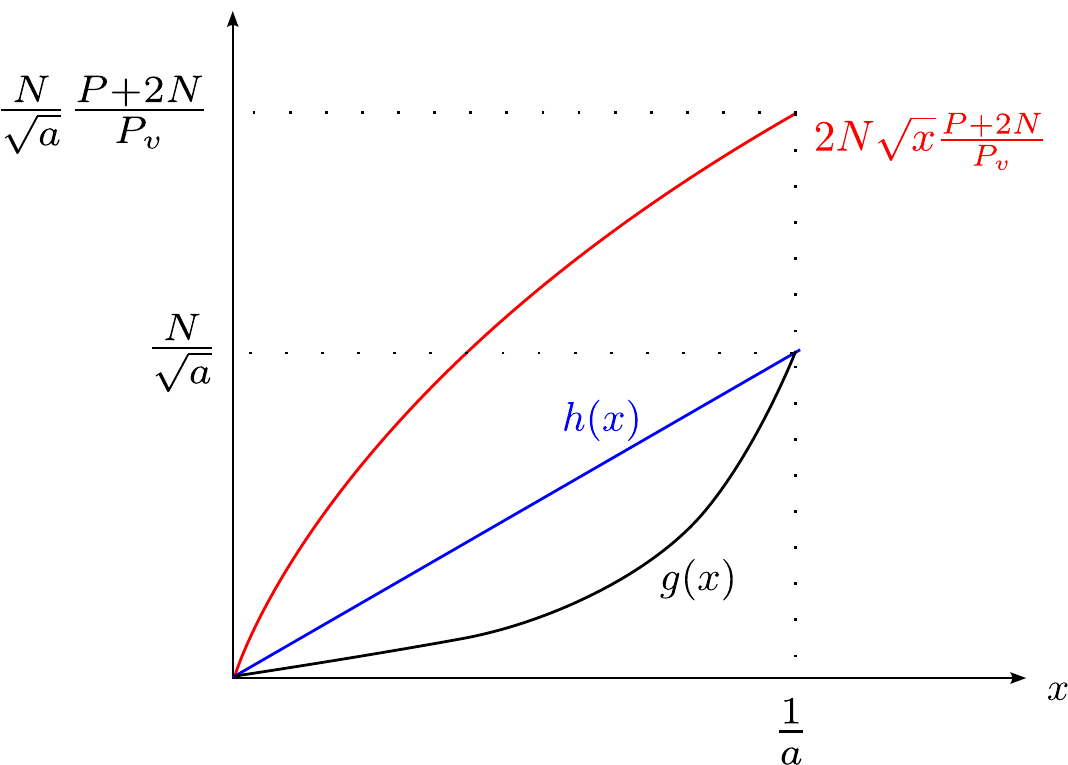} 
 \caption{Comparison of the functions $h$, $g$ and target upper bound.}
 \label{ImageSuSv} 
 \end{figure}

 To this end, we start by expressing the following: 
 \begin{equation}
  |c_1 - c_2|  >  A (\alpha_1 - \alpha_2)^2  
 \qquad \Leftrightarrow \qquad N |\cos(2 \theta_v)|  >  \dfrac{P_v P_u}{P+ 2N} \sin^2(\theta_u - \theta_v) \ .
 \end{equation}
 With the previous notations of the function $g$ and $x =\cos^2(2 \theta_v)$, we can rewrite this condition as: 
 \begin{equation}
 N \sqrt{x}  >  \dfrac{P_v  }{2(P+ 2N)}  g(x) \ .
 \end{equation}
 If $s_u s_v= 1$, then we can show easily that the above condition is always verified even with  $P_v \neq 0$ and $x\neq 0$. To see this note that for $x \in[0:1/a]$: 
 \begin{equation}
   g(x) = P_u (1 - x - \sqrt{(1-x)(1- a\star x)}) - 2N x \leq   P_u (1 - x  - 1 + a x ) - 2Nx \triangleq h(x) \ .
 \end{equation}
 Then, it is easy to show that for $x \in [0:1/a]$: 
 \begin{equation}
    N \sqrt{x} \geq h(x) \ ,
 \end{equation}
 since $h$ is linear and $h(0) = 0 $ and $h(1/a) = g(1/a) < N \sqrt{1/a}$. Fig. \ref{ImageSuSv} illustrates clearly our claim. 
 
  Thus, since the condition is always verified, the optimal solution is given by the rate pairs $(R_1, R_2) $ satisfying: 
  \begin{equation}
   \begin{array}{rcl}
   R_1 &\leq&  \dfrac{1}{2}\log_2 \left(  \dfrac{P_u + 2N}{2N} (1 - \sqrt{x})\right) \ , \\
   R_2 &\leq&   \dfrac{1}{2}  \log_2 \left( 1 + P_v \dfrac{1 - s_v\sqrt{1 -x }}{P_u( 1 - s_u\sqrt{1 -ax } )+ 2N }\right) \ .
   \end{array}
  \end{equation}
  If $s_u = s_v = -1$, then we show that the obtained rate region is included in the time sharing rate region. 
  To this end, we choose to show this claim on a larger rate region given by: 
     \begin{equation}
   \left\{\begin{array}{rcl}
   R_1 &\leq&  \dfrac{1}{2}\log_2 \left(  \dfrac{P_u + 2N}{2N} (1 - x)\right) \ , \\
   R_2 &\leq&   \dfrac{1}{2}  \log_2 \left( 1 + P_v \dfrac{1 - s_v\sqrt{1 -x }}{P_u( 1 - s_u\sqrt{1 -ax } )+ 2N }\right)\ .
   \end{array}\right.
  \end{equation}
  We proceed as follows to show that the obtained rate region for fixed $P_u, P_v $ and $N$, is concave. 
  
  Let $\alpha \in [0:1]$, such that: 
  \begin{equation}
    \dfrac{1}{2}\log_2 \left(  \dfrac{P_u + 2N}{2N} (1 - x)\right) \triangleq \alpha \log_2 \left( \dfrac{P_u + 2N}{2N} \right) +  (1 -\alpha)  \log_2 \left(  \dfrac{P_u + 2N}{2N} \left( 1 - \dfrac{1}{a} \right) \right)  \ .
   \end{equation} 
    Thus, we can show that: 
    \begin{equation}
   1 - x = \left( 1 - \dfrac{1}{a} \right) ^{1 - \alpha } \ .
\end{equation}
Thus, letting $ y \triangleq \left( 1 - \dfrac{1}{a} \right) $, the previous rate of user $2$ writes as: 
\begin{equation}
  R_2  \leq  \dfrac{1}{2}  \log_2 \left( 1 + P_v \sqrt{1-y}\dfrac{1  + \sqrt{ y ^{1 - \alpha}}}{P_u \left( 1 + \sqrt{ y^{1 - \alpha} - y } \right) }\right) \triangleq \dfrac{1}{2}  \log_2 \left( f(\alpha) \right) \ .
\end{equation}
Our aim is to show that $R_2$ is convex in $\alpha$, thus, we need to show that: 
\begin{equation} \label{SecondDeriv}
 f ^{\prime\prime} (\alpha) f(\alpha) - (f^\prime (\alpha) )^2 \geq 0  \ .
\end{equation}
We have that: 
\begin{equation}
 f^\prime (\alpha) = \dfrac{P_v   \log(y) \sqrt{1 - y} }{2 P_u \left( 1 + \sqrt{  y^{1 - \alpha} - y  }   \right)^2}   \left( - \sqrt{y^{1-\alpha}}+ \dfrac{ \sqrt{y^{1- \alpha}} + y }{ \sqrt{1 - y^{\alpha}}}\right)  \ .
\end{equation}
It is easy to see that thus  $R_2$ is decreasing in $\alpha $ since $ \log(y)\leq 0$.

As for the second derivative, one can show that it writes as: 
\begin{equation}
  f^{\prime\prime} (\alpha) = \dfrac{P_v  \log^2(y)\sqrt{1-y}}{4  P_u(1 - y^\alpha)\sqrt{1 - y^\alpha}} \dfrac{1}{\left(1 + \sqrt{  y^{1 - \alpha}  - y  }  \right)^3}   G(\alpha) \ ,
\end{equation}
where $G(\alpha)$ is given by:
\begin{IEEEeqnarray}{rCl}
 G(\alpha) &=& 2\sqrt {y ^{1-\alpha} - y} \left( \sqrt{y ^{1-\alpha} }(1 - \sqrt{1 - y^\alpha}) +y \right) \nonumber\\
 && \qquad + (1+ \sqrt {y ^{1-\alpha} - y}) \left( (1 - y^\alpha )  \sqrt{y^{1 - \alpha}}( \sqrt{1 - y^\alpha} -1) + y^\alpha (y + \sqrt{y^{1- \alpha}})\right) \ .\,\,\,\,\,\,\,\,
\end{IEEEeqnarray}
Showing that $R_2$ is convex in $\alpha$, i.e showing that \eqref{SecondDeriv} holds, amounts then to showing that: 
\begin{equation}
 \dfrac{P_v(1 + \sqrt{y^{1-\alpha }})\sqrt{1-y} + P_u (1 + \sqrt{y^{1-\alpha} -y} )}{\sqrt{1 - y^\alpha}\sqrt{1- y}}  G(\alpha) \geq P_v \left( \sqrt{y ^{1-\alpha} }(1 - \sqrt{1 - y^\alpha}) +y \right)^2 \ .
\end{equation}
We show the stronger result that consists in: 
\begin{equation}
 \dfrac{  G(\alpha) }{\sqrt{1 - y^\alpha} } \geq \left( \sqrt{y ^{1-\alpha} }(1 - \sqrt{1 - y^\alpha}) +y \right)^2 \ ,
\end{equation}
which would yield the desired inequality. 

  Note here that since: 
  \begin{IEEEeqnarray}{rCl}
 (1 - y^\alpha )  \sqrt{y^{1 - \alpha}}( \sqrt{1 - y^\alpha} -1) &+& y^\alpha (y + \sqrt{y^{1- \alpha}}) \nonumber \\
  &\geq&   \sqrt{y^{1 - \alpha}} (\sqrt{1 - y^\alpha} -1) + y^\alpha (y + \sqrt{y^{1- \alpha}})\\
  &\geq&  \sqrt{y^{1 - \alpha}} (1 - y^\alpha  -1 )+  y^\alpha (y + \sqrt{y^{1- \alpha}}) \\
  &\geq& 0 \ ,
  \end{IEEEeqnarray}
  then, 
  \begin{equation}
   G(\alpha) \geq 2\sqrt {y ^{1-\alpha} - y} \left( \sqrt{y ^{1-\alpha} }(1 - \sqrt{1 - y^\alpha}) +y \right) \ .
  \end{equation}
  Hence ,we can write: 
   \begin{IEEEeqnarray}{rCl}
 G(\alpha)  &-& \sqrt{1 - y^\alpha} \left( \sqrt{y ^{1-\alpha} }(1 - \sqrt{1 - y^\alpha}) +y \right)^2 \nonumber \\
 &\geq& \sqrt{1 - y^\alpha} \left( \sqrt{y ^{1-\alpha} }(1 - \sqrt{1 - y^\alpha}) +y \right)   
 \left( \sqrt{y ^{1-\alpha} }(1 + \sqrt{1 - y^\alpha}) +y \right)   \\
 &\geq& 0 \ ,
  \end{IEEEeqnarray} 
  this ends thus the proof. 
  Thus, $R_2$ being convex in $\alpha$ and $R_1$ linear in $\alpha$ the trajectory of $R_2(R_1)$ describes then a concave rate region. 
   
    If $s_u = s_v = 1$, we show that the obtained rate region is included in a suboptimal rate region compared to time sharing, studied earlier and given by: 
    \begin{equation}
     \left\{\begin{array}{rcl}
     R_1 &\leq& \dfrac{1}{2} \log_2 \left( \dfrac{P_u + 2N}{2N	} \right) \ , \\
     R_2 &\leq& \dfrac{1}{2} \log_2 \left( \dfrac{P +  N}{P_u +N} \right) \ .
     \end{array}\right.
    \end{equation}
    To show this, note that the bound on $R_1$ is trivial. However, concerning the second user's rate, note that as it writes: 
    \begin{equation}
     R_2 = \dfrac{1}{2}\log_2 \left( 1 +  P_v  \dfrac{1 - \sqrt{1-x}}{P_u (1- \sqrt{1-ax}) +2N}\right) \triangleq \log_2 \left( 1 +  g(x)\right) \ ,
    \end{equation}
    $R_2$ is not always strictly monotonic. The sign of its first derivative in $x$ is given by the sign of $g^\prime$: 
    \begin{equation} 
      (P_u + 2N)\sqrt{1 - a x} + P_u (a - 1 - a\sqrt{1-x})  \ ,
    \end{equation}
    that depends on the respective values of $P_u$ and $N$.
    If there exists any point for which the first derivative is null $x_{opt}$, then it will imply that: 
    \begin{IEEEeqnarray}{rcl}
     P_u \left( 1 - \sqrt{1 - ax}\right) +2N &=& (P_u + 2N) \left(1 - \sqrt{1-x} + 1 - \dfrac{1}{a}\right)\\
     &\geq& (P_u + 2N) \left(1 - \sqrt{1-x}  \right) \geq 0 \ .
    \end{IEEEeqnarray}
    Then, we can conclude that: 
    \begin{equation}
     R_2 \leq  \dfrac{1}{2}\log_2 \left( 1 +  \dfrac{P_v}{P_u  +2N}\right)   \ .
    \end{equation}
    It no such point exists such that $g^\prime(x)= 0$, then $R_2$ is increasing in $x$ and thus, the maximum value is obtained for $x = 1/a$, which clearly yields to the desired bound on $R_2$.
    
  Now, if $s_u s_v = -1$, then two cases unfold following the signs of $s_u$ and $s_v$. In both cases, the obtained rate region is shown not to outperform the optimal rate region we claim.
  If $s_u = -1$ and $s_v = 1$, then, we can show that the rate region: 
  \begin{equation}
   \left\{  \begin{array}{rcl}
     R_1 &\leq&  \dfrac{1}{2}\log_2 \left(   \dfrac{P_u+ 2N}{2N} (1 - \sqrt{x})\right)  \leq    \dfrac{1}{2} \log_2\left(   \dfrac{P_u+ 2N}{2N} \right)  \ ,\\
     R_2 &\leq&  \dfrac{1}{2}\log_2 \left( 1 +  P_v  \dfrac{1 - \sqrt{1-x}}{P_u (1+ \sqrt{1-ax}) +2N}\right)  \leq  \dfrac{1}{2}\log_2  \left(  \dfrac{P +2N}{P_u +2N}\right) \ ,
     \end{array}\right.
   \end{equation} 
   is included in the sub-optimal rate region given by \eqref{TimeSharingSuboptimal} which in turn does not outperform time sharing. 
   
   On the other side, if $s_u = 1$ and $s_v=-1$, then we will show that the second corner point of CD-DPC inner bound    contains the obtained rate region. In this case, it can be easily shown that both $R_1$ and $R_2$ are decreasing in $x$. However, not all values of $x $ are admissible due to the constraint. 
     
\begin{figure}[h!]
 \centering
 \includegraphics[scale=0.8]{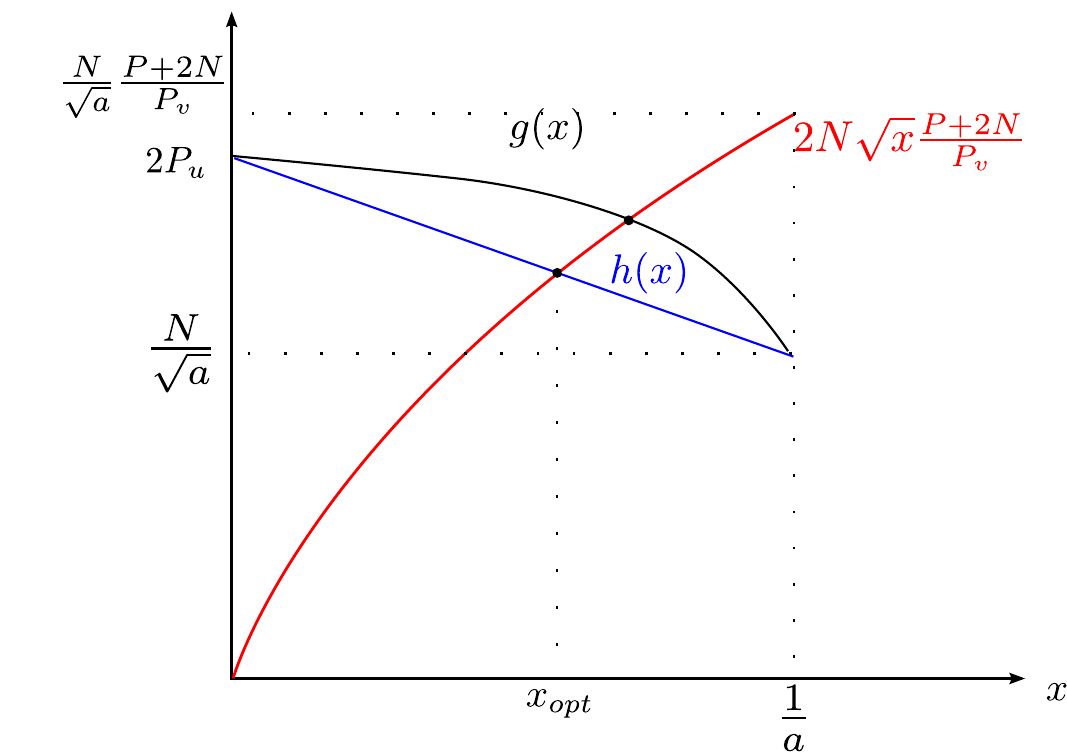} 
 \caption{Comparison of the functions $h$, $g$ and target upper bound.}
 \label{ImageSuSv-1} 
 \end{figure}

  Let us start by characterizing the set of values such that: 
    \begin{equation}
 N \sqrt{x}  >  \dfrac{P_v  }{2(P+ 2N)}  \left(P_u (1 - x +\sqrt{(1-x)(1- ax})) - 2Nx \right) \ .
 \end{equation}
 As done previously, we will solve only the simpler problem that yields a larger solution set and that is illustrated in Fig. \ref{ImageSuSv-1}:  
\begin{equation} \label{SimpleOptimization}
 N \sqrt{x}  >  \dfrac{P_v  }{2(P+ 2N)}  \left(P_u (1 - x +  1- ax  ) - 2Nx \right) \ .
\end{equation}
  Solving this problem yields the following value of the infinimum of all admissible beam directions: 
  \begin{IEEEeqnarray}{rCl}
   x_{opt} &=& \dfrac{P_u^2}{2 (P_u + N)^2 (P_u+ 2N)^2} \biggl[ 2(P_u+N )(P_u + 2N) + \dfrac{N^2(P+2N)^2}{P_v^2} \nonumber\\
   && \qquad \qquad \qquad \qquad  -\dfrac{N (P+2N) }{P_v } \sqrt{ \dfrac{N^2(P+2N)^2}{P_v^2}+ 4(P_u + N)(P_u + 2N)} \biggr] \ .\,\,\,\,\,\,\,\,
  \end{IEEEeqnarray}
  Since the solution of problem \eqref{SimpleOptimization} yields a smaller value of the inf of admissible solutions, the resulting rate region is wider. However, we show that it still remains included in the second corner point of MD-DPC inner bound given by 
  \begin{equation}
    \left\{\begin{array}{rcl}
    R_1 &\leq& \dfrac{1}{2} \log_2 \left(  \dfrac{P + 2N}{P_u + 2N}\right) \ ,  \\
    R_2 &\leq& \dfrac{1}{2} \log_2 \left(  \dfrac{P_v +  N}{ N}\right)   \ .
    \end{array} \right. 
   \end{equation} 
   The bound on the rate $R_2$ is quite trivial and requires no further proof. However, the bound on rate $R_1$ requires showing that: 
   \begin{equation}
    x_{opt} \geq \dfrac{P_u^2P_v^2}{(P_u +2N)^2(P_v + 2N)^2} \ ,
   \end{equation}
   which can be shown through evolved bounding techniques. 
 As a conclusion for these cases, no rate region outperforms the second corner point of CD-DPC inner bound. 
  
\subsection{case of $A_1 \neq A_2$ and $(B_1 - B_2)^2= (A_1 - A_2)(D_1 - D_2)$}
In this case, we start by showing that the above condition imposes: 
\begin{equation}
 \theta_u = \theta_v [\pi] \ .
\end{equation}
Let us first quickly denote: 
  \begin{IEEEeqnarray}{rCl}
 K_{Y_1} &\triangleq&  \cos^2(\theta_u) P_u + \cos^2(\theta_v) P_v + N \ , \\
  K_{Y_2} &\triangleq&  \sin^2(\theta_u) P_u + \sin^2(\theta_v) P_v + N \ .
  \end{IEEEeqnarray}
  Next, recall that: 
  \begin{IEEEeqnarray}{rCl}
  B_1 &=& \dfrac{P_v \cos(\theta_u) \cos(\theta_v)}{ \cos^2(\theta_u) P_u + \cos^2(\theta_v) P_v + N} =  \dfrac{P_v \cos(\theta_u) \cos(\theta_v)}{   K_{Y_1} } \ , \\
  B_2 &=& \dfrac{P_v \sin(\theta_u) \sin(\theta_v)}{ \sin^2(\theta_u) P_u + \sin^2(\theta_v) P_v + N} = \dfrac{P_v \sin(\theta_u) \sin(\theta_v)}{ K_{Y_2}} \ , \\
  A_1 &=&  \dfrac{P_v}{P_u}\dfrac{ \cos^2(\theta_u) P_u + N}{K_{Y_1}} = \dfrac{P_v}{P_u} \left(  1 - \dfrac{ \cos^2(\theta_v) P_v  }{K_{Y_1}}\right)   \ ,\\  
  A_2 &=&  \dfrac{P_v}{P_u}\dfrac{ \sin^2(\theta_u) P_u + N}{K_{Y_2}} = \dfrac{P_v}{P_u} \left(  1 - \dfrac{ \sin^2(\theta_v) P_v  }{K_{Y_2}}\right) \ , \\  
  D_1 &=& \dfrac{ \cos^2(\theta_v) P_v + N}{K_{Y_1}}  =  \left(  1 - \dfrac{ \cos^2(\theta_u) P_u  }{K_{Y_1}}\right)  \ ,\\
  D_2 &=&  \dfrac{ \sin^2(\theta_v ) P_v + N}{K_{Y_2}}  =  \left(  1 - \dfrac{ \sin^2(\theta_u) P_u  }{K_{Y_2}}\right)    \ .
  \end{IEEEeqnarray}
  And that $A_1 \neq A_2 \Rightarrow P_v \neq 0$. Thus,   
   \begin{IEEEeqnarray}{rCl}
   && (B_1 - B_2)^2= (A_1 - A_2)(D_1 - D_2) \nonumber \\  
   &\Leftrightarrow& \left( \dfrac{ \cos(\theta_u) \cos(\theta_v)}{   K_{Y_1} } - \dfrac{  \sin(\theta_u) \sin(\theta_v)}{ K_{Y_2}} \right)^2 \nonumber \\
   &=& \left(    \dfrac{ \sin^2(\theta_v) }{K_{Y_2}} - \dfrac{ \cos^2(\theta_v) }{K_{Y_1}}\right)  \left(  \dfrac{ \sin^2(\theta_u)   }{K_{Y_2}} -\dfrac{ \cos^2(\theta_u)  }{K_{Y_1}}\right) \quad  \\
   &\Leftrightarrow& K_{Y_1} K_{Y_2} \biggl( \sin(\theta_v)\cos(\theta_u ) - \sin(\theta_u)\cos(\theta_v ) \biggr)^2 = 0 \\
   &\Leftrightarrow& K_{Y_1} K_{Y_2}  \sin^2(\theta_u -\theta_v )  = 0 \\
   &\Leftrightarrow& \theta_u = \theta_v [\pi] \ .
   \end{IEEEeqnarray}
  
  The optimal solution of the system \eqref{MinOptimization}, is then given by: 
 \begin{IEEEeqnarray}{rCl}
 R_1 &=& \log_2 \left( \dfrac{ \min(\sin^2(\theta_v), \sin^2(\theta_v))P_u + N }{ N} \right) \\
 &=& \log_2 \left( \dfrac{ (1 - |\cos(2 \theta_v)| )P_u + 2 N }{ 2N} \right) \\
  &\leq&  \log_2 \left( \dfrac{  (1 - \sqrt{\cos^2(2 \theta_v) } ) P_u + 2 N }{ 2N} \right)\ ,
 \end{IEEEeqnarray}
 define then: 
 \begin{equation}
 x \triangleq \cos^2(2 \theta_v) \ .
 \end{equation}
 On the other side, note that: 
  \begin{IEEEeqnarray}{rCl}
 R_2 &=& \log_2 \left( \dfrac{  \cos^2(\theta_v + \pi/4)   P+   N }{ \cos^2(\theta_v + \pi/4) P_u +   N} \right) \\
 &=& \log_2 \left( \dfrac{ \sin( 2 \theta_v) P  + 2 N }{ \sin( 2 \theta_v) P_u+ 2N} \right) \\
  &\leq&  \log_2 \left( \dfrac{  (1 - s_v \sqrt{1 - x }) P  + 2 N }{ (1 - s_v \sqrt{1 - x }) P_u + 2N} \right) \ ,
 \end{IEEEeqnarray}
 if $s_v = -1$, then $R_2$ is decreasing in $x$, and thus the optimal value is given by:
 \begin{equation}
   R_2 =   \log_2  \left( \dfrac{   P  + 2 N }{   P_u + 2N}\right) \ .
 \end{equation} 
And since: 
\begin{equation}
R_1  \leq \log_2 \left( \dfrac{   P_u + 2 N }{ 2N} \right) \ ,
\end{equation}  
then the obtained region is included in the set of rate pairs such that: 
 \begin{equation}
 \left\{ \begin{array}{rcl}
 R_1  &\leq&  \log_2 \left( \dfrac{P_u + 2N}{2N} \right)  \ ,\\
 R_2  &\leq&  \log_2  \left( \dfrac{P+2N}{P_u + 2N}\right)\ .
  \end{array} \right.
 \end{equation} 
 which was shown to perform less than time sharing. 
 Now, if  $s_v = 1$, then $R_2$ is increasing in $x$ where $x\in [0:1] $, and hence, the maximum is obtained for $x=1$, which yields the same rate of user $2$. For similar reasons, the obtained rate pair does not outperform time sharing. 
%

Thus, the overall rate region obtained in this case, is included in mere time sharing. 
   \subsection{case of $A_1 \neq A_2$ and $(B_1 - B_2)^2 < (A_1 - A_2)(D_1 - D_2)$}
 Since we have that: 
 \begin{equation}
 (B_1 - B_2)^2 - (A_1 - A_2)(D_1 - D_2) = \dfrac{Pv^2}{K_{Y_1} K_{Y_2}}  \biggl( \sin(\theta_v -\theta_u)  \biggr)^2 \ ,
\end{equation}	  
 having $(B_1 - B_2)^2 < (A_1 - A_2)(D_1 - D_2)$ is impossible. 
  
  \subsection{case of $A_1 \neq A_2$ , $(B_1 - B_2)^2 > (A_1 - A_2)(D_1 - D_2)$ and $|c_1 - c_2| \leq \min(A_1, A_2) (\alpha_1 - \alpha_2)^2$}
  In this case, we can show that, the two possible optimum solutions are obtained for $\theta_u = \pi/4$ or $\theta_u = -\pi/4$.
  
  The case where $\theta_u = \pi/4$ is the claimed optimal rate region. 
  As for the case where $\theta_u = -\pi/4$, the resulting rate region consists of all rate pairs satisfying: 
  \begin{equation}
  \left\{ \begin{array}{rcl}
   R_1 &\leq& \dfrac{1}{2} \log_2 \left(  \dfrac{P_u + 2N}{P_u P_v\dfrac{1+y}{P + 2N + \sqrt{ P + 2N + (y^2-1)P_v}} + 2N}\right)\ , \\
   R_2 &\leq& \dfrac{1}{2} \log_2 \left(  1 + P_v \dfrac{1-y}{2(P_u + N)} \right) \ .
   \end{array}\right.
  \end{equation}
Note, that in this case, the maximum is achieved at both rates letting $y = -1$, thus, the resulting rate region can not outperform the rate region given by: 
 \begin{equation}
  \left\{ \begin{array}{rcl}
   R_1 &\leq& \dfrac{1}{2} \log_2 \left(  \dfrac{P_u + 2N}{ 2N}\right)  \ ,\\
   R_2 &\leq& \dfrac{1}{2} \log_2 \left(   \dfrac{ P +  N}{P_u +  N} \right)  \ ,
   \end{array}\right.
  \end{equation}
  which was clearly shown not to outperform time-sharing. 
 
 \subsection{case of $A_1 \neq A_2$ , $(B_1 - B_2)^2 > (A_1 - A_2)(D_1 - D_2)$ and $|c_1 - c_2| > \min(A_1, A_2) (\alpha_1 - \alpha_2)^2$} 
 
 In this peculiar last case, we show that the obtained rate region can not exceed time sharing neither.  In this case, the resulting rate region writes as: 
 \begin{equation}
 \left\{\begin{array}{rcl}
 R_1 &\leq& \dfrac{1}{2} \log_2 \left( \dfrac{(1 - |\cos(2 \theta_u)|)P_u + 2N }{2N } \right) \\
 R_2 &\leq& \dfrac{1}{2} \log_2 \left( 1 + P_v \dfrac{(1 - s_v \sqrt{1-\cos^2(2 \theta_v)})}{ P_u (1 - s_u \sqrt{1 - cos^2(2 \theta_u)})+ 2N}\right) 
 \end{array}\right.
 \end{equation}
 
 The case where $s_u = -1$ , then we can show resorting to the same tools used on the analysis of the concavity in the previous sections that the obtained rate region when  $ \cos^2(2 \theta_u)$spans the interval $[0:1]$, is concave for every value of $ \cos^2(2 \theta_v)$, thus, the resulting union can be at most concave. When $s_u = 1$, two cases unfold and $R_1$ and $R_2$ are both decreasing in $  \cos^2(2 \theta_u)$,  and for a fixed beam direction $\bB_v$, the optimal rate pair is given by: 
 \begin{equation}
 \left\{\begin{array}{rcl}
 R_1 &\leq& \dfrac{1}{2} \log_2 \left( \dfrac{ P_u + 2N }{2N } \right)\ , \\
 R_2 &\leq& \dfrac{1}{2} \log_2 \left( 1 + P_v \dfrac{(1 - s_v \sqrt{1-\cos^2(2 \theta_v)})}{2N}\right) \ .
 \end{array}\right.
 \end{equation}

 \section{Proof of Achievability of $\cR_{3-ARV}$}\label{Proof3ARV}
We fix a pmf $p_{QU_1U_2VX}$. Let $R_0, R_1, R_2$ denote the message rates and $T_{1,2}, T_{1,1} $ and $T_2$ denote the binning rates. Generate $2^{n\,R_0}$ sequences $q^n(w_0)$, $w_0 \in [1:2^{n\,R_0}]$ each following the pmf: $\prod^n_{i=1} P_{Q} (q_i(w_0))$. For each $w_0$, generate $2^{n\,T_2}$ sequences $v^n(l_2, w_0)$ following the pmf: $\prod^n_{i=1} P_{V|Q} (v_i(l_2, w_0)|q_i(w_0))$ and map them randomly in $2^{n\,R_2}$ bins $B^n(w_2, w_0)$. Generate similarly $2^{n\,T_{1,1}}$ sequences $u_1^n(l_{1,1}, w_0)$ and map them randomly in $2^{n\,R_1}$ bins $B^n_1(w_1,w_0)$ and $2^{n\,T_{1,2}}$ sequences $u_2^n(l_{1,2}, w_0)$ and map them in a distinct set of $2^{n\,R_1}$ bins $B^n_2(w_1,w_0)$. \\
\emph{Encoding:} for each message triple $(w_0, w_1, w_2)$ to be transmitted, find in the product of all bins $B^n(w_i,w_0)$, a triple of sequences $ u_1^n(l_{1,1}, w_0),u_2^n(l_{1,2}, w_0), v^n(l_1, w_0)$ such that: 
$$
 \Big( q^n(w_0), u_1^n(l_{1,1}, w_0),u_2^n(l_{1,2}, w_0), v^n(l_2, w_0) \Big) \in \typ{QU_1U_2V}\,.
$$
Send then a random mapping sequence: $x^n(w_0,l_{1,1}, l_{1,2} , l_2 )$. The encoding is error free if all inequalities in $\dsT$ are verified.\\
\emph{Decoding:} Each receiver decodes its intended messages $(w_0, w_j)$ by decoding the index $l_j$ and non-uniquely the common message, yielding the constraints stated in $\cM$.  
 
\bibliographystyle{IEEEtran}
\bibliography{FirstBibJabRef}
\end{document}